\documentclass[preprints,article,accept,moreauthors,LaTeX,dvi2pdf,10pt,a4paper]{Definitions/mdpi}

\firstpage{1} 
\pubvolume{xx}
\issuenum{1}
\articlenumber{5}
\pubyear{2020}
\copyrightyear{2020}
\externaleditor{}
\history{}

\pdfoutput=1

\usepackage{color}

\usepackage{slashed}
\usepackage{textcomp}
\usepackage{mathrsfs} 
\usepackage{amssymb}

\Title{Uniqueness Criteria for the Fock Quantization of Dirac Fields and Applications in Hybrid Loop Quantum Cosmology}

\Author{Jer\'onimo Cortez $^{1,\dagger}$, Beatriz Elizaga Navascu\'es $^{2,}$*$^{\dagger}$, Guillermo A. Mena Marug\'an $^{3,\ddagger,\dagger}$, Santiago Prado $^{3,\S,\dagger}$, and Jos\'e  M. Velhinho $^{4,\dagger}$}

\AuthorNames{Firstname Lastname, Firstname Lastname, Firstname Lastname, Firstname Lastname and Firstname Lastname}

\address{%
	$^{1}$ \quad Departamento de F\'isica, Facultad de Ciencias, Universidad Nacional Aut\'onoma de M\'exico, Ciudad de M\'exico~04510, Mexico; jacq@ciencias.unam.mx\\
	$^{2}$ \quad Institute for Quantum Gravity, Friedrich-Alexander University Erlangen-N{\"u}rnberg, Staudstra{\ss}e 7, 91058~Erlangen, Germany\\
	$^{3}$ \quad Instituto de Estructura de la Materia, IEM-CSIC, Serrano 121, 28006~Madrid, Spain\\
	$^{4}$ \quad Faculdade de Ci\^encias, Universidade da Beira Interior, R. Marqu\^es D'\'Avila e Bolama, 6201-001~Covilh\~a, Portugal; jvelhi@ubi.pt}

\corres{Correspondence: beatriz.b.elizaga@fau.de}

\firstnote{These authors contributed equally to this work.} 
\secondnote{E-mail: mena@iem.cfmac.csic.es}
\thirdnote{E-mail: santiago.prado@iem.cfmac.csic.es}

\abstract{In generic curved spacetimes, the unavailability of a natural choice of vacuum state introduces a serious ambiguity in the Fock quantization of fields. In this review, we study the case of fermions described by a Dirac field in non-stationary spacetimes, and present recent results obtained by us and our collaborators about well-motivated criteria capable to ensure the uniqueness in the selection of a vacuum up to unitary transformations, at least in certain situations of interest in cosmology. These criteria are based on two reasonable requirements. First, the invariance of the vacuum under the symmetries of the Dirac equations in the considered spacetime. These symmetries include the spatial isometries. Second, the unitary implementability of the Heisenberg dynamics of the annihilation and creation operators when the curved spacetime is treated as a fixed background. This last requirement not only permits the uniqueness of the Fock quantization but, remarkably, it also allows us to determine an essentially unique splitting between the phase space variables assigned to the background and the fermionic annihilation and creation variables. We first consider Dirac fields in 2+1 dimensions and then discuss the more relevant case of 3+1 dimensions, particularizing the analysis to cosmological spacetimes with spatial sections of spherical or toroidal topology. We use this analysis to investigate the combined, hybrid quantization of the Dirac field and a flat homogeneous and isotropic	background cosmology when the latter is treated as a quantum entity, and the former as a perturbation. Specifically, we focus our study on a background quantization along the lines of loop quantum cosmology. Among the Fock quantizations for the fermionic perturbations admissible according to our criteria, we discuss the possibility of further restricting the choice of a vacuum by the requisite of a finite fermionic backreaction and, moreover, by the diagonalization of the fermionic contribution to the total Hamiltonian in the asymptotic limit of large wave numbers of the Dirac modes. Finally, we argue in support of the uniquess of the vacuum state selected by the extension of this diagonalization condition beyond the commented asymptotic region, in particular proving that it picks out the standard Poincar\'e and Bunch-Davies vacua for fixed flat and de Sitter background spacetimes, respectively.}



\begin{document}

\section{Introduction}
\label{Intro}

Quantum Field Theory (QFT), namely the description of fields according to quantum rules, is one of the pillars of Modern Physics (see e.g. Refs. \cite{qft,qft2}). In this description, it is common to use a Fock formalism in which the physical processes are formulated  in terms of creation and annihilation of field excitations around a vacuum state. Typically, these excitations are interpreted as particles (or antiparticles) of the field. This kind of description has been adopted successfully for all fundamental physical interactions, except for gravity: a  fully satisfactory quantum field formulation of General Relativity, or more generally of the gravitational interaction, still remains as a defiant challenge. 

In standard QFT, the particle excitations develop in flat spacetime. The symmetries of this background spacetime (regarded as classical) are employed in the selection of a natural state for the Fock representation of the field: Poincar\'e invariance fixes a unique vacuum in Minkowski spacetime \cite{qftwald}.  However, it is well known that the generalization of the Fock quantization techniques to field theories in curved spacetimes is by no means straightforward, and in fact involves ambiguities. The most important of these ambiguities concerns the choice of a (quantum) representation of the algebra, obtained under Poisson brackets, of (the complex exponentiation of) the basic variables that describe the field, which are usually chosen to be canonical pairs. In traditional Non-Relativistic Quantum Mechanics, this algebra is known by the name of Weyl algebra (see e.g. Ref. \cite{gal}). Fortunately, in such case where the system has a finite number of degrees of freedom, the representation of the algebra in a Hilbert space is unique up to unitary equivalence, provided that certain mild conditions are imposed  (including continuity). This uniqueness result is known as the Stone-von Neumann theorem \cite{stone-vonneumann}. This means that two different representations of the same Weyl algebra in the same Hilbert space are necessarily related by a unitary operator. This property ensures the robustness of the theoretical predictions of Quantum Mechanics, and in particular of those derived from the evolution of quantum states, something essential for the viability of the probabilistic interpretation of the quantum theory. 

In QFT this uniqueness result for the quantum representation of the variables that describe the fields is  generally no longer valid.  It does not even apply to Fock representations of free field theories, which are typically governed by linear dynamical equations. In fact, it is well known (see e.g. \cite{BSZ}) that, for each given vacuum, there is an infinite number of linear canonical transformations, each of which provides a Fock representation of the field, that have no unitary correspondence in the Fock space. In short, this means that what in principle ought to be the corresponding unitary transformations just map the vacuum of the given representation to some vector which does not belong to its Hilbert space. Therefore, inequivalent quantum representations do exist in QFT, even in the simplest cases. This issue has been widely studied to be exclusive of systems, such as fields, with an infinite number of degrees of freedom, since a known mathematical criterion for the unitary implementability of a given linear canonical transformation involves a summability condition \cite{shale,derez}, which is trivially satisfied if the number of degrees of freedom is finite.  This fundamental obstacle translates into the existence of an infinity of quantum descriptions of the same physical system that, in general, are not equivalent. However, one may introduce physically motivated requirements in order to reduce this ambiguity of the quantum description, and even to eliminate it completely, in certain situations. A remarkable example is given by fields that propagate in stationary spacetimes. This contains the case of fields in flat, Minkowski spacetime that we mentioned above. In these types of scenarios, the possible representations of the analogue of the Weyl algebra [commonly known as the field canonical commutation relations (CCRs) for bosons, or canonical anticommutation relations (CARs) for fermionic fields] are restricted to only a single representation if one imposes that the quantum theory incorporate the symmetry displayed by the background  under timelike translations and that the evolution be generated by a positive Hamiltonian, that plays the role of an energy \cite{BSZ,kay}. At the quantum level, this implies that the vacuum state of the field is stationary. There is therefore a natural unitary implementation of the dynamics.

The situation is notably more complicated when one considers fields that propagate in non-stationary spacetimes. Such systems describe scenarios of great physical interest, such as processes of star collapse or the cosmological evolution of the Universe (essentially since its very beginning). In these cases there is no timelike symmetry of the field equations that one can try and impose in order to restrict the admissible quantizations. The situation gets even worse if one takes into account that the quantum representations of the CCRs or CARs at different times are not necessarily unitarily equivalent. As a consequence,  predictions based on the quantum evolution of the states lose  robustness. For this reason, it is especially relevant to determine some physical criterion that allows us to remove the ambiguity in the choice of a Fock quantization in non-stationary spacetimes (or at least in a convenient subset of them) and, at the same time, regain a notion of unitary quantum dynamics for the fields. 

Actually, this question has been investigated in recent years and the results indicate that the resolutions of the two problems are closely related. Indeed, for scalar fields in a multitude of non-stationary spacetimes of cosmological nature, it has been shown that a requirement of unitarity on the  dynamics of the basic field operators in the Heisenberg picture (henceforth referred to as Heisenberg dynamics or evolution) can be used to guarantee the uniqueness of the Fock representation of the CCRs (up to unitary equivalence) if, in addition, one imposes invariance under the symmetries of the field equations \cite{cmm,unigowdy2,gowdyqft,ccmv2,cmv1,cmvS2,cmv2,unit,cmsv,cmov1,cmov2,cmov3,cmov4,fmmov,flat,flat2,cfmmm,cmmv,cgmm,bianchiu}. These symmetries include the spatial isometries. In many cases, these isometries suffice to reach the desired uniqueness when combined with the demand of a unitary Heisenberg evolution. Recent discussions on the topic of unitary dynamics in QFT in curved spacetimes from the canonical perspective can be found in Ref. \cite{unit} (see also Ref. \cite{AA} for related investigations on this issue).  On the other hand, the nature of the vacuum state for fields in curved spacetimes and the Fock quantization of such fields from a covariant perspective have been widely investigated over the last decades, see e.g. Refs. \cite{adiabatic1,BD,Al,Fb}.

In this review, instead, we focus our attention on fermionic fields. Many of the most abundant elementary particles in standard matter are fermions, and in this respect one can say that fermionic fields describe more realistic matter contents than scalar fields. In addition, although there exist well-founded results about the selection of Fock representations of fermionic fields in cosmology, the literature on this topic is not as prolific as in the case of scalar fields. In this sense, a review of the recent results obtained by us and our collaborators about uniqueness criteria for the Fock representation of fermionic fields, based on a unitary Heisenberg dynamics or other related properties, appears especially useful. For the sake of concreteness, most of our discussion is devoted to the particular case of a Dirac fermion field, to which we will henceforth refer simply as  {\em Dirac field}.

The issue of determining a Heisenberg dynamics that can be realized as a unitary quantum transformation in fact involves a freedom in the splitting of the time dependence of the (fermionic) field. This time dependence can be separated in two parts: one that can be assigned to the quantum evolution of the creation and annihilation operators of the Fock representation and another that is due to the evolution of the background in which the field propagates. This second part can be treated as an explicit time dependence, via the background, when this is considered as a classical entity. Strictly speaking, this part of the evolution is not contained in the Heisenberg dynamics of the fermionic degrees of freedom. A fundamental idea in the search for a criterion to select a Fock quantization by imposing a notion of unitary dynamics is that the freedom in the splitting of the time dependence of the field can be employed to restrict the quantization in such a way that one ends up with a single family of equivalent representations while keeping nontrivial information about the fermionic evolution. 

The idea of using the aforementioned freedom to arrive at a preferred class of Fock representations has proven to be very fruitful in frameworks that surpass the scheme of QFT in a curved classical spacetime. This is the case of fields with a dynamics that can be viewed as a propagation in an auxiliary background, or even quantum geometries that present regimes in which they can be treated effectively. For instance, this idea has been applied in the framework of hybrid loop quantum cosmology (hLQC), in which the spacetime is no longer a classical entity, but a quantum object \cite{fmmov,flat,flat2,cgmm}. For cosmological systems of notable physical interest, hLQC combines a loop quantization of the zero modes that (classically) describe the Friedmann-Lema\^{\i}tre-Robertson-Walker (FLRW) spacetime that would correspond to a  cosmological universe, with a Fock quantization of the field degrees of freedom that propagate in such a cosmological spacetime, typically viewed as perturbations  \cite{hybridrev,hybridrev2}. The action of the gravitational system and its matter content is truncated at second order in these perturbations (generally assuming compact spatial sections). The combination of the loop and Fock techniques has to give rise to a consistent quantization of the total system, composed by the background cosmology and the perturbations. This consistency involves the imposition \`a la Dirac of a global Hamiltonian constraint, that interrelates the two types of representations as well as the evolution of the homogeneous universe and of its perturbations. For most of the relevant dynamical aspects of these perturbations, the information about the quantum geometry can be encapsulated in an effective geometry, that in many respects can be considered as emerging from a mean field approximation of the geometric degrees of freedom contained in the cosmological background. In this approximation, the fields that correspond to the (gauge invariant part of the) perturbations admit a QFT description where the background is the aforementioned effective geometry \cite{hybridrev,hybridrev2}. In this context, it is even clearer that building a formalism capable to maintain the unitarity of the Heisenberg dynamics in non-stationary spacetimes transcends the need to guarantee the robustness of the physical predictions of the theory. In hLQC, in particular, the additional advantage to pick out a unique family of equivalent Fock representations of the gauge invariant perturbations is that one can construct a Heisenberg evolution (with respect to some parameter of the complete quantum system) that behaves as a unitary quantum transformation in regimes where an effective background emerges.

Despite the attention that scalar (and tensor) perturbations deserve in cosmology, it is clear that a realistic matter content must include other types of fields, such as those that describe fermions, as we have already pointed out. Actually, in hLQC, recent works have introduced Dirac fermions and treated them as part of the perturbations \cite{fermihlqc}. The interest of contemplating the presence of these fields in the very early Universe goes beyond a formal question about the completeness of the description, because it is necessary to confirm that these fermionic fields do not affect substantially the otherwise well-established evolution of the primordial scalar perturbations, nor of the tensor ones. The results obtained in Refs. \cite{fermihlqc,backreaction} support the expectation that the possible effects are ignorable. 

It is worth remarking that there exists an inherent freedom to choose the splitting between the (fermionic) field variables and the degrees of freedom that describe the background, allowing one to change between different families of  annihilation and creation variables. This can be done by means of transformations that mix all these degrees of freedom while preserving the canonical symplectic structure of the combined system, including the field canonical (anti-)commutation relations. If the fields are treated as perturbations, it suffices that the canonical symplectic structure is preserved at the level of the perturbative truncation adopted in the system. Instead of considering this freedom a nuisance, one can try and exploit it in order to define variables for which the Hamiltonian of the selected fermionic degrees of freedom has certain nice quantum properties, desirable from the viewpoint of a good physical and mathematical behavior \cite{backreaction}. 

In fact, when fermions were studied for the first time within the hybrid approach to loop quantum cosmology in Ref. \cite{fermihlqc}, considering them as perturbations around a homogeneous and isotropic cosmological spacetime, the selection of fermionic variables for the corresponding Dirac field was restricted only by the requirements of invariance of the resulting Fock vacuum under the spatial isometries, a unitarily implementable Heisenberg evolution in the regime of QFT in a curved background, and a standard convention for particles and antiparticles. Nonetheless, with a rather reasonable choice made among the family restricted by these conditions, it was realized that the resulting Schr\"odinger equation for the fermionic degrees of freedom (after a sort of mean field approximation) involved ultraviolet divergences.  To solve these divergences, one either has to appeal to a regularization scheme with {\sl subtraction of infinities} or, alternatively, employ the remaining freedom in the choice of fermionic variables and restrict it even further by introducing additional requirements. Specifically, in Ref. \cite{backreaction} it was required that the fermionic backreaction be finite. In practice, this new restriction lowers the asymptotic order of the interaction part of the fermionic Hamiltonian at large wave numbers (defined for the Dirac field as the eigenvalues of the Dirac operator on the spatial sections of the background, in absolute value). As a consequence, the production of pairs of particles and antiparticles decreases for large wave numbers, becoming negligible asymptotically.

Actually, it is possible to go one step beyond in the same direction and, by taking advantage again of the freedom to split the degrees of freedom, demonstrate that one can absorb all the interaction terms of the fermionic Hamiltonian so as to make them identically zero order by order in the asymptotic regime of large wave numbers \cite{fermidiagonalization}. In this way, one clearly improves the quantum behavior of the (fermionic) field contribution to the Hamiltonian of the gravitational system. Moreover, one also greatly reduces the surviving ambiguity in the choice of Fock representation and vacuum for the field. Besides, since the resulting (fermionic) field Hamiltonian contribution is diagonal by construction on states with a definite number of particle (and antiparticle) excitations, at least asymptotically, the dynamics ruled by it is very simple. The vacuum of the naturally associated Fock representation changes only by a rotating phase. In this sense, one can interpret that this vacuum and the corresponding splitting of degrees of freedom in the hybrid quantization approach are those that get best adapted to the cosmological evolution. Finally, it is remarkable that this criterion of asymptotic diagonalization reproduces the standard choices of vacuum state in well-understood situations, within the scheme of QFT in a curved classical background \cite{fermidiagonalization,diagonalization}. This happens e.g. in the case of a flat spacetime, as well as for a de Sitter cosmology, scenario where the Bunch-Davies state is a natural vacuum \cite{bunchdavies}.

The rest of this review is organized as follows. In Section \ref{Sec:Unit-DF} we provide the basics for the construction of a Fock representation of the CARs for a Dirac field, within the framework of QFT in a curved spacetime. We also  summarize the results about the use of symmetries and of the unitarity of the Heisenberg evolution as criteria to select a preferred family of equivalent Fock representations. We then explicitly apply these criteria in Section \ref{Sec:DF-2+1}, that deals with the case of a Dirac field in a non-stationary spacetime in 2+1 dimensions. The more interesting case of Dirac fields in a cosmological spacetime in 3+1 dimensions is reviewed in Section \ref{FLRW}. After reviewing these aspects and results of QFT in curved spacetimes, in Section \ref{LQC} we consider the use of canonical transformations to introduce a suitable splitting between the degrees of freedom of the cosmological background and of the Dirac field, treated in principle as a perturbation. In that section, we show how to use this freedom in the splitting to improve the quantum properties of the fermionic contribution to the Hamiltonian of the system, and in particular to make finite the backreaction that appears in it. The possibility of further employing this freedom to diagonalize the fermionic contribution to the Hamiltonian in the asymptotic limit of large wave numbers is reviewed in Section \ref{diagonalization}. There, we also explain that this diagonalization requirement can pick out a unique vacuum state under reasonable conditions, and that this state coincides with the natural one in situations of interest in QFT, like for Minkowski and de Sitter backgrounds. In addition, we also comment on the relation of adiabatic states with the vacuum selected by our criterion. Finally, we present the conclusions and some additional remarks in Section \ref{conclu}. We set the speed of light in vacuo, the Newton gravitational constant, and the reduced Planck constant equal to the unit.

\section{Fock quantization of the Dirac field}
\label{Sec:Unit-DF}

This section contains some background material about the Fock quantization of a Dirac field in a curved spacetime. Special emphasis is put on the inherent ambiguity in the representation of the CARs associated with the infinitely many inequivalent complex structures available to construct the quantum theory, as well as on the combined criteria of symmetry invariance and of unitary implementability of the dynamics, that have been successfully employed to remove this ambiguity (and, even more, the ambiguity in the choice of basic field variables) in diverse, physically interesting fermionic (and bosonic) systems.

For the sake of clarity, let us begin our discussion by introducing the classical setting.

\subsection{Background spacetime and Dirac equation}
\label{2.2}

As backgrounds for the propagation of the field, we consider globally hyperbolic spacetimes, or just globally hyperbolic regions, $({\cal{M}}\approx \mathbb{I}\times \Sigma,g_{\mu\nu})$, in either three or four dimensions. Here, $\mathbb{I}$ denotes an interval of the real line, and $\Sigma$ is a Riemannian, Cauchy (hyper-)surface of dimension $d$, with $d=2,3$. For mathematical convenience, we restrict this surface to be topologically compact. In order to ensure that the spacetime (or region) admits a spin structure \cite{sgeom}, we additionally require that $\cal{M}$ admit a global orthonormal frame \cite{geroch2}. So, the spacetime metric can be globally written as 
\begin{align}\label{tetrads}
g_{\mu\nu}=e_{\mu}^{a}e_{\nu}^{b}\eta_{ab},
\end{align}
where $\eta_{ab}$ is the Minkowski metric in $(d+1)$-dimensions, with signature $\{-,+,...,+\}$, and $e_{\mu}^{a}$ is a(n orthonormal) coframe field, with dual frame $e^{\mu}_{a}$. Throughout this work, Greek indices from the middle of the alphabet denote spacetime indices ($\mu,\nu,...=0,...,d$), whereas Latin indices from the beginning of the alphabet account for the internal Lorentz gauge introduced by the frame ($a,b,...=0,...,d$). Besides, Greek indices from the beginning of the alphabet denote spatial indices ($\alpha,\beta,...=1,...,d$).

By employing an Arnowitt-Deser-Misner (ADM) decomposition of the considered spacetime (region) \cite{Grav}, we introduce a coordinate system in $\cal{M}$, say $\{x^{\mu}\}=\{x^{0},x^{\alpha}\}$, with $x^{0}=t\in \mathbb{I}$ being the time parameter and $\{x^{\alpha}\}$ coordinatizing $\Sigma$. The line element in coordinates $\{x^{\mu}\}$ reads 
\begin{align}\label{ADM}
\text{d}s^{2}=g_{\mu\nu}\text{d}x^{\mu}\text{d}x^{\nu}=-(N^{2}-N_{\alpha}N^{\alpha})\text{d}t^{2}+2N_{\alpha}\text{d}x^{\alpha}\text{d}t+h_{\alpha\beta}\text{d}x^{\alpha}\text{d}x^{\beta},
\end{align}
where $N$ and $N^{\alpha}$ are, respectively, the lapse function and the shift vector, and $h_{\alpha\beta}$ is the induced metric on the Cauchy surface $\Sigma$.

Let then $\Psi$ be a free, complex, and anticommuting Dirac spinor with mass $m$, propagating in $({\cal{M}},g_{\mu\nu})$. The dynamics of $\Psi$ is governed by the first-order linear equation
\begin{align}\label{Deq}
e^{\mu}_{a}\gamma^{a}\nabla^{S}_{\mu}\Psi-m\Psi=0.
\end{align}
Here, the operator $\nabla^{S}_{\mu}$ stands for the spin lifting of the Levi-Civit\`{a} covariant derivative \cite{sgeom}, and $\gamma^{a}$ are the constant Dirac matrices that generate the Clifford algebra of a flat spacetime in $(d+1)$ dimensions:
\begin{align}\label{Cliff}
\gamma^{a}\gamma^{b}+\gamma^{b}\gamma^{a}=2\eta^{ab}I,
\end{align}
where $I$ is the identity matrix and $\eta^{ab}$ is the (inverse of the) Minkowski metric. 

On account of the global hyperbolicity of $({\cal{M}},g_{\mu\nu})$, the Dirac equation (\ref{Deq}) has a well-posed Cauchy formulation \cite{dimock}. So, given any smooth initial value $\Psi(\vec{x})$ of the spinor field on a certain (compact) Cauchy surface $\Sigma_{0}$, say at $t=t_{0}$ (where $t_{0}\in \mathbb{I}$ is a fixed, but arbitrary, reference time), there exists a unique smooth solution $\Psi(t,\vec{x})$ to Equation (\ref{Deq}) which is defined on all of ${\cal{M}}$ and such that $\Psi(t,\vec{x})\vert_{\Sigma_{0}}=\Psi(\vec{x})$. The solution $\Psi(t,\vec{x})$, restricted to the domain of dependence of an arbitrary closed subset $S$ of $\Sigma_0$, depends only upon $\Psi(\vec{x})\vert_{S}$. Henceforth, we fix $\Sigma_0$ (i.e., the section at $t=t_0$) as the Cauchy reference surface. Let $\mathcal{S}$ be the complex linear space of (smooth) solutions to the Dirac equation (\ref{Deq}) which arises from the complex vector space $\mathcal{P}=\{\Psi(\vec{x})\}$ of (smooth) initial conditions at time $t_0$. Note that, by construction, the map $\mathcal{S}\ni \Psi\mapsto I_{t_{0}}(\Psi) = \Psi\vert_{\Sigma_{0}}$ is an isomorphism between the linear spaces $\mathcal{S}$ and $\mathcal{P}$. 

The space of Cauchy data $\mathcal{P}$ is naturally equipped with the product \cite{dimock}
\begin{align}\label{innerd}
(\Psi_{1},\Psi_{2})_D=\int_{\Sigma_{0}}\text{d}^{d}\vec{x}\,\sqrt{\text{h}}\,\Psi_{1}^{\dagger}\gamma^{0}n^{\mu}e_{\mu}^{a}\gamma_{a}\Psi_{2},
\end{align}
where $\gamma_{a}=\eta_{ab}\gamma^{b}$, $\text{h}$ is the determinant of $h_{\alpha\beta}$, the dagger denotes the Hermitian adjoint, and $n^{\mu}$ are the spacetime components of the  future-directed unit normal to the Cauchy surface $\Sigma_{0}$. The space of solutions $\mathcal{S}$ is endowed with an inner product of the form (\ref{innerd}), though now at an arbitrary Cauchy surface $\Sigma_{t'}$; this is so because of the independence of the mapping $(\,\cdot\, , \, \cdot\,)_{D}:\mathcal{S}\times \mathcal{S}\to \mathbb{C}$ upon the spatial section on which it is evaluated.

The exact meaning of what we understand by classical dynamical evolution in the space $\mathcal{P}$ is as follows. Given any solution $\Psi(t,\vec{x})$ in $\mathcal{S}$, the evaluation  $\Psi(t,\vec{x})\vert_{\Sigma_{t'}}$ at a fixed time $t'$ defines the spinor $\Psi'(\vec{x})=\Psi(t',\vec{x})$ on $\Sigma_0$. Namely, we can take the induced spinor field on $\Sigma_{t'}$ as initial condition on $\Sigma_0$. Clearly, the mapping $I_{t'}:\mathcal{S}\to\mathcal{P}$ defined by $\Psi(t,\vec{x})\mapsto I_{t'}\big(\Psi(t,\vec{x})\big)=\Psi(t',\vec{x})=\Psi'(\vec{x})$ is an isomorphism. Then, by considering the entire interval $\mathbb{I}$, we get a one-parameter family of isomorphisms $I_{t}:\mathcal{S}\to\mathcal{P}$. The action of this family of mappings on a solution $\Psi\in \mathcal{S}$ gives the dynamical orbit of the Cauchy initial datum $\Psi(t,\vec{x})\vert_{\Sigma_{0}}=\Psi(\vec{x})$ in $\mathcal{P}$; that is, time evolution in the space of Cauchy data $\mathcal{P}$ is given by the one-parameter family of linear transformations $T_{(t,t_0)}=I_{t}\circ I^{-1}_{t_{0}}$. 

\subsection{Fock quantization, unitarity, and uniqueness}

Let us next discuss the Fock quantization of Dirac fields using an approach based on the space of Cauchy data. It is worth remarking that an analogous approach constructed from the space of solutions is readily available, given the isomorphism $I_{t_0}$ between both spaces.

We start by equipping the space of Cauchy data $\mathcal{P}$ with a complex structure, namely a real linear automorphism $J:\mathcal{P}\to \mathcal{P}$ with the property $J^{2}=-I$, and such that it leaves the inner product (\ref{innerd}) invariant. The complex structure allows for a natural splitting of $\mathcal{P}$ into two mutually complementary orthogonal subspaces (with respect to the considered inner product) $\mathcal{P}_{J}^{\pm}=(\mathcal{P} \mp i J\mathcal{P})/2$, that are eigenspaces of $J$ with eigenvalue $\pm i$. Let $\bar{\mathcal{P}}$ be the complex conjugate of $\mathcal{P}$, endowed with the complex conjugate of Equation (\ref{innerd}) as its inner product. The complex structure $J$ is naturally defined by linearity on $\bar{\mathcal{P}}$, and we similarly have that the corresponding $\pm i-$eigenspaces,  $\bar{\mathcal{P}}_{J}^{\pm}=(\bar{\mathcal{P}} \mp i J\bar{\mathcal{P}})/2$, provide a decomposition of $\bar{\mathcal{P}}$ into mutually orthogonal subspaces. Note that  $\bar{\mathcal{P}}_{J}^{\pm}$ and $\mathcal{P}_{J}^{\pm}$ are related by $\bar{\mathcal{P}}_{J}^{\pm}=\overline{\mathcal{P}_{J}^{\mp}}$.

By performing the Cauchy completion of $\mathcal{P}_{J}^{+}$ and $\bar{\mathcal{P}}_{J}^{+}$ in their inner products, we get the one-particle Hilbert spaces $\mathcal{H}^{p}_{J}$ and $\mathcal{H}^{ap}_{J}$ of, respectively, {\em particles} and {\em antiparticles}. The Hilbert space of the quantum theory is taken to be the antisymmetric Fock space \begin{align}\label{fockd}
\mathcal{F}_{J}=\oplus_{n=0}^{\infty}(\otimes^{a}_{n}\mathcal{H}_{J}),
\end{align}
where $\mathcal{H}_{J}=\mathcal{H}^{p}_{J}\oplus\mathcal{H}^{ap}_{J}$ is the one-particle Hilbert space associated with the complex structure $J$, and $\otimes^{a}_{n}\mathcal{H}_{J}$ denotes the $n$-fold antisymmetric tensor product of $\mathcal{H}_{J}$, with $\otimes^{a}_{0}\mathcal{H}_{J}=\mathbb{C}$ \cite{qftwald,Smatrix}. 

Let $\{\psi^{p}_{n}(\vec{x})\}$ and $\{\psi^{ap}_{n}(\vec{x})\}$ be complete orthonormal bases for, respectively, $\mathcal{H}^{p}_{J}$ and $\mathcal{H}^{ap}_{J}$. Then, the quantum field is (formally) represented in $\mathcal{F}_{J}$ by
\begin{align}\label{ovdf}
\hat{\Psi}(\vec{x})=\sum_{n}[\hat{a}_{n}\psi^{p}_{n}(\vec{x})+\hat{b}^{\dagger}_{n}\bar\psi^{ap}_{n}(\vec{x})],
\end{align}
where $\hat{a}_{n}$ is the annihilation operator associated with the spinor $\bar{\psi}^{p}_{n}$, whereas $\hat{b}^{\dagger}_{n}$ is the creation operator associated with the spinor $\psi^{ap}_{n}$. The adjoint operators $\hat{a}^{\dagger}_{n}$ and $\hat{b}_{n}$ correspond to the creation and annihilation operators of, respectively, particles and antiparticles. For a detailed discussion about the definition of fermionic annihilation and creation operators, see for instance Ref. \cite{Smatrix}. For now, let us stress that the annihilation and creation operators here displayed correspond to the mode expansion projections of the (smeared) annihilation and creation operators specified in Ref. \cite{Smatrix}. The basic operators $\{\hat{a}_{n},\hat{a}^{\dagger}_{n},\hat{b}_{n}, \hat{b}^{\dagger}_{n}\}$ satisfy the anticommutation relations,
\begin{align}\label{car}
[\hat{a}_{n},\hat{a}^{\dagger}_{m}]_{+}=\delta_{nm}, \qquad [\hat{b}_{n},\hat{b}^{\dagger}_{m}]_{+}=\delta_{nm},
\end{align}
with the remaining anticommutators being null. The vacuum state of the theory corresponds to the unique (up to a phase) normalized cyclic vector in $\mathcal{F}_{J}$ which vanishes under the action of all the annihilation operators, $\hat{a}_{n}$ and $\hat{b}_{n}$.

Let us emphasize that the choice of a complex structure for the Fock quantization determines the annihilation and creation operators of the theory. Thus, in general, different complex structures define different representations of the CARs. Furthermore, there exist infinitely many of these representations that fail to be unitarily equivalent \cite{qftwald}. This is where the ambiguity in the Fock representation of the Dirac field resides. Let us be more specific. Let $\mathcal{F}_{J}$ and $\mathcal{F}_{J'}$ be two distinct Fock spaces, constructed from the different complex structures $J$ and $J'$. It can then be shown that, on the Fock space $\mathcal{F}_{J}$, the annihilation and creation operators defined by $J'$ (and which are naturally associated with the basis of $\mathcal{F}_{J'}$) are given by expressions of the form
\begin{align}\label{bog1f}
\hat{a}'_{n}=\sum_{m}(\alpha^{f}_{nm}\hat{a}_{m}+\beta^{f}_{nm}\hat{b}^{\dagger}_{m}), \qquad \hat{a}'^{\dagger}_{n}=\sum_{m}({\bar\alpha}^{f}_{nm}\hat{a}^{\dagger}_{m}+{\bar\beta}^{f}_{nm}\hat{b}_{m}),  \\ \label{bog2f}
\hat{b}'_{n}=\sum_{m}(\bar\alpha^{g}_{nm}\hat{b}_{m}+\bar\beta^{g}_{nm}\hat{a}^{\dagger}_{m}), \qquad \hat{b}'^{\dagger}_{n}=\sum_{m}({\alpha}^{g}_{nm}\hat{b}^{\dagger}_{m}+{\beta}^{g}_{nm}\hat{a}_{m}).
\end{align}
Here, $\alpha^{f}_{nm},\beta^{f}_{nm},\alpha^{g}_{nm}$, and $\beta^{g}_{nm}$ are (complex) coefficients satisfying the relationships
\begin{align}\label{bogcondf}
\sum_{l}(\alpha^{h}_{nl}\bar{\alpha}^{h}_{ml}+\beta^{h}_{nl}\bar{\beta}^{h}_{ml})=\delta_{nm}, \qquad \sum_{l}(\alpha^{f}_{nl}\bar\beta^{g}_{ml}+\beta^{f}_{nl}\bar\alpha^{g}_{ml})=0, \qquad h=f,g.
\end{align}
That is, the annihilation and creation operators defined by the two distinct complex structures $J$ and $J'$ are related by a Bogoliubov transformation. 

By definition, unitary equivalence between the representations defined by $J$ and $J'$ means that there exists a unitary operator $\hat U:\mathcal{F}_{J}\to \mathcal{F}_{J}$ intertwining the two representations, i.e.\ such that $\hat{a}'_{n}=\hat{U}^{-1}\hat{a}_{n}\hat{U}$ and $\hat{b}'_{n}=\hat{U}^{-1}\hat{b}_{n}\hat{U}$. 
The transformation defined in Equations \eqref{bog1f} and \eqref{bog2f} is unitarily implementable then, in the sense that the transformation defined by the coefficients $\alpha^{f}_{nm}$, $\beta^{f}_{nm}$, $\alpha^{g}_{nm}$, and $\beta^{g}_{nm}$ is a {\em bona fide} canonical transformation between the classical annihilation and creation variables corresponding to the considered operators.

A well-known result \cite{derez} states that unitary equivalence is achieved if and only if
\begin{align}\label{ueqf}
\sum_{n,m}(|\beta^{f}_{nm}|^{2}+|\beta^{g}_{nm}|^{2})<\infty.
\end{align}
In general, given any two arbitrary complex structures, this condition is not  satisfied. In fact, infinitely many inequivalent Fock representations of the CARs are possible, just as it happens with bosonic fields and their corresponding CCRs. The usual strategy to remove these type of ambiguities and to arrive at a (hopefully) unique Fock representation is to exploit the symmetries of the system.  One typically  requires that the complex structure (or the vacuum, in more physical terms) be invariant under some natural existing  symmetries. As already mentioned in the Introduction, a crucial role is played here by time-translation invariance, and therefore that strict strategy fails to produce a unique representation in non-stationary scenarios, including very familiar and cosmologically relevant spacetimes.
 
Notice that a  complex structure that remains invariant under time evolution immediately gives rise to a unitary implementation of (the canonical transformations generated by) the dynamics, therefore allowing the standard probabilistic interpretation of the quantum theory. It is therefore natural that, in non-stationary settings, one should try to preserve the unitary implementation of dynamical transformations, though giving up on (non-available) fully time-translation invariant complex structures, taking into account that this invariance is a sufficient, but by no means necessary, condition  for such a unitary implementation. Let us make this more explicit. Suppose we are given a complex structure $J$ on $\cal{P}$, and construct the corresponding Fock representation, with associated operators $\hat{a}_{n}$ and $\hat{b}_{n}$ as above. Since the field equations are linear, the transformations that correspond to time evolution from initial time   $t_0$ to arbitrary time $t$ are linear, and we obtain, for each $t$, new operators of the general form
\begin{eqnarray}\label{t-a}
\hat{a}_{n}(t)&=&\sum_{m}\Big(\alpha^{f}_{nm}(t,t_{0})\hat{a}_{m}+\beta^{f}_{nm}(t,t_{0})\hat{b}^{\dagger}_{m}\Big),  \\ \label{t-b} \hat{b}^{\dagger}_{n}(t)&=&\sum_{m}\Big({\alpha}^{g}_{nm}(t,t_{0})\hat{b}^{\dagger}_{m}+{\beta}^{g}_{nm}(t,t_{0})\hat{a}_{m}\Big),
\end{eqnarray}
with $\hat{a}_{n}^{\dagger}(t)$ and $\hat{b}_{n}(t)$ being supplied by the adjoint expressions of, respectively, Equations (\ref{t-a}) and (\ref{t-b}). It should be clear that, since the field evolution is a canonical transformation, the new operators satisfy the CARs, and we therefore have a family of new Fock representations. In fact, the new operators are simply those associated with the transformed complex structures $T_{(t,t_{0})}JT^{-1}_{(t,t_{0})}$ \cite{unit}, where $T_{(t,t_{0})}$ is the evolution map introduced at the end of Section \ref{2.2}. Then, a unitary implementation of the dynamical transformations implies the unitary equivalence between all the new representations,
for all $t$, and the original one defined by $J$. Of course, for a complex structure $J$ that remains invariant under time evolution, the fulfillment of the unitary equivalence condition \eqref{ueqf} is trivial, since all the nondiagonal beta coefficients of the associated Bogoliubov transformations are null.

On the other hand, there seems to be no compelling reason to relax the requirement of invariance under other natural remaining symmetries, such as isometries of the spatial manifold $\Sigma$, since invariant complex structures under these type of  symmetries typically exist.  So, the strategy that we adopt to deal with the ambiguity of the Fock quantization in non-stationary settings is the following. We require that the complex structure be invariant under the spatial isometries (and possibly other remaining symmetries of the system) and that it allows a unitary implementability of the dynamics. These combined criteria have been shown to be viable and effective in addressing the issue of the uniqueness of the quantization for a large class of field systems. The criteria were introduced for the first time in the context of midisuperspace models\footnote{The term \emph{midisuperspace}, originally introduced by K. Kucha\v{r} \cite{K1,K2}, refers to models that, even though possessing some symmetry, retain an infinite number of degrees of freedom, e.g. certain inhomogeneous models.}, concretely to specify a unique preferred quantization of the inhomogeneous fields in Gowdy cosmological models \cite{unigowdy2,ccm1,gowdyqft,cmv1,ccmv2,cmvS2}, and since then they have been profusely and successfully employed to address the uniqueness of the quantization of (test) scalar fields in various, physically relevant cosmological backgrounds \cite{cmsv,cmv2,cmov1,cmov2,cmov3,cmov4,flat,MenaMarugan:2013tba,flat2,cmmv,bianchiu} (for a review, see Ref. \cite{Cortez:2019orm}). Concerning fermionic fields and CARs, the same criteria  have been successfully applied to single out a unique preferred quantum description for (test) Dirac fields in $2+1$ dimensions \cite{fermi3} and in FLRW spacetimes \cite{uf1,uf2,uf3,Cortez:2018esi}, as we discuss in the next two sections.

It is worth pointing out that, as discussed in the Introduction, to achieve unitarily implementable dynamics in the type of non-stationary scenarios here considered, it is inevitable to explore the freedom in the splitting of the time dependence of the field between a genuine quantum Heisenberg  evolution and an explicit dependence on the spacetime background. Typically, superimposed on the intrinsic dynamical evolution of the field variables, there is an explicitly time-dependent part coming from  the non-stationary background spacetime itself. This last contribution to the total time dependence may be viewed as classical in nature, and effectively obstructs the possibility of a unitary quantum evolution. The solution is to extract the latter part by means of a time-dependent canonical transformation (performed at the classical level). This type of modification of the quantum notion of the field evolution is unavoidable in all of the cosmological systems analyzed so far,  in order to recover a unitary implementability of the dynamics.  Crucial in this approach is to pinpoint exactly the correct splitting between the intrinsic time dependence of the field and the time dependence coming from external factors, such as a non-stationary background. It is of the utmost importance to stress that this splitting is far from being arbitrary. It is guided, and to a great extent determined, by the requirement of a unitarily implementable dynamics. 

\section{Dirac fields in 2+1 dimensions}
\label{Sec:DF-2+1}

This section is devoted to discuss the applicability of the criteria of symmetry invariance and of unitary implementability of the dynamics in the Fock quantization of a concrete class of field systems, namely the case of a free Dirac field in $2+1$-dimensional spacetimes which are conformally ultrastatic, with a time-dependent conformal factor. We show that, under rather non-stringent conditions on the time dependence of the cosmological background, and once a convention on the notions of particle and antiparticle has been established, a unique family of equivalent Fock representations is singled out by imposing (i) invariance under the unitary transformations that implement the symmetries of the equations of motion, and (ii) a nontrivial and unitarily implementable dynamics \cite{fermi3}.

\subsection{Dirac spinor in conformally ultrastatic spacetimes}

Let us consider a fermionic field  coupled to a  globally hyperbolic, smooth manifold (or region) $\cal{M}$, with the topology of $\mathbb{I}\times \Sigma$, where (as before)  $\mathbb{I} \subseteq \mathbb{R}$ is an interval of the real line, and $\Sigma$ is a connected, compact, and orientable two-dimensional Riemannian manifold. Since, in particular, $\cal{M}$ is an orientable three-dimensional manifold, it is stably parallelizable \cite{stiefel}. We consider here conformally ultrastatic background geometries, so that the metric can be written as
\begin{align}\label{metric}
\text{d}s^{2}= a^{2}(\eta)\left(-\text{d}\eta^{2}+{}^{0}h_{\alpha\beta}(\vec{x})\text{d}x^\alpha\text{d}x^\beta\right).
\end{align}
Up to the scale factor $a(\eta)$, which contains the non-stationary information of the metric, ${}^{0}h_{\alpha\beta}$ is the metric induced on the spatial surfaces $\Sigma_{\eta}$ defined at each fixed value of the conformal time $\eta$.

The Dirac field couples to the geometry by means of the global (co)frame \eqref{tetrads} defined, up to $SO(2,1)$ (orthochronous) gauge transformations, by the metric (\ref{metric}). Since in three dimensions any of the two irreducible complex representations of the Clifford algebra \eqref{Cliff}  are generated by $2 \times2$ Dirac matrices, complex fermionic fields  are locally represented by two-component spinors $\Psi$. In turn, we describe the components of these spinors by Grassmann variables, in order to encode the anticommuting nature of the fermionic field. We represent the Dirac matrices by
\begin{align}
\gamma^0=i\begin{pmatrix}
-1 & 0 \\ 0 & 1
\end{pmatrix},\quad \gamma^1=i\begin{pmatrix}
0 & -1 \\ 1 & 0
\end{pmatrix}, \quad \gamma^2=\begin{pmatrix}
0 & 1 \\ 1 & 0
\end{pmatrix}.
\end{align}

The action for a fermionic field of mass $m$ is given by
\begin{align}\label{action}
I_f=-i\int \text{d}\eta\,\text{d}^2\vec{x} \sqrt{-\text{g}}\left[\frac12(\Psi^\dagger \gamma^0 e^\mu_a \gamma^a \nabla^S_\mu \Psi-\text{h.c.})-m\Psi^\dagger \gamma^0\Psi\right],
\end{align}
where $\text{g}$ is the determinant of the spacetime metric $g_{\mu\nu}$ and h.c. stands for Hermitian conjugate. 
Besides, using the spin connection one-form 
\begin{align}\label{omegas}
\omega_\mu^{ab}=\frac12\left(e^{\nu a}\partial_\mu e^b_\nu + e^{\nu a}e^{\lambda b}\partial_\lambda g_{\mu\nu}-e^{\nu b}\partial_\mu e^a_\nu -e^{\nu b} e^{\lambda a}\partial_\lambda g_{\mu\nu}\right),
\end{align}
the spin lifting of the Levi-Civit\`a covariant derivative is locally defined on the spinors as  \cite{sgeom}
\begin{align}\label{nablas}
\nabla^S_\mu \Psi=\partial_\mu\Psi-\frac14 \omega_\mu^{ab} \gamma_b\gamma_a\Psi.
\end{align}

It is convenient to partially fix the internal Lorentz gauge by choosing $n^{\mu}e^{a}_{\mu}=\delta^{a}_{0}$, where $n^{\mu}$ is the future-directed unit vector field normal to the spatial surfaces $\Sigma_{\eta}$. This leads to a reduction of the structure group of the bundle of oriented frames from $SO(2,1)$ (orthochronous) to $SO(2)$ \cite{isham}, restricting the spin structure to the  double cover of the reduced frame bundle. For each of the considered spacetime manifolds and choice of spin structure on them, this restriction is well defined and provides an identical spin structure on each of all the two-dimensional leaves that foliate the spacetime manifold  \cite{sgeom}. Thus, for each value of the time parameter, the field behaves as a spinor geometrically defined on each of the two-dimensional spatial manifolds $\Sigma_{\eta}$ that foliate the background. Equivalently, within the Cauchy data approach, the field can be described by one-parameter families of spinors, parametrized by the conformal time $\eta$, defined on an {\em initial} Cauchy reference surface $\Sigma_0$, specified by $\eta=\eta_{0}$. Let us denote by $\cal{P}$ the space of Cauchy data, namely the space of spinors at $\Sigma_0$. The aforementioned one-parameter families on $\cal{P}$ are nothing but the result of evolving the Cauchy data in time (see Section \ref{Sec:Unit-DF}). Thanks to the adopted gauge fixing, we can make a direct use of the spectral analysis of the Dirac operator defined on the two-dimensional Cauchy surface $\Sigma_0$, instead of the Dirac operator on the whole Lorentzian geometry. In fact, in this gauge, the covariant derivative on spinors becomes simply
\begin{eqnarray}
\nabla^S_0 \Psi&=&\partial_0 \Psi,\quad \nabla^S_\alpha \Psi={}^{(2)}\nabla^S_\alpha\Psi-\frac14 \tilde\omega_\alpha^{ab} \gamma_b\gamma_a \Psi,
\\
\tilde\omega_\alpha^{ab}&=&\frac12\left(e^{\beta a}e^{0 b}\partial_0 g_{\alpha\beta}- e^{\beta b}e^{0 a}\partial_0 g_{\alpha\beta}\right),
\end{eqnarray}
where ${}^{(2)}\nabla^S_\alpha$ is the spin covariant derivative on the spatial leaf with metric ${}^{0}h_{\alpha\beta}$. It can then be checked that
\begin{align}
e^\mu_a \gamma^a \nabla^S_\mu \Psi=\frac{\gamma^0}{a}\left(\partial_0+\frac{a'}{a}\right)\Psi-\frac{i}{a}\slashed D\Psi,
\end{align}
where the prime stands for the derivative with respect to the conformal time $\eta$, and $\slashed D$ denotes the Dirac operator on $\Sigma_0$.

Once the partial gauge fixing is performed, the inner product \eqref{innerd} on the space of Cauchy data $\cal{P}$ simplifies to
\begin{align}\label{inner2d}
( \Psi_1,\Psi_2 )_{D}=a_0^2\int_{\Sigma_{0}} \text{d}^{2}\vec{x} \sqrt{{}^{0}\text {h}} \Psi_1^\dagger(\vec{x}) \Psi_2(\vec{x}),
\end{align}
where $a_{0}=a(\eta_{0})$ and ${}^{0}\text {h}$ is the determinant of ${}^{0}h_{\alpha\beta}$. 

The Dirac operator $\slashed D$ is essentially self-adjoint with respect to the inner product \eqref{innerd} and, since $\Sigma_0$ is compact, it necessarily has a discrete spectrum, with eigenvalues $\pm \omega_n$, labeled by natural numbers $n$, with  $\omega_n$ ($\geq 0$) growing with $n$  \cite{sgeom}. Then, the space of Cauchy data $\cal{P}$ can be endowed with a basis formed by a set of eigenspinors of the Dirac operator. Let $a_0^{-1}\rho^{np}(\vec{x})$ be the eigenspinors with positive eigenvalue $\omega_n$ and orthonormal  with respect to the inner product \eqref{innerd}, where the index $p$ accounts for the degeneracy. Since $\slashed D$ anticommutes with $\gamma^{1}\gamma^{2}=\gamma^{0}$, we can chose as eigenspinors with negative eigenvalue $-\omega_n$ those defined as $\bar{\sigma}^{np}(\vec{x})=-\gamma^0\rho^{np}(\vec{x})$. Like $\rho^{np}(\vec{x})$, these eigenspinors form an orthonormal set when the product is rescaled by the factor $a^{-2}_{0}$. With this rescaling, the set $\{\rho^{np}(\vec{x}),\bar{\sigma}^{np}(\vec{x})\}$ provides a complete, orthonormal basis for $\cal{P}$.

Let $g_n$ be the degeneracy of the eigenspace labeled by $n$, so that $p=1,...,g_{n}$. The explicit form of $g_n$ depends on the spectral details of the Dirac operator $\slashed D$ and, consequently, on the particular $2$-manifold considered. Nevertheless, for our purposes, we do not need the actual value of $g_n$, but only to know its behavior in the ultraviolet regime of large eigenvalues $\omega_n$. So, let us introduce the counting function $\chi_{\slashed D}(\omega)$ of the Dirac operator on a $d$-dimensional compact Riemannian manifold; that is, $\chi_{\slashed D}(\omega)$ is the function that counts the number of positive eigenvalues of $\slashed D$ that are not greater than  $\omega$ (including degeneracy). From the  Weyl asymptotic formula \cite{JRoe}, it follows that $\chi_{\slashed D}(\omega)$ grows at most as $\omega^{d}$ when $\omega$ goes to infinity. Using this result with $d=2$, we conclude that the degeneracy behaves in the large $n$ limit as $g_n=o(\omega_n^2)$, where the symbol $o(\omega_n^2)$ means negligible with respect to $\omega_n^2$.

In terms of the basis  $\{\rho^{np}(\vec{x}),\bar{\sigma}^{np}(\vec{x})\}$, the {\it{dynamical}} families of spinors in $\cal{P}$ (each one of them parametrized by the conformal time $\eta$) are given by
\begin{align}\label{expan}
\Psi(\eta,\vec{x})=\frac{1}{a(\eta)}\psi(\eta,\vec{x}),\quad  \psi(\eta,\vec{x})=\sum_{n=0}^{\infty}\sum_{p=1}^{g_{n}}\left[ s_{np}(\eta)\rho^{np}(\vec{x})+\bar{r}_{np}(\eta)\bar{\sigma}^{np}(\vec{x})\right].
\end{align}
Here, we use overlined symbols to indicate complex conjugation. Apart from the global factor $a^{-1}(\eta)$, the time dependence (equivalently, the $\eta$-parameterization) is captured by the time-dependent coefficients $s_{np}$ and $\bar{r}_{np}$ of $\psi$, which take care of the Grassmannian nature of the fermionic field. The auxiliary field $\psi$ shows symmetric canonical Dirac brackets with its corresponding adjoint field that do not depend on the background \cite{casal,T-N}. We represent the algebra generated by these brackets in a Fock space, with the brackets being replaced with anticommutators , thus obtaining  a Fock representation of the CARs and therefore a quantization of both the auxiliary and the original field, $\psi$ and $\Psi$.

By writing the field anticommutation relations in terms of the modes $s_{np}$, $\bar{r}_{np}$, and their complex conjugates, one finds that the only nonvanishing Dirac brackets are $\{s_{np},\bar{s}_{np}\}=-i$ and $\{r_{np},\bar{r}_{np}\}=-i$, which are symmetric due to the anticommutativity of our Grassmann variables. In the quantum theory, they become anticommutators of the corresponding operators \cite{casal}. 

From Equations \eqref{action} and  \eqref{expan}, the equations of motion for the fermionic modes are given by \cite{fermi3}
\begin{align}\label{1order}
s_{np}^{\prime}=i(\omega_{n}+ima)\bar{r}_{np}, \qquad r_{np}^{\prime}=-i(\omega_{n}+ima)\bar{s}_{np},
\end{align}
and their complex conjugates. These equations only couple the modes $s_{np}$ and $\bar{r}_{np}$ (respectively $\bar{s}_{np}$ and $r_{np}$) with the same labels $n$ and $p$, and do not depend on the degeneracy label $p$. They can be combined into the second-order differential equation
\begin{align}\label{2order}
z_{np}^{\prime\prime}=-(\omega_{n}^2+m^{2}a^2)z_{np}+i\frac{m a^{\prime}}{\omega_n+ima}z_{np}^{\prime},
\end{align}
where $z_{np}$ denotes either $s_{np}$ or $r_{np}$. The general solution to this equation does not depend on the label $p$, except through the initial conditions, and is a linear combination of two complex independent solutions that we write in the form 
$\exp{[(-1)^{l+1}i\Theta^{l}_{n}(\eta)]}$ with $l=1,2$. Let $\Theta^{l}_{n}(\eta_{0})=\Theta_{n,0}^{l}$ and $(\Theta^{l}_{n})^{\prime}(\eta_{0})=\Theta^{l}_{n,1}$ be the initial conditions at the initial reference time $\eta_0$, and let us call 
$\Omega^{l}_{n,0}=\exp{[(-1)^{l+1}i\Theta^{l}_{n,0}]}$. A simple inspection shows that the integration constants of the general solution relate the initial conditions on $\Theta^{l}_{n}$ and their derivatives to the initial conditions $s^{0}_{np}$ and $r^{0}_{np}$ for the modes (and their complex conjugates), via Equations  \eqref{1order}. One can then deduce that time evolution in the complex linear space of spinors $\psi$ is dictated by the linear transformation \cite{fermi3}
\begin{eqnarray} \label{evol1}
\begin{pmatrix}
s_{np}  \\ \bar{r}_{np} 
\end{pmatrix}_{\!\!\eta}&=&\mathcal{V}_n(\eta,\eta_0)\begin{pmatrix}s_{np}  \\ \bar{r}_{np} 
\end{pmatrix}_{\!\!\eta_0},
\end{eqnarray}
\begin{eqnarray}
 \label{evol2}
\mathcal{V}_n(\eta,\eta_0)&=&\begin{pmatrix}
\Delta_{n}^{2}e^{i\Theta^{1}_{n}(\eta)}+\Delta^{1}_{n}e^{-i\Theta_{n}^{2}(\eta)} &  \zeta^{1}_{n}e^{i\Theta_{n}^{1}(\eta)}-\zeta_{n}^{2}e^{-i\Theta_{n}^{2}(\eta)} \\  \bar{\zeta}^{2}_{n}e^{i\bar{\Theta}_{n}^{2}(\eta)}-\bar{\zeta}_{n}^{1}e^{-i\bar{\Theta}_{n}^{1}(\eta)} &
\bar{\Delta}_{n}^{2}e^{-i\bar{\Theta}^{1}_{n}(\eta)}+\bar{\Delta}^{1}_{n}e^{i\bar{\Theta}_{n}^{2}(\eta)}
\end{pmatrix},
\end{eqnarray}
where the subindex $\eta$ in column-vectors denotes evaluation at the given value of the conformal time, and the constants $\Delta^{l}_{n}$ and  $\zeta_{n}^{l}$ are
\begin{align}\label{icconsts}
\Delta^{l}_{n}=\frac{\Theta_{n,1}^{l}}{\Omega^{\tilde{l}}_{n,0}(\Theta^{1}_{n,1}+\Theta^{2}_{n,1})}, 
\qquad \zeta_{n}^{l}=\frac{\omega_n+im a_0}{\Omega^{l}_{n,0}(\Theta^{1}_{n,1}+\Theta^{2}_{n,1})},
\end{align} 
where $\tilde{l}$ is the complementary of $l$ in $\{1,2\}$, namely $\{l,\tilde{l}\}=\{1,2\}$.

In order to analyze whether the quantum theory admits a unitarily implementable dynamics, we do not really need to obtain the solution for $\mathcal{V}_n(\eta,\eta_0)$. It is sufficient to know its behavior in the ultraviolet regime of large $\omega_{n}$. Under the mild condition that the scale factor be twice differentiable and with a second derivative that is integrable over each compact subinterval of the time domain $\mathbb{I}$, a careful asymptotic analysis of the dynamics of the modes $\{z_{np}\}=\{s_{np},r_{np}\}$ shows that two independent solutions to Equation \eqref{2order} can be specified as follows \cite{fermi3}:
\begin{align}\label{phases}
\Theta^{l}_{n}=\omega_{n}\Delta\eta +\int_{\eta_{0}}^{\eta}\text{d}\tilde\eta\Sigma^{l}_{n}(\tilde\eta),\qquad \Sigma^{l}_{n}(\tilde\eta)=\Lambda^{l}_{n}(\tilde\eta)-(-1)^l\frac{ ma^{\prime}(\tilde\eta)}{2[\omega_n+i m a(\tilde\eta)]},
\end{align}
for $l=1,2$, where $\Delta\eta=\eta-\eta_{0}$, and $\Lambda^{l}_{n}(\eta)$ is a function with $\Lambda^{l}_{n}(\eta_0)=0$ that, in the ultraviolet regime, is at most of order $\omega^{-1}_{n}$. With this choice, the constants  \eqref{icconsts} turn out to be
\begin{align}\label{constants}
\Delta^{l}_{n}=\frac12-(-1)^l \frac{m a^{\prime}_0 }{4\omega_n(\omega_n+i m a_0)},\qquad \zeta^{l}_{n}=\frac12+i\frac{ma_0}{2\omega_n}.
\end{align} 

\subsection{Fock quantization and unitary evolution}

Let us now discuss the quantization of our fermionic system. Concretely, in this section we present the unique, preferred Fock quantization singled out by the criteria of symmetry invariance and of unitary implementability of the dynamics introduced in Section \ref{Sec:Unit-DF}. The construction is performed in three steps. (1) We first focus on determining the family of invariant complex structures; namely those that commute with the action of the group of symmetries  of the equations of motion for the modes. By construction, these complex structures lead to Fock vacuum states that are invariant under the unitary transformations generated by the symmetry group. We then consider time-dependent families of annihilation and creation variables associated with the invariant complex structures. By interpreting these families as dynamical trajectories, a specific redistribution of the implicit and explicit time dependence of the field is made. The dynamics that we wish to implement quantum mechanically is that of the implicitly time-dependent part, corresponding to the evolution of the annihilation and creation variables. (2) Each of the families of annihilation and creation variables defines an invariant Fock representation and a specific quantum evolution in the corresponding Fock space. We impose the criterion of unitary implementability of the dynamics, together with the requirement that the evolution be not trivialized. This leads us to set (or better said, characterize) all invariant Fock quantizations with a nontrivial and unitarily implementable dynamics. (3) Finally, we show that all such Fock quantizations turn out to be, in fact, unitarily equivalent, up to conventions in the notions of particles and antiparticles.

On account of Equation \eqref{1order}, it is clear that the field equations are invariant under the set of transformations that interchange eigenmodes of the Dirac operator with the same value of $\omega_n$. Since the Dirac operator is built from the spatial metric ${}^{0}h_{\alpha\beta}$, these symmetries include the isometries (if any) of the Cauchy surface $\Sigma_0$. It should be clear that linear transformations commuting with the action of the symmetry group of   the Dirac equation are composed of $2\times 2$ blocks which, at most, can mix the modes $s_{np}$ and $\bar{r}_{np}$ with the same value of $p$. In addition, using all the available symmetries one can reason that the blocks are necessarily the same  for all the modes corresponding to the same eigenvalue of the Dirac operator (in norm) \cite{fermi3}. So, in particular, invariant complex structures are completely determined by a series of $2\times 2$ matrices labeled by  $n\in\mathbb{N}$.

Let us remark that, given a complex structure, their associated annihilation and creation variables, namely the classical counterparts of the corresponding operators in an expansion of the type \eqref{ovdf}, diagonalize its action. Since invariant complex structures can mix only  modes $s_{np}$ and $\bar{r}_{np}$ with the same labels, the corresponding annihilation and creation variables must be linear combinations of these modes. The annihilation variables of particles and antiparticles are denoted by $a_{np}$ and $b_{np}$, respectively, while the creation variables are their complex conjugates $\bar{a}_{np}$ and $\bar{b}_{np}$. These variables must satisfy the usual Dirac brackets for annihilation and creation sets, 
$\{a_{np},\bar{a}_{np}\}=\{b_{np},\bar{b}_{np}\}=-i$ and $\{a_{np},b_{np}\}=0$, which lead to the CARs \eqref{car} in the quantum theory.

Now, let us consider time-dependent families of annihilation and creation variables associated with invariant complex structures. Specifically, 
\begin{align}\label{blockcs12d}
\begin{pmatrix}
a_{np}  \\ \bar{b}_{np} 
\end{pmatrix}_{\!\!\eta}=\mathcal{F}_n(\eta)
\begin{pmatrix}
s_{np}  \\ \bar{r}_{np} 
\end{pmatrix}_{\!\!\eta},\qquad \mathcal{F}_n(\eta)=\begin{pmatrix}
f_1^n(\eta) & f_2^n(\eta) \\ g_1^n(\eta) & g_2^n(\eta)
\end{pmatrix}.
\end{align}
In order that we get the required Dirac brackets for $a_{np}$, $b_{np}$, and their complex conjugates, the time-dependent functions $f_{l}^{n}$ and $g_{l}^{n}$ ($l=1,2$) must satisfy
\begin{align}\label{sympl}
|f_{1}^{n}|^{2}+|f_{2}^{n}|^{2}=1, \qquad |g_{1}^{n}|^{2}+|g_{2}^{n}|^{2}=1, \qquad f_{1}^{n}\bar{g}^{n}_{1}+f_{2}^{n}\bar{g}^{n}_{2}=0.
\end{align}
Combining these conditions, we can write
\begin{align}\label{fgrel2d}
g_{1}^{n}=\bar{f}_{2}^{n}e^{iG^{n}}, \qquad g_{2}^{n}=-\bar{f}^{n}_{1}e^{iG^{n}},\qquad f_{1}^{n}g^{n}_{2}-g^{n}_{1}f^{n}_{2}=-e^{iG^{n}},
\end{align}
where $G^n$ is a certain phase. Thus, it suffices just one complex function and two (real) phases for each $n$ to characterize one family of annihilation and creation variables of the form \eqref{blockcs12d}.

If we interpret the variables in each of these families as conforming to dynamical trajectories, their evolution differs (from each other, in general, and) from that of the modes $s_{np}$ and $\bar{r}_{np}$ that dictate the evolution of the auxiliary spinor field $\psi$, because of the explicit dependence on $\eta$ of the matrices $\mathcal{F}_n$. By substituting the inverse of Equation \eqref{blockcs12d} in Equation \eqref{expan}, we obtain the expression of $\Psi$  in terms of the introduced variables $a_{np}(\eta)$ and $\bar{b}_{np}(\eta)$. The result makes it clear that the dynamics of $\Psi$ is determined by the evolution of the annihilation and creation variables, and by the explicitly time-dependent contributions coming from $\mathcal{F}_n$ and the scale factor. It is only the implicitly time-dependent part, that is, the part corresponding to the evolution of the annihilation and creation variables, the one that we want to implement as a quantum Heisenberg evolution. In the following, we restrict our attention to quantizations that are not only invariant, but also admit a unitary implementability of this dynamics.

For any of the allowed families of variables given in Equation \eqref{blockcs12d}, the dynamical evolution is a Bogoliubov transformation relating  those  variables at two different times, say the arbitrary time $\eta$ and the initial time $\eta_{0}$. By using Equations \eqref{evol1} and \eqref{blockcs12d}, we get 
\begin{align} \label{bog2d}
\begin{pmatrix} a_{np} \\ \bar{b}_{np} \end{pmatrix}_{\!\!\eta}=\mathcal{B}_{n}(\eta,\eta_{0})\begin{pmatrix} a_{np} \\ \bar{b}_{np} \end{pmatrix}_{\!\!\eta_{0}}, \qquad \mathcal{B}_{n}(\eta,\eta_{0})=\begin{pmatrix} \alpha_{n}^{f}(\eta,\eta_{0}) & \beta_{n}^{f}(\eta,\eta_{0}) \\ \beta_{n}^{g}(\eta,\eta_{0}) & \alpha_{n}^{g}(\eta,\eta_{0}) \end{pmatrix},
\end{align}
where $\mathcal{B}_{n}(\eta,\eta_{0})=\mathcal{F}_{n}(\eta)\mathcal{V}_{n}(\eta,\eta_{0})\mathcal{F}^{-1}_{n}(\eta_{0})$. This Bogoliubov transformation introduces  the family of evolved complex structures $J_{\eta}= \mathcal{B}_{n}(\eta,\eta_{0})J_{\eta_0} \mathcal{B}^{-1}_{n}(\eta,\eta_{0})$ on the space of initial Cauchy data, where $J_{\eta}$ is the invariant complex structure associated with (i.e., has a diagonal action on) the annihilation and creation variables at time $\eta$. As we have seen in Section \ref{Sec:Unit-DF}, the transformations \eqref{bog2d} are implementable as unitary operators on the Fock space defined by $J_{\eta_0}$ if and only if
\begin{align}\label{ucond2d}
\sum_{n}g_{n}|\beta_{n}^{f}(\eta,\eta_0)|^{2}<\infty \qquad \text{and} \qquad \sum_{n}g_{n}|\beta_{n}^{g}(\eta,\eta_0)|^{2}<\infty, \quad \forall \eta\in \mathbb{I}.
\end{align}

Let us recall that the number of eigenstates of the Dirac operator with positive eigenvalue not greater than $\omega$ grows as $\omega^{2}$ in the ultraviolet regime. Using this asymptotic behavior, it is possible to show \cite{fermi3} that, for large $N_1$ and arbitrary $N_2>N_1$, and with $N_1\geq \omega_{n_1}>N_1-1$ and $N_2\geq \omega_{n_2}>N_2-1$, 
\begin{align}\label{omegatwo} 
\sum_{n=n_1}^{n_2} \frac{g_n} {\omega_n^4} \leq \sum_{N=N_1}^{N_2} \frac{K}{N^3},
\end{align}
where $K$ is some positive constant. This identity is important for the subsequent analysis because it implies that the sequence $\{\sqrt{g_n}\omega_n^{-2}\}_{n\in {\mathbb{N}}}$ is square summable.

Employing Equations \eqref{phases} and \eqref{constants}, as well as relations \eqref{sympl} and \eqref{fgrel2d}, one can check that, in the asymptotic regime of large $\omega_n$,
\begin{align}
|\beta_{n}^{h}|=\frac{1}{2}
\bigg|&\big(h_{2}^{n,0}-h_{1}^{n,0}\big)\left[h_{1}^{n}\left(1+i\int \Sigma^1_n\right)+h_{2}^{n}\left(1+i\int \bar\Sigma^2_n\right)\right] e^{i\omega_{n}\Delta\eta}\nonumber \\+&\big(h_{2}^{n,0}+h_{1}^{n,0}\big)\left[h_{1}^{n}\left(1-i\int \Sigma^2_n\right)-h_{2}^{n}\left(1-i\int \bar\Sigma^1_n\right)\right]e^{-i\omega_{n}\Delta\eta}\nonumber \\
+&\frac{2ma_0}{\omega_n}\left(h_{1}^{n,0}h_{1}^{n}+h_{2}^{n,0}h_{2}^{n}\right)\sin(\omega_n\Delta\eta)\bigg|+\mathcal{O}(\omega_n^{-2}),\label{beta3}
\end{align}
where the integrals are over conformal time from  $\eta_{0}$ to $\eta$, the symbol $\mathcal{O}$ stands for asymptotic order, and $h$ can be set equal to either $f$ or $g$. In order to lighten the notation, we have omitted the dependence of these functions on $\eta$ and denoted the evaluation at $\eta_0$ with the superscript $0$, preceded by a comma. It is worth noticing that relations \eqref{sympl} and \eqref{fgrel2d} guarantee that $|\beta^{f}_{n}|=|\beta^{g}_{n}|$. So, for the purpose of unitarity, it suffices to analyze just one of these types of coefficients.

It is convenient to employ again the notation $\{l,\tilde{l}\}=\{1,2\}$. Then, given that  $|h_{\tilde{l}}^{n}|^{2}+|h_{l}^{n}|^{2}=1$, we get $h_{\tilde{l}}^{n}=e^{iH^{n}_{\tilde{l}}}\sqrt{1-|h_{l}^{n}|^{2}}$, where $H^{n}_{\tilde{l}}$ denotes some phase that may depend on time.
	
It is worth remarking that, in addition to conditions \eqref{ucond2d} which severely restrict the behavior of the coefficients $h^{n}_{l}$, both in their mode and time dependence, one should naturally demand that $h^{n}_{l}$ be such that the dynamics of the annihilation and creation variables is not trivialized when compared with the original Dirac evolution. Otherwise, the criterion of a unitary implementation of the dynamics would be useless, since it would pose no restriction on the Fock representation. Indeed, one may always extract all of the asymptotically dominant time dependence of the fermionic field by means of explicitly time-dependent canonical transformations, and trivialize in this way the requirement of a unitarily implementable evolution. More specifically, by examining Equation \eqref{beta3}, it can be seen that the dominant contribution of the Dirac dynamics dictated by $\mathcal{V}_{n}(\eta,\eta_{0})$ is given by imaginary exponentials of the phases $\pm \omega_{n}\Delta \eta$. Thus, in order to avoid a trivial evolution, we rule out the possibility that these dynamical contributions are counterbalanced with a specific choice of (time and mode-dependent) phases in the linear combinations that determine the annihilation and creation variables. 

All together, taking $h$ as either $f$ or $g$, the requirements of a nontrivial and unitarily implementable dynamics impose, as a necessary condition, that asymptotically \cite{fermi3} 
\begin{align}\label{hh}
h_{l}^{n}=\frac{e^{iH^{n}_{l}}}{\sqrt{2}}+\vartheta^{n}_{h,l}, \qquad h_{\tilde{l}}^{n}=\pm e^{iH^{n}_{l}}\sqrt{1-|h_{l}^{n}|^{2}}=\pm e^{iH^{n}_{l}} \left[\frac{1}{\sqrt{2}}-\text{Re}(e^{-iH^{n}_{l}}\vartheta^{n}_{h,l})\right]+\mathcal{O}(|\vartheta^{n}_{h,l}|^{2}),
\end{align}
for a subset of the natural numbers, $n\in\mathbb{N}^{\pm}_{l}$, and with $\vartheta^{n}_{h,l}$ being some 
mode-dependent and time-dependent  complex function that goes to zero in the limit of large $\omega_n$. Here, the union of the four subsets $\mathbb{N}^{\pm}_{l}$ gives (up to a finite number of elements) the natural numbers, allowing for the possibility that up to three of these subsets be empty, and having identified $h_{l}^{n}$ with $h_{1}^{n}$ for $n\in\mathbb{N}^{\pm}_{1}$ and with $h_{2}^{n}$ for $n\in\mathbb{N}^{\pm}_{2}$, with the $\pm$ superscripts indicating the relative sign for $h_{\tilde{l}}^{n}$ in 
the second and third identitites of Equation \eqref{hh}. By substituting this equation into relation \eqref{beta3}, as well as using that the integral of $(\bar\Sigma^{1}_{n}-\Sigma^{2}_{n})$ behaves as $m\left(a-a_0\right)/\omega_{n}+\mathcal{O}(\omega_n^{-2})$ in the ultraviolet regime \cite{fermi3}, one gets that the asymptotic behavior of the complex norm of the beta coefficients, for all $n\in\mathbb{N}^{\pm}_{l}$, is
\begin{align}\label{betass}
|\beta_{n}^{h}|=
\frac{1}{\sqrt{2}}\bigg|& \left[\vartheta^{n,0}_{h,l}e^{-iH^{n,0}_{l}}+\text{Re}(e^{-iH^{n,0}_{l}}\vartheta^{n,0}_{h,l})-i(-1)^{l}\frac{ma_{0}}{\sqrt{2}\omega_n}\right]e^{ \pm i \omega_{n}\Delta\eta }\nonumber\\
&
- \left[\vartheta^{n}_{h,l}e^{-iH^{n}_{l}}+\text{Re}(e^{-iH^{n}_{l}}\vartheta^{n}_{h,l})-i (-1)^{l}\frac{ma}{\sqrt{2}\omega_n}\right]e^{\mp  i \omega_{n}\Delta\eta}
\bigg|,
\end{align}
up to certain terms that are negligible compared with the largest order between $\omega_{n}^{-1}$ and $\vartheta^{n}_{h,l}$. 

So far, no condition has been set on $\vartheta^{n}_{h,l}$ other than it must go to zero in the ultraviolet limit. However, by adding the requirement that the sequence $\{g_{n} |\beta_{n}^{h}|^{2}\}_{n\in\mathbb{N}^{\pm}_{l}}$ be summable, one gets a restriction on how fast $\vartheta^{n}_{h,l}$ must tend to zero. The line of reasoning is the following. One assumes that the beta coefficients are square summable in the subsets $\mathbb{N}^{\pm}_{l}$, including degeneracy, and then one looks for the functions $\vartheta^{n}_{h,l}$ (if any) that solve the resulting conditions. From Equation \eqref{betass}, one finds that there are two different situations that lead to distinct conditions on $\vartheta^{n}_{h,l}$, namely, either the sequence $\{\sqrt{g_{n}}\omega_{n}^{-1}\}_{n\in\mathbb{N}^{\pm}_{l}}$ is square summable, or not. In the first situation, it is not difficult to check that $\{\sqrt{g_{n}}\vartheta^{n}_{h,l}\}_{n\in\mathbb{N}^{\pm}_{l}}$ must be square summable. For the alternative situation (that is when $\{\sqrt{g_{n}}\omega_{n}^{-1}\}_{n\in\mathbb{N}^{\pm}_{l}}$ is not a square summable sequence), by using the implications of Equation \eqref{omegatwo} and recalling that any trivialization of the fermionic dynamics is excluded, one can see that the functions \cite{fermi3}
\begin{align}\label{theta}
\tilde{\vartheta}^{n}_{h,l}=\vartheta^{n}_{h,l}+e^{iH^{n}_{l}} \text{Re}(e^{-i H^{n}_{l}}\vartheta^{n}_{h,l})-i (-1)^{l}\frac{ma}{\sqrt{2}\omega_n}e^{iH^{n}_{l}}
\end{align}
must form a sequence that is square summable, including degeneracy, in the subsets $\mathbb{N}^{\pm}_{l}$ where $\{\sqrt{g_{n}}\omega_{n}^{-1}\}_{n\in\mathbb{N}^{\pm}_{l}}$ fails to satisfy such square summability.

In total, given an invariant family of complex structures $J_{\eta}$ characterized by the annihilation and creation variables \eqref{blockcs12d}, the necessary and sufficient conditions for the corresponding Fock representations of the CARs to be unitarily equivalent, and therefore to support a unitarily implementable nontrivial dynamics, are the following. (1) The functions $h^{n}_{l}$ and $h^{n}_{\tilde{l}}$ must be asymptotically of the form \eqref{hh} for $n\in \mathbb{N}^{\pm}_l$. (2) The terms $\vartheta^{n}_{h,l}$ must be such that either (2a) if $\{\sqrt{g_{n}}\omega_{n}^{-1}\}_{n\in\mathbb{N}^{\pm}_{l}}$ is square summable, they form a sequence that is square summable (including over the degeneracy), or otherwise, (2b) the sequence $\{\sqrt{g_{n}}\tilde \vartheta^{n}_{h,l}\}_{n\in\mathbb{N}^{\pm}_{l}}$ is square summable, with $\tilde{\vartheta}^{n}_{h,l}$ given in Equation \eqref{theta}. 

Up to now, we have characterized all families of annihilation and creation variables that (i) share the symmetries of the equations of motion, and (ii) evolve according to nontrivial dynamics that are unitarily implementable at the quantum level. Each of these families determines an invariant Fock representation (e.g. that associated with the choice of an invariant complex structure at the initial time $\eta_0$) with a nontrivial, unitary quantum evolution in the corresponding Hilbert space. By referring to the combination of a Fock representation and a specific quantum dynamics as a {\emph{Fock quantization}} of the system, the question now is whether the invariant Fock quantizations with unitarily implementable nontrivial dynamics are equivalent quantum theories or not. Before we address this issue of uniqueness, let us make some remarks about the preceding results.

Note that the criterion of unitary implementability, together with the requirement of a nontrivial dynamics, fix (up to phases) the leading order behavior of the coefficients $h^{n}_{l}$ in the asymptotic regime of large $\omega_n$, as it is shown in 
Equation \eqref{hh}. In addition, for all those cases where $\{\sqrt{g_{n}}\omega_{n}^{-1}\}_{n\in\mathbb{N}^{\pm}_{l}}$ fails to be a square summable sequence, the imaginary part of $e^{-iH_{l}^n}\vartheta^{n}_{h,l}$ must have its dominant asymptotic contribution of order $\omega_{n}^{-1}$, and equal to the function $ma/(\sqrt{2}\omega_n)$ in absolute value. This follows simply by realizing that the square summability of the sequence $\{\sqrt{g_{n}}\tilde \vartheta^{n}_{h,l}\}_{n\in\mathbb{N}^{\pm}_{l}}$ implies that $\tilde{\vartheta}^{n}_{h,l}=o(\omega^{-1}_{n})$ in Equation \eqref{theta}. Hence, for a nonzero fermionic mass $m$, the coefficients $h_{l}^n$ asymptotically depend in a very specific way on the eigenvalue of the Dirac operator and on the mass of the field. Most importantly, there is also a specific dependence  on time, by means of a dependence on the background where the field propagates. Finally, let us emphasize that the analysis about unitarity holds not only for a positive value of the mass $m$, but also when this mass is zero. That is, the families of annihilation and creation variables for the massless Dirac field, selected by the criteria of symmetry invariance and of unitarity, are fully characterized by coefficients $(h^{n}_{l},h^{n}_{\tilde l})$ with an asymptotic form given by Equation \eqref{hh}, and such that the sequences $\{\sqrt{g_n} \vartheta^{n}_{h,l}\}_{n\in\mathbb{N}^{\pm}_l}$ are square summable. It is worth noticing that, within this family, the choice $f_1^n=f_2^n=g_{1}^n=-g_2^n=1/\sqrt{2}$ provides a representation which corresponds to the Fock quantization constructed from the (celebrated) conformal vacuum, i.e., the natural vacuum specified by imposing the conformal symmetry of the massless Dirac equation in the quantum theory.

\subsection{Uniqueness of the quantization}

Let us now address the issue of uniqueness. With this aim, we proceed as follows. First, as a reference, we adopt a certain Fock quantization that is invariant and possesses a nontrivial and unitarily implementable dynamics. Next, we consider any other invariant Fock quantization that admits a nontrivial, unitarily implementable dynamics, and we examine whether it is unitarily related with the reference Fock quantization or not. If the answer is in the affirmative, then the uniqueness is proven.

A simple choice of reference quantization is the Fock quantization characterized by annihilation and creation variables with 
$f_{1}^{n}=1/\sqrt{2}-i a m/(\sqrt{2} \omega_n)$, $f_{2}^{n}=\sqrt{1-|f_{1}^{n}|^{2}}$, $g_{1}^{n}=f_{2}^{n}$, and $g_{2}^{n}=-\bar{f}_{1}^{n}$. Note that, in the case of the massless field, this choice defines the natural quantization with conformal vacuum. Let $\{\bar{\tilde{a}}_{np},\bar{\tilde{b}}_{np},\tilde{a}_{np},\tilde{b}_{np}\}$ be {\emph{any}} other choice of annihilation and creation variables that defines a Fock quantization with a nontrivial, unitarily implementable dynamics. The coefficients $\tilde{f}_{l}^{n}$ and $\tilde{g}_{l}^{n}$ that characterize these variables then satisfy all of the conditions stipulated in the previous subsection. 

Given our reference Fock quantization and this other arbitrary one allowed by our criteria, it follows from Equation \eqref{blockcs12d} that
the annihilation and creation variables associated with them are related via the time-dependent Bogoliubov transformation $\mathcal{K}_{n}(\eta)=\tilde{\mathcal{F}}_n(\eta)\mathcal{F}^{-1}_n(\eta)$, so that
\begin{align}\label{bogK2d}
\begin{pmatrix} \tilde a_{np} \\  \bar{\tilde{b}}_{np} \end{pmatrix}_{\!\!\eta}=\mathcal{K}_{n}(\eta)\begin{pmatrix} a_{np} \\ \bar{b}_{np} \end{pmatrix}_{\!\!\eta},\qquad  {\rm{with}} \quad \mathcal{K}_{n}=\begin{pmatrix} \kappa_{n}^{f} & \lambda_{n}^{f} \\ \lambda_{n}^{g} & \kappa_{n}^{g} \end{pmatrix}.
\end{align}
It is not difficult to check that the norm of the off-diagonal coefficients is $|\lambda^{h}_{n}|=|\tilde{h}_{1}^{n}h_{2}^{n}-\tilde{h}_{2}^{n}h_{1}^{n}|$. Besides, using Equation \eqref{fgrel2d}, one gets that $|\lambda^{f}_{n}|=|\lambda^{g}_{n}|$. Thus, the square summability conditions on $\lambda_{n}^{f}$ and $\lambda_{n}^{g}$, that must be satisfied in order for the Bogoliuvov transformation to be unitarily implementable, turn out to be just one (and the same) condition. Then, the transformation determined by the sequence of matrices $\mathcal{K}_{n}$ is implementable on the reference Fock space as a unitary operator if and only if 
\begin{align}\label{ucond22d}
\sum_{n}g_{n}|\lambda_{n}^{f}(\eta)|^{2}<\infty, \quad \forall \eta\in\mathbb{I}, 
\end{align}
where $|\lambda_{n}^{f}|=\big\vert\tilde{f}_{1}^{n}\sqrt{1-|f_{1}^{n}|^{2}}-\tilde{f}_{2}^{n}f_{1}^{n}\big\vert$. In case this condition is satisfied, we can consider the two analyzed quantizations as physically equivalent. Note that the above condition ensures  that the Fock representations defined for every value of the conformal time $\eta$ are unitarily equivalent. 

Taking e.g. $h=f$ in the formulas of the preceding subsection, we have that the coefficients $\tilde{f}^n_l$ and $\tilde{f}^n_{\tilde l}$ are, respectively, of the asymptotic form \eqref{hh}. Then, one gets that, for $n\in\mathbb{N}^{+}_{l}$ \cite{fermi3},   
\begin{align}
	|\lambda^{f}_{n}|=\frac{1}{\sqrt{2}}\left| \vartheta^{n}_{\tilde{f},l}+e^{i\tilde{F}^{n}_{l}}\text{Re}(e^{-i\tilde{F}^{n}_{l}}\vartheta^{n}_{\tilde{f},l})\right| ,
\end{align}
up to terms $\mathcal{O}(|\vartheta^{n}_{h,l}|^{2})$ in the asymptotic limit of large $\omega_{n}$ if the field is massless, or up to terms $\mathcal{O}(\omega_{n}^{-1})$ if the mass does not vanish and $\{\sqrt{g_{n}}\omega_{n}^{-1}\}_{n\in\mathbb{N}^{+}_{l}}$ happens to be square summable. Alternatively, if this sequence is not square summable and the field is massive, one must have
\begin{align}
	|\lambda^{f}_{n}|=\frac{1}{\sqrt{2}}|\tilde \vartheta^{n}_{\tilde{f},l}|+\mathcal{O}(\omega_{n}^{-2}),
\end{align}
for $n\in\mathbb{N}^{+}_{l}$. Since the sequences formed by $\vartheta^{n}_{\tilde{f},1}$ and $\vartheta^{n}_{\tilde{f},2}$ in the 
two first cases above, and by $\tilde\vartheta^{n}_{\tilde{f},1}$ and $\tilde\vartheta^{n}_{\tilde{f},2}$ in the last case, are square summable by hypothesis in their respective subsets, including degeneracy, it follows that the unitary equivalence condition \eqref{ucond22d} is immediately satisfied for all $n\in\mathbb{N}_{l}^{+}$.

On the other hand, it can be shown that Equation \eqref{ucond22d} fails to be satisfied if any of the considered subsets of integers $\mathbb{N}^{-}_{l}$ has infinite cardinality. The reason for this failure is rooted at the difference in the relative sign of the coefficients in the pair $(\tilde{f}^{n}_{1},\tilde{f}^{n}_{2})$ with respect to that in $(f^{n}_{1},f^{n}_{2})$ \cite{fermi3}. Therefore, if one then insists and interchanges the roles of $\tilde f^n_l$ and $\tilde g^n_l$ in the definition of the annihilation and creation variables for the subsets $\mathbb{N}^{-}_{l}$, something that in practice amounts to an interchange between the relative signs of the pair  $(\tilde f^n_1,\tilde f^n_2)$ and the signs for the pair $(\tilde g^n_1,\tilde g^n_2)$ [see Equation \eqref{fgrel2d}], one gets that both pairs would display the same relative signs as the coefficients of the reference quantization, and then condition \eqref{ucond22d} would be satisfied for $n\in \mathbb{N}^{-}_{l}$. According to Equations \eqref{blockcs12d}, the exchange of $\tilde f^n_l$ and $\tilde g^n_l$ can be interpreted as a change in the convention of what are particles and what are antiparticles. In this sense, the inequivalence between quantizations before one performs the explained interchange can be understood as a spurious result coming from the the fact that we are just considering two  Fock quantizations with the opposite convention for the concept of particle and antiparticle in an infinite number of modes.

In summary, the criterion of invariance under the symmetries of the equations of motion and the unitary implementability of a nontrivial quantum dynamics removes the ambiguities in the representation of the CARs, both for the massive and for the massless Dirac fields in $2+1$ dimensions, selecting a unique family of unitarily equivalent Fock representations, together with a notion of quantum evolution, up to conventions about the concept of particles and antiparticles.

\section{Fock quantization of Dirac fields in  FLRW cosmologies}\label{FLRW}

We now discuss the Fock quantization of Dirac fields in cosmological spacetimes of the FLRW type. In particular, in this section we show that one can achieve results about the uniqueness of the quantization very much like in the previous section.  More precisely, we consider minimally coupled massive Dirac fields, propagating in homogeneous and isotropic FLRW spacetimes, with 3-dimensional spatial hypersurfaces that can be either spherical or toroidal.  The spherical case was notably analyzed in great depth by  D'Eath and Halliwell  \cite{H-D}, within the context of the Wheeler-DeWitt approach to quantum cosmology. In that seminal treatment, a special time-dependent family of Fock representations was chosen for the Dirac field, by means of an instantaneous diagonalization of the Dirac Hamiltonian. In particular, such family is associated with vacua which are invariant under the symmetries of the Dirac equation. Moreover, it was shown in Ref. \cite{H-D} that particle production over time remains finite for those vacua, a fact that is an exclusive characteristic of quantizations that admit an unitary implementation of the dynamics. We now show that this family of Fock representations is in fact uniquely selected, up to unitary equivalence, by the criteria of invariance under spatial symmetries and unitary implementability of the dynamics. Also, we present a similar result concerning the Dirac field in flat (compact) FLRW spacetime. In this case, besides spatial translations, the symmetry group includes as well helicity generated spin rotations.

\subsection{Dirac spinors in FLRW cosmologies}

As before, we consider spacetime manifolds with topology of the type   $\mathbb{I}\times\Sigma$, where $\mathbb{I}$ is a connected  interval of the real line and $\Sigma$ is certain spatial  Cauchy surface. In the case of a spherical universe, $\Sigma$ is isomorphic to the 3-sphere, $S^{3}$, whereas for universes with (compact) flat spatial sections  $\Sigma$ is isomorphic  to the 3-torus, $T^{3}$. In the following, we often refer to the above two situations as the spherical case and the flat case, for $S^3$ and $T^3$ respectively. The metrics associated with such universes can be written as
\begin{align}\label{flrw}
\text{d}s^{2}=a^{2}(\eta)(-\text{d}\eta^{2}+{}^{0}h_{\alpha\beta}\text{d}x^{\alpha}\text{d}x^{\beta}),
\end{align}
where ${}^{0}h_{\alpha\beta}$ is either the 3-dimensional spherical metric or the flat metric, and $a(\eta)$ is the scale factor.

It follows from previous comments in Section \ref{Sec:Unit-DF} that spin structures can always be defined in the above types of cosmological spacetimes \cite{sgeom,geroch2}. Dirac fields $\Psi$, of mass $m$, correspond therefore to sections of the associated spinor bundle. Explicitly, we adopt the Weyl representation of the Dirac matrices
\begin{align}
\gamma^{a}=i\begin{pmatrix}
	0 & \sigma^{a} \\ \widetilde{\sigma}^{a} & 0
	\end{pmatrix},
\end{align}
where $\sigma^{0}=\widetilde{\sigma}^{0}$ is the identity  $2\times 2$ matrix and $\sigma^{i}=-\widetilde{\sigma}^{i}$ (with $i=1,2,3$) are the Pauli matrices. Such representation of the generators of the  Clifford algebra allows us to describe  Dirac fields by means of a pair of two-component spinors $\phi^{A}$ and ${\bar{\chi}}_{A'}$ possessing well-defined and opposite chirality. We take $\phi^{A}$ to be the left-handed projection of $\Psi$, while $\bar{\chi}_{A'}$ is the right-handed one. Moreover, we adopt the same conventions as in Ref. \cite{H-D} concerning the treatment of spinor indices.

In these cosmological spacetimes, the action for the Dirac field of mass  $m$ is
\begin{align}\label{action3d}
I_f=-i\int \text{d}\eta\,\text{d}^3\vec{x}\, a^{4}\sqrt{^0\text{h}}\left[\frac12(\Psi^{\dagger}\gamma^{0} e^\mu_a \gamma^a \nabla^S_\mu \Psi-\text{h.c.})-m\Psi^{\dagger}\gamma^{0}\Psi\right],
\end{align}
where the spin covariant derivative $\nabla^S_\mu$ is given by relations \eqref{omegas} and \eqref{nablas}, adapted to the models in question. 

Let us perform again a partial fixing of the internal Lorentz gauge, with the purpose of providing a rigorous treatment of the spatial dependence of spinors, as well as to analyze their properties under the symmetry groups associated with the considered FLRW cosmologies. To be specific, the gauge group of the  orthonormal and oriented frame bundle can be reduced from $SO(3,1)$ to $SO(3)$ \cite{isham}. In the considered spacetime models, this procedure gives rise to a well-defined restriction of the spin structures to the double cover of the reduced bundles, having $SU(2)$ as gauge group. Finally, this restriction provides one (and the same) spin structure on each of the spatial hypersurfaces that constitute the foliation of the considered cosmologies. In the case of $S^{3}$, the spin structure turns out to be unique \cite{bar,yasushi}. In the flat case, on the other hand, there are eight possible spin structures associated with distinct periodic conditions for the spinors in $T^{3}$ \cite{dtorus,ginoux}. In practice, this partial gauge fixing is obtained again by imposing the conditions $n^{\mu}e_{\mu}^{a}=\delta^{a}_0$ which, moreover, greatly simplify the Hamiltonian analysis of the system \cite{T-N}. In particular, one can check that the Dirac operator (defined over left and right-handed spinors) on the reference spatial hypersurface $\Sigma_0$ takes the form 
\begin{align}\label{dop3d}
ia \sqrt{2}\,e^{\alpha AA'}{}^{(3)}D_{\alpha}, 
\end{align} 
where $e^{\alpha AA'}$ is the spinor version of the triad and ${}^{(3)}D_{\alpha}$ is the spin lifting of the Levi-Civit\`a covariant derivative with respect to the 
metric ${}^{0}h_{\alpha\beta}$ \cite{H-D}.

Once the gauge fixing has been performed, we introduce  the following inner product on the space of left-handed spinors defined on the spatial hypersurface $\Sigma_0$ (as well as the corresponding definition for right-handed spinors): 
\begin{align}\label{inner3d}
\int_{\Sigma_0} \text{d}^3\vec{x}\, \sqrt{{}^{0}\text{h}}\,\bar{\chi}_{A'}I^{AA'}\phi_{A},
\end{align}
where summation over repeated indices is assumed, $I^{AA'}$ denotes the components of the identity matrix, and $\phi_{A}$ and $\chi_{A}$ are arbitrary spinors. Since both the spherical and the toroidal hypersurfaces are geodesically complete \cite{hopf}, it follows that the Dirac operator \eqref{dop3d} is essentially self-adjoint with respect to the inner product \eqref{inner3d}, with discrete spectrum in both cases.  

The spectrum of the  Dirac operator on $S^{3}$ consists of the following sequence of eigenvalues \cite{H-D,yasushi}:
\begin{align}\label{eigenvsphere}
\pm\omega_{n}=\pm\left(n+\frac{3}{2}\right),\qquad n\in\mathbb{N},
\end{align}
each of which has a corresponding degeneracy  $g_{n}=(n+1)(n+2)$. Using a notation similar to that employed in the previous section, the left-handed eigenspinors with eigenvalues $\omega_n$ and $-\omega_n$ are denoted by $\rho_{A}^{np}$ and ${\bar{\sigma}}^{np}_{A}$, respectively, where the label $p=1,...,g_{n}$ accounts for the degeneracy. Once normalized, the set of all such elements forms an orthonormal basis -- with respect to the inner product \eqref{inner3d} -- for the space of left-handed spinors in $S^{3}$. An analogous basis for right-handed spinors is readily obtained by complex conjugation.

In the spherical case, the chiral projections of the Dirac field can then be written as
\begin{align}\label{harm1s}
\phi_A(\eta,\vec{x})=& a^{-3/2}(\eta)\sum_{npp'}\breve{\alpha}_{n}^{pp'}[m_{np}(\eta)\rho_{A}^{np'}(\vec{x})+{\bar{r}}_{np}(\eta){\bar\sigma}_{A}^{np'}(\vec{x})],\\ \label{harm2s}
{\bar\chi}_{A'}(\eta,\vec{x})=& a^{-3/2}(\eta)\sum_{npp'}\breve{\beta}_{n}^{pp'}[{\bar{s}}_{np}(\eta){\bar\rho}_{A'}^{np'}(\vec{x})+t_{np}(\eta)\sigma_{A'}^{np'}(\vec{x})],
\end{align}
with analogous expressions for the complex conjugate versions, and with
\begin{align}
\sum_{npp'}=\sum_{n=0}^{\infty}\sum_{p=1}^{g_n}\sum_{p'=1}^{g_n}.\nonumber
\end{align}
The anticommutative nature of the fermionic field $\Psi$ is encoded in the Grassmann variables $m_{np}$, $r_{np}$, $t_{np}$, and $s_{np}$ (and their complex conjugate versions), that moreover carry the time dependence of the Dirac field. Finally, the constant coefficients  $\breve{\alpha}_{n}^{pp'}$ and $\breve{\beta}_{n}^{pp'}$ are included for convenience, to avoid the dynamical coupling of modes with different values of the label $p$.

Concerning now the flat FLRW model, the spectrum of the corresponding Dirac operator -- with  compactification period $l_{0}$ -- consists of the following sequence of eigenvalues \cite{dtorus,ginoux}:
\begin{align}\label{eigenvtorus}
\pm\omega_{k}=\pm \frac{2\pi}{l_{0}} \left|\vec{k}+\vec{\tau}\right|, \qquad \vec{\tau}=\frac{1}{2}\sum_{j=1}^{3}\epsilon^{j}\vec{v}_{j}, \qquad \vec{k}\in\mathbb{Z}^{3},
\end{align}
where the three $\vec{v}_{j}$'s form the standard  orthonormal basis for the lattice  $\mathbb{Z}^{3}$, and the three numbers  $\epsilon^{j}\in\{0,1\}$ characterize each of the possible choices of spin structure\footnote{These spin structures determine the periodicity or antiperiodicity of the Dirac field in each of the orthogonal directions that define $T^3$. If, in harmony with spatial isotropy, one imposes the same global behavior for the field in all these directions, the choice of spin structure is restricted to either $\epsilon_j=0$ or $\epsilon_j=1$ for all $j$.}. Given any such spin structure, we identify the label  $k$ in $\omega_{k}$ (or equivalently in $-\omega_{k}$) with the norm of any of the wave vectors  $\vec{k}\in\mathbb{Z}^{3}$ corresponding to $\omega_{k}$. The degeneracy $g_{k}$ associated with each eigenvalue $\omega_{k}$ (or  $-\omega_k$) does not possess in this case a closed expression. However, well-known results in Riemannian geometry  allow us to conclude that $g_{k}$ grows asymptotically  as $\mathcal{O}(\omega_k^2)$, for unboundedly large $\omega_k$ \cite{flat,eisberg}. Let us then fix a spin structure on $T^{3}$ and choose a set of triads associated with the flat metric. The eigenspinors of the Dirac operator then form a basis of the space of spinors in $T^{3}$, basis that can be made orthonormal with respect to the inner product \eqref{inner3d}. In particular, if one chooses a diagonal triad such that the spin connection 1-form becomes null, the Dirac operator turns out to be the standard one for flat Euclidean space.  Its (left-handed) eigenspinors corresponding to eigenvalues $\pm\omega_{k}$ are
\begin{align}\label{eigens}
w_{\vec{k}\,A}^{(\pm)}(\vec{x})=u^{(\pm)}_{\vec{k}\,A}\exp\left[{i\frac{2\pi}{l_{0}}\big( \vec{k}+\vec{\tau}\big)\cdot\vec{x}}\right], 
\end{align}
where $u^{(\pm)}_{\vec{k}\,A}$ are some $\vec{x}$-independent two-component spinors, subject to the condition that the eigenvalue equation must hold. They can be chosen so that the spinors $w^{(\pm)}_{\vec{k}\,A}$ are normalized in the inner product \eqref{inner3d} and such that
\begin{align}\label{norm}
\int_{\Sigma_0} \text{d}^3\vec{x}\, w^{(+)}_{\vec{k}'A}\epsilon^{AB}w^{(-)}_{\vec{k}B}=0,\qquad \int_{\Sigma_0} \text{d}^3\vec{x}\, w^{(\pm)}_{\vec{k}'A}\epsilon^{AB}w^{(\pm)}_{\vec{k}B}=e^{iC^{(\pm)}_{\vec{k}}}\delta_{\vec{k}',-\vec{k}-2\vec{\tau}} ,
\end{align}
for all $\vec{k},\vec{k}'\neq -\epsilon^{j}\vec{v}_{j}/2$. Summation over repeated indices is assumed\footnote{Except for the index $\vec{k}$ on the right-hand side of the second relation in Equation \eqref{norm}.}. Finally, the constants $C^{(\pm)}_{\vec{k}}$ are some phases that can be chosen conveniently by modifying those of $u^{(\pm)}_{\vec{k}\,A}$. Just like in the case of $S^3$, the chiral projections $\phi_{A}$ and ${\bar{\chi}}_{A'}$ in the current flat case can be expanded in Dirac modes in an analogous fashion as in Equations \eqref{harm1s} and \eqref{harm2s}, with corresponding   Grassmann variables $m_{\vec{k}}$, $r_{\vec{k}}$, $t_{\vec{k}}$, and $s_{\vec{k}}$. 

Returning to the spherical FLRW case, notice that upon introduction of the mode expansions \eqref{harm1s} and \eqref{harm2s} in the Dirac action, and once the second-class constraints of the fermionic system are solved \cite{Dirac}, one ends up with the following symmetric Dirac brackets for the mode variables  \cite{T-N,casal}:
\begin{align}\label{canonicalbrackets}
\{x_{np},\bar{x}_{np}\}=-i, \qquad \{y_{np},\bar{y}_{np}\}=-i,
\end{align}
where the ordered pair  $(x_{np},y_{np})$ stands for  $(m_{np},s_{np})$ or $(t_{np},r_{np})$.  Using Grassmann variational derivatives and requiring stationarity of the action, one obtains Dirac equations for the modes:
\begin{align}\label{1orders}
x_{np}'=i\omega_{n}x_{np}-ima\bar{y}_{np}, \qquad y_{np}'=i\omega_{n}y_{np}+ima\bar{x}_{np},
\end{align}
as well as the complex conjugate versions. One can combine these dynamical equations to obtain decoupled second-order equations, that are actually the same for all modes   $\{x_{np},y_{np}\}$ with the same label $n$. Denoting the mode variables generically  by $\{z_{np}\}$, the resulting equation for given $\omega_{n}$ and (nonzero\footnote{The case $m=0$ is slightly different, although straightforward to handle.}) mass $m$ is the following:
\begin{align}\label{2orders}
z_{np}''=\frac{a'}{a}z_{np}'-\left(\omega_{n}^2+m^{2}a^{2}+i\omega_{n}\frac{a'}{a}\right)z_{np}.
\end{align}

The general solution to this equation is a linear combination of two independent complex solutions, which we write again in the form  $\exp[{i\Theta^{1}_{n}(\eta)}]$ and $\exp[{-i\Theta^{2}_{n}(\eta)}]$. The general expression of the fermionic modes at arbitrary time $\eta$ can then be written as a linear transformation of the corresponding initial values, that assigns a different weight to the two independent solutions of Equation \eqref{2orders}:
\begin{align}\label{evmodes3ds}
&x_{np}(\eta)=\left[\Delta_{n}^{2}e^{i\Theta^{1}_{n}(\eta)}+\Delta^{1}_{n}e^{-i\Theta_{n}^{2}(\eta)}\right]x^{0}_{np}-\left[\zeta^{1}_{n}e^{i\Theta_{n}^{1}(\eta)}-\zeta_{n}^{2}e^{-i\Theta_{n}^{2}(\eta)}\right]\bar{y}^{0}_{np},\nonumber\\ &
y_{np}(\eta)=\left[\Delta_{n}^{2}e^{i\Theta^{1}_{n}(\eta)}+\Delta^{1}_{n}e^{-i\Theta_{n}^{2}(\eta)}\right]y^{0}_{np}+\left[\zeta^{1}_{n}e^{i\Theta_{n}^{1}(\eta)}-\zeta_{n}^{2}e^{-i\Theta_{n}^{2}(\eta)}\right]\bar{x}^{0}_{np},
\end{align}
where $\Delta^{l}_{n}$ and $\zeta_{n}^{l}$, $l=1,2$, are constants that depend on the initial conditions of the independent solutions $\exp[{i\Theta^{1}_{n}(\eta)}]$ and $\exp[{-i\Theta^{2}_{n}(\eta)}]$, and of their derivatives, at the reference time $\eta_0$ (see Ref. \cite{uf1} for explicit expressions and details). 

Just as in the previous section, one can obtain the  asymptotic behavior of the linear transformations \eqref{evmodes3ds} in the ultraviolet regime of large $\omega_n$, and subsequently study the unitary implementability of such transformations at the quantum level. For that matter, let us impose the initial conditions $\Theta_{n}^{l}(\eta_{0})=0$. One obtains the following expressions for the (exponents of the) solutions to Equation \eqref{2orders} \cite{uf1}:
\begin{align}\label{z2}
\Theta^{l}_{n}(\eta)=\omega_{n}\Delta\eta+\frac{i}{2}[1+(-1)^{l}]\ln{\left(\frac{a}{a_0}\right)}+\int_{\eta_{0}}^{\eta}\text{d}\tilde\eta\,\Lambda^{l}_{n}(\tilde\eta),
\end{align}
where $a_{0}=a(\eta_{0})$, $\Delta\eta=\eta-\eta_0$, and the functions $\Lambda^{l}_{n}$ are solutions of an equation of Riccati type which become negligible in the limit of unboundedly large $\omega_n$. In particular, the functions $\Lambda^{l}_{n}$ have a behavior of the type 
$\mathcal{O}(\omega_{n}^{-1})$. The asymptotic values obtained for the constants $\Delta^{l}_{n}$ and $\zeta_{n}^{l}$ are the following \cite{uf1}:
\begin{align}
\Delta^{1}_{n}=& 0,\qquad \Delta^{2}_{n}=1,\\ \zeta^{1}_{n}=& \zeta^{2}_{n}=\zeta_{n}=\frac{ma^2_{0}}{2\omega_{n}a_0+ia_{0}'}=\frac{ma_{0}}{2\omega_{n}}+\mathcal{O}(\omega_{n}^{-2}).
\end{align}

The analysis concerning the fermionic dynamics in the flat case is quite similar, leading to mode solutions that are the analogues of Equations (\ref{evmodes3ds}): 
\begin{align}\label{evmodes3dt}
&x_{\vec{k}}(\eta)=e^{i\Theta^{1}_{k}(\eta)}x^{0}_{\vec{k}}-\zeta_{k}\left[e^{i\Theta_{k}^{1}(\eta)}-e^{-i\Theta_{k}^{2}(\eta)}\right]\bar{y}^{0}_{-\vec{k}-2\vec{\tau}}\, ,\nonumber\\ &
y_{\vec{k}}(\eta)=e^{i\Theta^{1}_{k}(\eta)}y^{0}_{\vec{k}}+\zeta_{k}\left[e^{i\Theta_{k}^{1}(\eta)}-e^{-i\Theta_{k}^{2}(\eta)}\right]\bar{x}^{0}_{-\vec{k}-2\vec{\tau}}\,.
\end{align}
The corresponding asymptotic expressions, in the limit of large $\omega_k$, are
\begin{align}
\zeta_{k}=\frac{ma^2_{0}}{2\omega_{k}a_0+ia_{0}'}=\frac{ma_{0}}{2\omega_{k}}+\mathcal{O}(\omega_{k}^{-2})
\end{align}
and
\begin{align}
\Theta^{l}_{k}(\eta)=\omega_{k}\Delta\eta+\frac{i}{2}[1+(-1)^{l}]\ln{\left(\frac{a}{a_0}\right)} +\int_{\eta_{0}}^{\eta}\text{d}\tilde\eta\,\Lambda^{l}_{k}(\tilde\eta), 
\end{align}
where the functions $\Lambda^{l}_{k}$ are solutions of a Riccati equation with an asymptotic behavior of the type 
$\mathcal{O}(\omega_{k}^{-1})$. 

\subsection{Fock quantization and unitary evolution}

Considering both the spherical and the flat FLRW cosmologies, we proceed now to characterize all Fock representations for the Dirac field which satisfy the following requirements. First, the associated vacua must be invariant under the action of the natural symmetries of the system, among them the spatial isometries (and helicity generated spin rotations, in the flat case). Secondly, the Fock quantizations are required to admit a (nontrivial) unitary implementation of the dynamics at the quantum level. 

Let us start by analyzing the behavior of Dirac spinors in the  spherical case, when the isometry transformations of $S^3$ are applied. The transformation group is then $SO(4)$, or equivalently the double cover $\text{Spin}(4)=SU(2)\times SU(2)$, with action in  $S^{3}$ defined by means of a Clifford multiplication. As a consequence, $\text{Spin}(4)$ acts on the cross-sections of the spinor bundle on $S^{3}$ \cite{yasushi}. Notice that this action, when viewed on the four-component Dirac spinor, is reducible to two blocks: the action of $\text{Spin}(4)$ over spinors  $\phi^{A}$, and the complex conjugate action over spinors $\bar{\chi}_{A'}$. Both such representations of $\text{Spin}(4)$ are unitary with respect to the inner product \eqref{inner3d}, and it follows that each block is further decomposable in a direct sum of irreducible representations.

Invariant vacua are associated with invariant complex structures, and we therefore seek complex structures which commute with the action of  $\text{Spin}(4)$ over spinors. To begin with, given the decompositions  \eqref{harm1s} and \eqref{harm2s}, any complex structure can be seen as an infinite dimensional matrix in the basis  formed by the modes $\{m_{np},\bar{r}_{np},t_{np}, \bar{s}_{np}\}$. From this point on, we follow the analysis carried out in Ref. \cite{yasushi}, concerning the eigenspaces of the Dirac operator on $S^{3}$. Consider the action of $\text{Spin}(4)$ on spinors of the type $\phi^{A}$. Using  Frobenius  theorem \cite{frob}, it is shown in Ref. \cite{yasushi} that each eigenspace corresponds exactly  to a representation space of one of the irreducible representations of  $\text{Spin}(4)$ on spinors. Moreover, each such irreducible component shows up in the direct sum with  multiplicity equal to one. This means that the representation spaces generated for each  $n$ by the sets
\begin{align}\label{eigensp}
\{\rho^{np}_{A}\}_{p=1,...,g_{n}} \qquad \text{and} \qquad \{{\bar{\sigma}}^{np}_{A}\}_{p=1,...,g_{n}}
\end{align}
provide two irreducible representations which are necessarily inequivalent. Likewise, the representations associated with the sets
\begin{align}\label{eigenspz}
\{\bar{\rho}^{np}_{A'}\}_{p=1,...,g_{n}} \qquad \text{and} \qquad \{\sigma^{np}_{A'}\}_{p=1,...,g_{n}}
\end{align}
simply reproduce those generated by the sets \eqref{eigensp}, and therefore each irreducible representation associated, for each $n$, with one of the sets \eqref{eigenspz} is  unitarily equivalent to a corresponding one coming from the sets \eqref{eigensp}. All this information (combined with an inspection of the dynamical mode equations \cite{uf2}) allows us to apply Schur's lemmas \cite{Reps} to conclude that a complex structure that commutes with the action of the group of isometries of $S^{3}$ over the space of Dirac spinors cannot mix modes $m_{np}$, $\bar{r}_{np}$, $t_{np}$, and $\bar{s}_{np}$ with different values of $n$. Moreover, within the subspace associated with a fixed value of $n$, such complex structures cannot mix the modes $\{m_{np},\bar{s}_{np}\}$ with $\{t_{np},\bar{r}_{np}\}$, since the two sets provide inequivalent irreducible representations of  $\text{Spin}(4)$. Let us consider first, for each given $n$, the subspace generated by the modes $\{m_{np},\bar{s}_{np}\}$. The restriction of an invariant complex structure to any such subspace can then be characterized by means of four linear maps, relating in all possible pairings the two subspaces generated by $\{m_{np}\}$ and by $\{\bar{s}_{np}\}$. Each of these  four maps has to be proportional to the identity, as ensured by Schur's lemma. Finally, similar considerations can be applied to the subspace generated by the modes $\{t_{np},\bar{r}_{np}\}$ \cite{uf2}.

We now turn to the flat FLRW model, and consider the action of isometries of $T^{3}$ on the Dirac field. Restricting our attention to continuous transformations, the isometry group is generated by constant translations along each of the orthogonal  directions of the 3-torus. A general translation on the torus is thus $\vec{x}\rightarrow \vec{x}+\vec{\theta}$, where for each component we have $2\pi\theta_{\alpha}/l_{0}\in S^1$. For any given choice of spin structure on $T^3$, one can easily check that a general translation simply results into the following transformation:
\begin{align}
w_{\vec{k}\,A}^{(\pm)}(\vec{x})\longrightarrow e^{i2\pi\vec{k}\cdot\vec{\theta}/l_{0}}e^{i2\pi\vec{\tau}\cdot\vec{\theta}/l_{0}}w_{\vec{k}\,A}^{(\pm)}(\vec{x})
\end{align}
in each of the elements \eqref{eigens} of the basis of (left-handed) eigenspinors. 
For each $\vec{k}\in\mathbb{Z}^{3}$ there are therefore two copies of the same 1-dimensional complex irreducible representation, with  different $\vec{k}$'s giving rise to   inequivalent representations \cite{uf3}. Again, one can perform the same analysis for the spinors of opposite chirality. Then, taking into account the mode decomposition of the Dirac field $\Psi$, and using again  Schur's lemma, one concludes the following. A complex structure that commutes with the action of translations on $T^3$ can at most mix the modes   $(m_{\vec{k}},\bar{s}_{-\vec{k}-2\vec{\tau}},t_{-\vec{k}-2\vec{\tau}},\bar{r}_{\vec{k}})$ among themselves, for each fixed  $\vec{k}\in\mathbb{Z}^{3}$, and is trivial otherwise.

In the flat case there is an additional symmetry of the Dirac system, following from the conservation of helicity in the evolution of the Dirac field in conformal time  $\eta$. In fact, one can consider the projection of the spin angular momentum in the direction of the linear momentum of the particle, a projection that (except for the subspace generated by the modes with $\omega_{k}= 0$) defines the helicity operator 
$\mathfrak{h}$ \cite{halzen}:
\begin{align}
\mathfrak{h}=[-\vec{\nabla}^{2}]^{-1/2}\begin{pmatrix} -i\vec{\sigma}\cdot\vec{\nabla} & 0 \\ 0 & -i\vec{\sigma}\cdot\vec{\nabla} \end{pmatrix}.
\end{align}
Here, $\vec{\nabla}$ is the standard 3-dimensional Euclidean  gradient and $\vec{\sigma}$ denotes a vector with components given by the Pauli matrices. Eigenspinors of $\mathfrak{h}$ with eigenvalues $+1$ or $-1$ are said to have positive or negative helicity, respectively. With the choice of gauge such that the spin connection vanishes,  it turns out that the matrix blocks of  $\mathfrak{h}$ (apart from the factor $[-\vec{\nabla}^{2}]^{-1/2}$) correspond exactly to the  Dirac operator on $T^{3}$. One can then  check that the  positive helicity part of the  Dirac field $\Psi$ is generated  by the coefficients   $m_{\vec{k}}$ and $\bar{s}_{\vec{k}}$, for all $\vec{k}\in\mathbb{Z}^{3}$ different from $\vec{\tau}$. On the other hand, the negative helicity contribution is generated by the modes $t_{\vec{k}}$ and $\bar{r}_{\vec{k}}$, for all $\vec{k}\in\mathbb{Z}^{3}$ different from $\vec{\tau}$. A simple inspection of the equations of motion  \eqref{1orders} shows that helicity is indeed a conserved quantity. Therefore, one can include, as an additional symmetry of the fermionic system, the 1-parameter group of spin rotations generated by helicity, by means of the complex exponentiation of $\mathfrak{h}/2$ multiplied by the angle of rotation. Such group is immediately  unitary with respect to the inner product \eqref{innerd}, since the operator  $\mathfrak{h}$ is essentially self-adjoint. It follows that the unitary implementation of this symmetry at the quantum level is ensured whenever the complex structure that defines the quantization  does not mix positive helicity modes with  negative helicity ones.

The complex structures characterized above define the sets of creation and annihilation operators that provide invariant Fock representations of the CARs for the Dirac field in the considered homogeneous and isotropic scenarios. In the spherical case, let us denote the classical counterparts of the annihilation operators for particles and antiparticles by $a^{(x,y)}_{np}$ and $b^{(x,y)}_{np}$, respectively. The corresponding creation variables are the complex conjugate ones,  $\bar{a}_{np}^{(x,y)}$ and $\bar{b}_{np}^{(x,y)}$. In the case of $T^3$, we denote the annihilation variables by  $a_{\vec{k}}^{(x,y)}$ and $b_{\vec{k}}^{(x,y)}$, respectively for particles and antiparticles. We recall that the  pairs $(x,y)$ (with the appropriate labels) denote any of the ordered pairs of mode coefficients $(m,s)$ or $(t,r)$. In the following, we consider all the possible (time-dependent) families of fermionic creation and annihilation variables selected by invariant complex structures.

In the case of $S^{3}$, the creation and annihilation variables in question can then be written as
\begin{align}\label{anni3ds}
\begin{pmatrix} a_{np}^{(x,y)} \\ \bar{b}^{(x,y)}_{np} \end{pmatrix}_{\!\!\eta}=\begin{pmatrix}f_{1}^ {n}(\eta) & f_{2}^{n}(\eta) \\ g_{1}^ {n}(\eta) & g_{2}^{n}(\eta) \end{pmatrix}\begin{pmatrix} x_{np} \\ \bar{y}_{np} \end{pmatrix}_{\!\!\eta}.
\end{align}
Once more, the label $\eta$ denotes dependence on conformal time. Notice that the time-dependent functions $f_{l}^{n}$ and $g_{l}^{n}$ (with $l=1,2$) may differ for the pairs of modes $(m_{np},\bar{s}_{np})$ and $(t_{np},\bar{r}_{np})$, although this is not explicit in the notation. The following relations must again be satisfied:
\begin{align}\label{sympl3ds}
|f_{1}^{n}|^{2}+|f_{2}^{n}|^{2}=1, \qquad |g_{1}^{n}|^{2}+|g_{2}^{n}|^{2}=1, \qquad f_{1}^{n}\bar{g}^{n}_{1}+f_{2}^{n}\bar{g}^{n}_{2}=0,
\end{align}
such that anticommutators  of the type \eqref{car} are obtained at the quantum level. 

Turning to the flat case, the relations  \eqref{anni3ds} are replaced with 
\begin{align}\label{anni3dt}
\begin{pmatrix} a_{\vec{k}}^{(x,y)} \\ \bar{b}_{\vec{k}}^{(x,y)} \end{pmatrix}_{\!\!\eta}=\begin{pmatrix}f_{1}^{\vec{k}}(\eta) & f_{2}^{\vec{k}}(\eta) \\ g_{1}^{\vec{k}}(\eta) & g_{2}^{\vec{k}}(\eta) \end{pmatrix}\begin{pmatrix} x_{\vec{k}} \\ \bar{y}_{-\vec{k}-2\vec{\tau}} \end{pmatrix}_{\!\!\eta},
\end{align}
where conditions analogous to Equation (\ref{sympl3ds}) again apply.

When evaluated at different times, the sets of variables  \eqref{anni3ds} are related to each other by means of dynamical Bogoliubov transformations. The  general form of such linear transformations can be obtained taking into account Equations \eqref{evmodes3ds} for the evolution of the fermionic modes  in the spherical case. In general terms, the Bogoliubov transformation relating the annihilation and creation variables at the initial time $\eta_{0}$ with those at any time  $\eta$ is given by a sequence of blocks $\mathcal{B}_{n}$ such that
\begin{align}\label{bog3ds}
\begin{pmatrix} a_{np}^{(x,y)} \\ \bar{b}^{(x,y)}_{np} \end{pmatrix}_{\!\!\eta}=\mathcal{B}_{n}(\eta,\eta_{0})\begin{pmatrix} a_{np}^{(x,y)} \\ \bar{b}^{(x,y)}_{np} \end{pmatrix}_{\!\!\eta_{0}}, \qquad \mathcal{B}_{n}=\begin{pmatrix} \alpha_{n}^{f} & \beta_{n}^{f} \\ \beta_{n}^{g} & \alpha_{n}^{g} \end{pmatrix}.
\end{align}
The absolute values of the  coefficients $\beta_{n}^{f}$ and $\beta_{n}^{g}$ have the following expression 
\cite{uf1,uf2}:
\begin{align}\label{beta3ds}
|\beta_{n}^{h}(\eta,\eta_0)|=&\Bigg|\left[-h_{1}^{n}\bigg(h_{2}^{n,0}+\zeta_n h_{1}^{n,0}\bigg)e^{i\int \Lambda^{1}_{n}}+\bar\zeta_n h^{n}_{2}h_{2}^{n,0}\frac{a}{a_0} e^{i\int\bar{\Lambda}^{2}_{n}}\right] e^{i\omega_{n}\Delta\eta}\nonumber\\& +\left[h_{2}^{n}\bigg(h_{1}^{n,0} -\bar\zeta_n h_{2}^{n,0}\bigg)e^{-i\int \bar{\Lambda}^{1}_{n}}+\zeta_n h^{n}_{1}h_{1}^{n,0}\frac{a}{a_0}e^{-i\int\Lambda^{2}_{n}}\right] e^{-i\omega_{n}\Delta\eta}\Bigg|,\end{align}
where the integrals are over conformal time from $\eta_0$ to $\eta$, $h$ denotes either $f$ or $g$, and the superscript $0$ stands for evaluation at the initial time.

Analogous considerations, of course, apply to the flat case, with dynamical Bogoliubov transformations between variables of the type \eqref{anni3dt} that are characterized by matrices $\mathcal{B}_{\vec{k}}$, with corresponding coefficients $\beta_{\vec{k}}^{f}$ and $\beta_{\vec{k}}^{g}$, for which the explicit expressions can be found in Ref. \cite{uf3}.

Considering for instance the Fock representation defined by the annihilation and creation variables at the initial time $\eta_{0}$ (or equivalently by the associated complex structure), the  dynamical Bogoliubov transformations \eqref{bog3ds} can be implemented on the corresponding Fock space by means of unitary quantum operators if and only if \cite{shale,derez}
\begin{align}\label{ucond3ds}
\sum_{n}g_{n}|\beta_{n}^{h}(\eta,\eta_0)|^{2}<\infty \qquad \text{for} \qquad h=f, g.  
\end{align}
Actually, it is sufficient  to ensure the above condition for either $h=f$ or $h=g$, since it follows again from relations \eqref{sympl3ds} that         $|\beta_{n}^{g}(\eta,\eta_0)|=|\beta_{n}^{f}(\eta,\eta_0)|$ \cite{uf1}. Let us then fix $h$ to be either $f$ or $g$. The unitary implementability of the dynamics therefore depends on the asymptotic behavior of the beta coefficients in the limit of large $\omega_n$, that in turn depends on the behavior  of the sequences $h_{l}^{n}$. A detailed analysis, carried out in Ref. \cite{uf2}, shows that, apart from an uninteresting alternative which would effectively trivialize the quantum dynamics in a similar way as it was discussed in the previous section, the fulfillment of the unitarity condition  \eqref{ucond3ds} requires that the functions $h_{l}^{n}$ behave asymptotically as
\begin{align}\label{unith}
h^{n}_{l}=(-1)^{l+1}\frac{ma}{2\omega_n}e^{iH^{n}_{\tilde{l}}}+\vartheta^{n}_{h,l},\qquad h^{n}_{\tilde{l}}=e^{iH^{n}_{\tilde{l}}}+\mathcal{O}(\omega_k^{-2}),
\end{align}
where $\{l,\tilde{l}\}$ is the set $\{1,2\}$. Moreover, the sequences   $\vartheta^{n}_{h,l}$ must be  square summable, including degeneracy. 

Before continuing with our discussion, a comment is in order. Since we have not put any restriction  on the global asymptotic behavior of the sequences $h^{n}_{l}$ for fixed $l$, it is possible that neither $h^{n}_{1}$ nor $h^{n}_{2}$ actually converges over $\mathbb{N}$. In fact, the sum \eqref{ucond3ds} can be made finite with  $h^{n}_{1}$ taking the role of  $h^{n}_{l}$ in Equation \eqref{unith} for $n$ in a subset of $\mathbb{N}$ and $h^{n}_{2}$ taking that role over a complementary subset (modulo finite subsets of $\mathbb{N}$). Hence, the above behavior of  $h^{n}_{l}$ is required only for $n\in\mathbb{N}_{l}\subset\mathbb{N}$, with $\mathbb{N}_{1}\cup \mathbb{N}_{2}=\mathbb{N}$ modulo finite subsets (including the possibility of one of the subsets $\mathbb{N}_{l}$ being empty). 

The analysis concerning the flat case  follows similar lines, apart from a careful handling of the already mentioned issue of the accidental degeneracy of the Dirac eigenspaces (for full details, see Ref. \cite{uf3}). The conclusion is that the requirement of (nontrivial) unitary implementation of the dynamics completely fixes the explicit dependence on time in the dominant part of the Dirac field, in a very similar way as in the spherical case [see Equation \eqref{unith}], for an infinite number of modes in the asymptotic sector of large $\omega_k$. Hence, the time-dependent scaling that must be introduced in the mode dynamics of the Dirac field, such that the remaining part of the evolution can be unitarily implemented, is completely  fixed (in absolute value) at dominant order. In particular, the scaling factor for each mode is similarly determined in terms of $ma/(2\omega_{k})$, and also includes the global term $a^{-3/2}$, introduced in the mode expansions \eqref{harm1s} and \eqref{harm2s} adapted to the case of $T^3$.

\subsection{Uniqueness of the quantization}

In the previous subsection we have characterized the Fock quantizations of the Dirac field which allow a unitary implementation of the dynamics and which possess vacua that are invariant under the natural symmetries of the considered cosmological models. We now show that, for each of these models, all such quantizations are unitarily equivalent. Direct consequences are the removal of quantization ambiguities typically present in QFT and the selection of a very specific, well-defined notion of quantum evolution.

Let us start with the case of spherical sections. One of the simplest choices of functions that satisfy the conditions derived above for a unitarily implementable quantum dynamics is 
\begin{align}\label{reference3ds}
f_{1}^{n}=\frac{ma}{2\omega_{n}},\qquad f_{2}^{n}=\sqrt{1-(f_{1}^{n})^2},\qquad g_{1}^{n}=f_{2}^{n}, \qquad g_{2}^{n}=-f_{1}^{n},
\end{align}
for all $n\in\mathbb{N}$ and for both pairs $(m_{np},\bar{s}_{np})$ and $(t_{np},\bar{r}_{np})$. We take this choice as defining our  \emph{reference}
family of complex structures. Let us then consider any other family of invariant complex structures allowing a unitary implementation of the dynamics. Such family is defined by certain annihilation and creation variables, $\tilde{a}_{np}^{(x,y)}$ and $\bar{\tilde{b}}_{np}^{(x,y)}$, with coefficients $\tilde{h}_{l}^{n}$ (for $h$ identified with $f$ or $g$ and for $l=1,2$) that have the asymptotic behavior  described at the end of the preceding subsection. In particular, the subdominant sequences $\vartheta^{n}_{\tilde{h},l}$ appearing in Equation \eqref{unith} are square summable (degeneracy included) over subsequences $\mathbb{N}_l$. The relation between this family and the reference one  is given by a Bogoliubov transformation determined by a sequence of matrices  $\mathcal{K}_n$  such that
\begin{align}\label{bogK3ds}
\begin{pmatrix} \tilde a_{np}^{(x,y)} \\ \bar{\tilde{b}}^{(x,y)}_{np} \end{pmatrix}_{\!\!\eta}=\mathcal{K}_{n}(\eta)\begin{pmatrix} a_{np}^{(x,y)} \\ \bar{b}^{(x,y)}_{np}  \end{pmatrix}_{\!\!\eta}, \qquad \mathcal{K}_{n}=\begin{pmatrix} \kappa_{n}^{f} & \lambda_{n}^{f} \\ \lambda_{n}^{g} & \kappa_{n}^{g} \end{pmatrix}.
\end{align}
One can check that the absolute values of the nondiagonal elements of these matrices have the following expression  \cite{uf1,uf2}:
\begin{align}\label{kl3ds}
|\lambda^{h}_{n}|=|\tilde{h}_{1}^{n}h_{2}^{n}-\tilde{h}_{2}^{n}h_{1}^{n}|,
\end{align}
where $h$ stands again for both types of functions $f$ and $g$. The two Fock quantizations, namely the reference one associated with the family \eqref{reference3ds} and the one defined by the coefficients $\tilde{h}_{l}^{n}$, are unitarily equivalent if and only if the Bogoliubov transformation \eqref{bogK3ds} is itself unitarily implementable, a demand that is equivalent to the fulfillment of the conditions
\begin{align}\label{ucond23ds}
\sum_{n}g_{n}|\lambda_{n}^{f}(\eta)|^{2}<\infty \qquad \text{and} \qquad \sum_{n}g_{n}|\lambda_{n}^{g}(\eta)|^{2}<\infty,
\end{align}
for all $\eta$ of interest. Once more, one of the above conditions is redundant, since the equality $|\lambda^{g}_{n}|=|\lambda^{f}_{n}|$ is again ensured by relations \eqref{sympl3ds}. 

Let us then focus on the first condition in  Equation \eqref{ucond23ds} and consider the situation such that the asymptotic behavior  \eqref{unith} applies to the functions  $\tilde{f}_{l}^{n}$. The alternative, with Equation \eqref{unith} applying to the functions $\tilde{g}_{l}^{n}$, can be treated in a completely analogous way.
One can show from  Equation \eqref{kl3ds} that, in the limit  of unboundedly large $\omega_n$, the coefficients $\lambda^{f}_{n}$ have the following behavior on the subsequence $\mathbb{N}_{1}$:
\begin{align}\label{lambdafg3ds}
|\lambda^{f}_{n}|=|\vartheta^{n}_{\tilde{f},1}|+\mathcal{O}(\omega_{n}^{-2}).
\end{align}
Since by hypothesis $\vartheta^{n}_{\tilde{f},1}$ is square summable (including the degeneracy $g_n$) over $\mathbb{N}_{1}$, it follows immediately that the condition for unitary equivalence is satisfied if the set $\mathbb{N}_{2}$ is finite (or empty). Suppose now that $\mathbb{N}_{2}$ is an infinite set. It is clear that $\lambda_{n}^{f}$ behaves asymptotically like $\mathcal{O}(1)$ for $n\in\mathbb{N}_{2}$, and therefore the summability required in Equation \eqref{ucond23ds} cannot be attained, leading to apparently  inequivalent quantizations. However, this inequivalence stems, as it happened in the previous section, from the fact that the quantization associated with the coefficients $\tilde{h}_{l}^{n}$ defines a convention for the concept of particles and antiparticles which is completely the opposite, for an infinite number of modes, of the convention corresponding to the reference quantization. Once both conventions are reconciled, the quantizations are seen to be physically equivalent. In fact, suppose that, in our reference quantization \eqref{reference3ds}, we switch the convention concerning particles and antiparticles for all modes corresponding to  $\mathbb{N}_{2}$. This redefinition is effectively attained with the interchange $f_{l}^{n}\leftrightarrow g_{l}^{n}$ ($n\in\mathbb{N}_{2}$), as follows from the definition \eqref{anni3ds}. Then, the  behavior of the new coefficients $\lambda_{n}^{f}$  would no longer be $\mathcal{O}(1)$ on $\mathbb{N}_{2}$, but they would behave (in norm) as $|\vartheta^{n}_{\tilde{f},2}|+\mathcal{O}(\omega_{n}^{-2})$ instead. Since, also by the hypothesis of a unitarily implementable evolution, $\vartheta^{n}_{\tilde{f},2}$ is square summable (degeneracy included) over $\mathbb{N}_{2}$, one concludes  that conditions \eqref{ucond23ds} are now satisfied, confirming therefore the unitary equivalence between the two Fock quantizations, after the two conventions concerning  particles and antiparticles have been harmonized.

The analysis of the uniqueness of the Fock quantization of the Dirac field in the flat FLRW case proceeds in a similar fashion. One can again choose  a reference complex structure allowing a unitary implementation of the dynamics, namely the one characterized by the following matrix elements in Equation \eqref{anni3dt}, for all $\omega_k\neq 0$:
\begin{align}\label{ref}
f_{1}^{k}=\frac{ma}{2\omega_{k}}, \qquad f_{2}^{k}=\sqrt{1-\left(f_{1}^{k}\right)^{2}}, \qquad g_{1}^{k}=f_{2}^{k},\qquad g_{2}^{k}=-f_{1}^{k}.
\end{align}
Applying the same type of arguments as above, one can prove \cite{uf3} that, as it happened in the case of $S^3$, once a convention concerning particles and antiparticles is fixed, the condition that there exist a nontrivial unitary implementation of the dynamics is sufficient to ensure the unitary equivalence of all Fock quantizations associated with invariant vacua. 

Let us conclude with a brief comment on a key difference between the current study and previous analogous works on quantum scalar fields in FLRW cosmologies \cite{cmv2,cmov1,cmov2,flat,MenaMarugan:2013tba,flat2}. Like in the scalar field case, our criteria uniquely fix not only  the Fock representation once a suitable set of variables for the field has been chosen, but actually they greatly reduce as well the ambiguity in this choice of variables that arises from time-dependent linear redefinitions. Considering, for instance, a spherical spatial topology, this affects the global scaling introduced in the decompositions \eqref{harm1s} and \eqref{harm2s}, as it does for the scalar field, but now it affects also the scalings in the particle and antiparticle contributions that are induced by the time dependence of the functions $f_{l}^{n}$ and  $g_{l}^{n}$ subject to the unitarity conditions \eqref{unith}. As such, the Dirac field presents specific and different time-dependent scalings in its particle and antiparticle parts, that are also different for each of the two chiralities, introducing an aspect which is absent in the scalar field analysis. For the scalar field, the requirement of unitary dynamics imposes a global scaling of the original field variable, such that the scaled field in practice behaves like a conformally coupled field, and one might  wrongly believe that the possibility of attaining a unitarily implementable dynamics is somehow constrained by the availability of a conformal  symmetry in the scaled theory (at least in the ultraviolet regime). The work on Dirac fields definitely puts aside such type of misconceptions.

\section{Hamiltonian backreaction of Dirac perturbations in hLQC}\label{LQC}

The previous uniqueness results about the Fock quantization of Dirac fields in FLRW cosmologies are of special importance beyond the context of QFT. Actually, they can be used (and further developed) to confer great robustness to the full quantization of a homogeneous and isotropic cosmological spacetime perturbed with small matter inhomogeneities described by a Dirac field, in the context of hybrid quantum cosmology. As mentioned in the Introduction, this framework for the quantum description of cosmological systems employs techniques from a theory of quantum cosmology (here we focus our attention on the case of loop quantum cosmology) for quantizing the spatially homogeneous zero modes of the geometry, while the inhomogeneous fields are quantized using standard Fock representations. In this setting, the features of the resulting quantum theory and its predictions are strongly affected by the precise knowledge obtained not only about a unique preferred Fock space for the fields, but also about which part of their degrees of freedom displays a genuine unitary quantum evolution when one reaches regimes where the cosmological background behaves classically. This knowledge serves in hLQC to separate in a specific way this homogeneous background from the variables that describe the fermionic perturbations, and such splitting can be refined by further imposing some physically sound properties on the full quantum system, such as a proper definition of the fermionic part of the Hamiltonian operator. Along these lines, in this section we are going to revisit the main results in the hybrid loop quantization of a flat homogeneous and isotropic cosmology with fermionic perturbations, in particular in what concerns the study of the Hamiltonian operator and the consequences on the quantum backreaction of the fermionic matter on the cosmological background.

\subsection{Fermionic perturbations in flat FLRW: Splitting of the phase space}

Let us start by considering the Einstein-Hilbert action restricted to symmetry-reduced universes with a metric given in Equation \eqref{flrw}, taking ${}^{0}h_{\alpha\beta}$ as the Euclidean metric in coordinates adapted to the spatial homogeneity, and letting the lapse function, that we call $N_0$, be arbitrary. We particularize again the discussion to a topology of the flat spatial sections equal to the compact $T^3$-topology. In order to include standard inflationary scenarios in our system, we minimally couple a homogeneous scalar field $\phi$ with potential $V(\phi)$, that plays the role of an inflaton. In the canonical ADM framework, the degrees of freedom of this cosmological model can be described with the scale factor $a$, the inflaton $\phi$, and their canonical momenta, respectively denoted by $\pi_{a}$ and $\pi_{\phi}$. On classical FLRW solutions, these variables are subject only to one constraint, arising from the zero mode of the Hamiltonian constraint, that generates global time reparameterizations:
\begin{align}
\label{H02}
H_{|0} = \frac{1}{2l_0^3 a^3}\left[\pi_\phi^2-\frac{4\pi}{3}a^2\pi_a^2+2l_0^6 a^6 V(\phi)\right],
\end{align}
where we recall that $l_0$ is the compactification length of $T^3$.

To include inhomogeneous fermionic content in this cosmological model, we minimally couple a Dirac field $\Psi$ of mass $m$ and treat it entirely as a perturbation (including its purely homogeneous part, if it had any). For physical completeness, one may also introduce purely inhomogeneous and anisotropic perturbations of the metric and the inflaton field\footnote{The zero modes of the metric and scalar field can be conveniently isolated and accounted for in the scale factor and homogeneous inflaton, owing to the compactness of the spatial sections.}. Within this perturbative hierarchy, we conduct the analysis at the lowest nontrivial order and thus we truncate the whole action of the system (and its associated symplectic structure) at quadratic order in all of the perturbations. Since the Dirac action is precisely quadratic in the fermionic field, at this order it only couples with the homogeneous sector of the cosmology\footnote{Henceforth, to simplify the terminology and shorten the notation we will refer to the variables that describe this homogeneous and isotropic background as FLRW variables, even when they are not evaluated on classical solutions. In fact, in this section they are rather generic canonical variables, subject to being represented as quantum operators in a Hilbert space.}. Therefore, in practice, we can treat its associated spinor bundle as if it were defined on a pure (spatially flat) FLRW universe. Furthermore, this coupling implies that the Dirac field is a perturbative gauge invariant at our order of truncation, namely it is independent of perturbatively linear coordinate redefinitions that respect the manifest homogeneity of the background.

In order to take advantage of the previous results about the uniqueness of the Fock quantization of the Dirac field, and to allow for a convenient spatial mode expansion of the fermionic fields, we partially fix again the local Lorentz gauge imposing the condition $n^{\mu}e_{\mu}^{a}=\delta^{a}_0$ on the tetrad of our background. This allows us to generically perform a decomposition of the two chiral components of the fermionic perturbations as that given in Equations \eqref{harm1s} and \eqref{harm2s} (after replacing the eigenspinors of the Dirac operator on $S^3$ by their analogues for $T^3$). Recall that these behave under local gauge transformations as $SU(2)$ spinors defined in $T^3$, with a spin structure that is given by the choice of the vector $\vec{\tau}$ [c.f. Equation \eqref{eigenvtorus}]. Of equal importance is the fact that this partial gauge fixing eliminates all the nontrivial canonical brackets between the homogeneous FLRW geometry and the (rescaled) Dirac field \cite{T-N}. Then, the only nonvanishing brackets between the fermionic mode coefficients $m_{\vec{k}}$, $\bar{r}_{\vec{k}}$, $t_{\vec{k}}$, and $\bar{s}_{\vec{k}}$ (and their complex conjugates) are
\begin{align}\label{canonicalbracketst3}
\{x_{\vec{k}},\bar{x}_{\vec{k}}\}=-i, \qquad \{y_{\vec{k}},\bar{y}_{\vec{k}}\}=-i,
\end{align}
where $(x,y)$ again denotes any of the two possible ordered pairs of mode coefficients $(m,s)$ or $(t,r)$ (omitting their associated wave vector labels).
The introduction of this fermionic mode expansion in the Dirac Hamiltonian coupled to our considered background gives rise to the following contribution to the total Hamiltonian of the system:
\begin{align} &N_0H_D=N_0\left[\delta_{\vec{0}}^{\vec{\tau}} H_{\vec{0}} + \sum_{\vec{k}\neq\vec{\tau}}\sum_{(x,y)} H^{(x,y)}_{\vec{k}}\right] ,\\
\label{H0F}
&H_{\vec{0}}= m  \big( {s}_{\vec{0}} {\bar r}_{\vec{0}} + r_{\vec{0}} {\bar s}_{\vec{0}} + m_{\vec{0}} {\bar t}_{\vec{0}} + t_{\vec{0}} {\bar m}_{\vec{0}} \big),\\
\label{HkF}
&H^{(x,y)}_{\vec{k}}= 
m  \big( {y}_{-\vec{k}-2\vec\tau} x_{\vec{k}} + {\bar{x}}_{\vec{k}} {\bar y}_{-\vec{k}-2\vec\tau} \big) -  \frac{\omega_k}{a} \big( {\bar{x}}_{\vec{k}} x_{\vec{k}} - y_{\vec{k}} {\bar y}_{\vec{k}} \big).
\end{align}
As we have explained above, $N_0$ is the homogeneous lapse function of the FLRW background. It follows that the Dirac perturbations, at our considered order of quadratic truncation in the action, contribute only to the global zero mode of the Hamiltonian constraint of the entire system. Explicitly, if one ignores the rest of perturbative fields in our model (which do not couple to the fermionic field at quadratic order), the zero mode of the Hamiltonian constraint is given by the sum of $H_{|0}$ and $H_D$.

At this point in the discussion, it is worth remarking that there exists an inherent freedom in the description of the cosmological model at hand. Treating the system formed by the FLRW  universe and its perturbations as a whole entity, one can always mix the different sectors of the phase space by means of canonical transformations. Even if this mixing does not affect the physical behavior of the system at the classical level at the end of the day, the freedom in identifying the sets of basic variables that describe each sector can strongly affect the properties of the hybrid quantization, given that the perturbative fields are quantized with a different type of representation (\`a la Fock) than the homogeneous background (with loop techniques). Focusing the attention on the choice of splitting between the FLRW sector and the fermionic one, one can understand this as the specific assignment of how each of these two types of degrees of freedom contributes to the dynamics of the entire system. To take into account this panorama, and with the aim put on characterizing the Fock representation for the fermionic perturbations, we introduce general families of annihilation and creation variables of the form \eqref{anni3dt} where the coefficients $f_1^{\vec{k}},f_2^{\vec{k}},g_1^{\vec{k}},$ and $g_2^{\vec{k}}$ are now taken as functions of the canonical variables that describe the geometric sector of the cosmological background, understood hereafter as the sector that corresponds to the scale factor and its momentum\footnote{One can generalize the analysis to coefficients that are functions also of the inflaton and its momentum, thus allowing for a dependence on all the degrees of freedom of the FLRW background. However, this generalization is not necessary for our discussion, except at some point in Section \ref{diagonalization}, where we comment on it explicitly.}. Furthermore, these functions are subject to the conditions
\begin{align}
\label{g1,g2,f2}
g^{\vec{k}}_{1}= e^{i G^{\vec{k}}}{\bar f}^{\vec{k}}_{2},\qquad g^{\vec{k}}_{2} = - e^{i G^{\vec{k}}} {\bar f}^{\vec{k}}_{1} ,
\qquad
f^{\vec{k}}_{2}= e^{iF^{\vec{k}}_{2}} \sqrt{1- | f^{\vec{k}}_{1} |^2},
\end{align}
where $G^{\vec{k}}$ and $F^{\vec{k}}_{2}$ are phases given by generic real functions of the FLRW geometry. These conditions are nothing but the equivalent of Equation \eqref{sympl3ds} (adapted to the case of $T^3$). Therefore, they guarantee that the definition of the fermionic annihilation and creation variables is a canonical transformation in the fermionic sector of the phase space (namely, if one freezes the FLRW background variables and ignores their Poisson brackets). 

Each of these new families of fermionic variables can codify in different ways the possible splittings in the dynamical behavior, between the homogeneous cosmological geometry and the fermionic perturbations, that preserve the linearity in the perturbative sector. However, when the cosmological system is viewed as a whole dynamical entity, and unless the coefficients $f_1^{\vec{k}}$, $f_2^{\vec{k}}$, $g_1^{\vec{k}}$, and $g_2^{\vec{k}}$ are constant, the new fermionic variables do not form a canonical set with the scale factor and its canonical momentum. These variables must be modified if one wishes to restore the canonical algebra fulfilled by the original basic set for the description of the system. In other words, the canonical transformation that started with the above definition of families of fermionic annihilation and creation variables must be completed. This can be readily done, at our order of quadratic perturbative truncation, by demanding that the symplectic potential of the FLRW and fermionic sectors remain unchanged after the transformation, up to contributions that are of higher perturbative order than quadratic. This procedure leads to the new corrected variables for the scale factor and its momentum:
\begin{align}
\label{modifiedscale}
&\tilde{a}=a +\frac{i}{2}\sum_{\vec{k},(x,y)}[
(\partial_{\pi_{a}}x_{\vec{k}})  {\bar x}_{\vec{k}}+(\partial_{\pi_{a}}{\bar x}_{\vec{k}})  x_{\vec{k}}+(\partial_{\pi_{a}}y_{\vec{k}})  {\bar y}_{\vec{k}}+(\partial_{\pi_{a}}{\bar y}_{\vec{k}})  y_{\vec{k}} ],
\\
&\label{modifiedpi}
{\tilde{\pi}}_{a}=\pi_{a}-\frac{i}{2}\sum_{\vec{k},(x,y)}[
(\partial_{a}x_{\vec{k}})  {\bar x}_{\vec{k}}+(\partial_{a}{\bar x}_{\vec{k}})  x_{\vec{k}}+(\partial_{a}y_{\vec{k}})  {\bar y}_{\vec{k}}+(\partial_{a}{\bar y}_{\vec{k}})  y_{\vec{k}} ].
\end{align} 
Here, the partial derivatives affect only the explicit dependence of the fermionic modes on the cosmological background geometry, via the functions $f_1^{\vec{k}}$, $f_2^{\vec{k}}$, $g_1^{\vec{k}}$, and $g_2^{\vec{k}}$. Thus, the new FLRW variables $\tilde{a}$, $\tilde{\pi}_{a}$, $\phi$, and $\pi_{\phi}$ form a canonical set with the fermionic annihilation and creation variables defined by means of Equations \eqref{anni3dt} and \eqref{g1,g2,f2}.

In the following, for convenience, we restrict our attention to families of annihilation and creation variables defined by coefficients $f_1^{k}$, $f_2^{k}$, $g_1^{k}$, and $g_2^{k}$ that depend on the wave vector $\vec{k}$ only through the corresponding Dirac eigenvalue $\omega_k$. Actually, this restriction comes from the symmetry of the fermionic equations of motion, that ultimately can be related to the isotropy of the background spacetime in the limit where the spatial sections become noncompact. We refer the reader to Ref. \cite{backreaction} for an extended version of the subsequent analysis, including the possibility of a general dependence of the functions $f_1^{\vec{k}}$, $f_2^{\vec{k}}$, $g_1^{\vec{k}}$, and $g_2^{\vec{k}}$ on $\vec{k}$.

\subsection{Fermionic Hamiltonian: Restrictions on the quantization}

Every new set of canonical variables for the FLRW background and the fermionic perturbations, namely every new choice of phase space splitting, naturally contributes to the total Hamiltonian of the system in a different way. In particular, if one writes the original zero mode of the Hamiltonian constraint (that we recall is the sum of $H_{|0}$ and $H_D$) in terms of an arbitrary new canonical set of variables as defined above, and truncates it at quadratic order in the perturbations, one obtains \cite{backreaction,fermidiagonalization}
\begin{align}\label{hamconstr}
H_{|0}(\tilde{a},\tilde{\pi}_{a},\phi,\pi_{\phi})+\tilde{H}_D(\tilde{a},\tilde{\pi}_{a},\phi),
\end{align}
where the round brackets indicate functional evaluation of the corresponding, preceeding function on the new set of FLRW variables, namely, the direct replacement of its dependence on the old untilded set by the new one. The term $\tilde{H}_D$ is the contribution of the new fermionic variables to the zero mode of the Hamiltonian constraint. As our notation indicates, it can be made independent of the canonical momentum of the inflaton. This is always possible by means of a suitable redefinition of the homogeneous lapse function at our perturbative truncation order \cite{fermihlqc}. It is given by
\begin{align}\label{newham}
\tilde{H}_D=&\sum_{\vec{k}\neq\vec{\tau},(x,y)}\Bigg[h_D^{k}\left(\bar{a}^{(x,y)}_{\vec{k}}a^{(x,y)}_{\vec{k}}-a^{(x,y)}_{\vec{k}}\bar{a}^{(x,y)}_{\vec{k}}+\bar{b}^{(x,y)}_{\vec{k}}b^{(x,y)}_{\vec{k}}-b^{(x,y)}_{\vec{k}}\bar{b}^{(x,y)}_{\vec{k}}\right)\nonumber \\
&+h_G^{k}\left(\bar{b}^{(x,y)}_{\vec{k}}b^{(x,y)}_{\vec{k}}-b^{(x,y)}_{\vec{k}}\bar{b}^{(x,y)}_{\vec{k}}\right) +\bar{h}_I^{k}\left(a^{(x,y)}_{\vec{k}}b^{(x,y)}_{\vec{k}}\right)-h_I^{k}\left(\bar{a}^{(x,y)}_{\vec{k}}\bar{b}^{(x,y)}_{\vec{k}}\right)\Bigg],
\end{align}
where we have ignored the contribution from the fermionic zero modes, since they can be isolated and quantized separately without obstructions, and
\begin{align}\label{hd}
&h_D^{k}=\frac{\omega_k}{2a}\left(| f_2^{k}| ^2-| f_1^{k} | ^2\right) + m\text{Re}\left(f_1^{k}\bar{f}^{k}_2\right)+\frac{i}{2}\left(\bar{f}^{k}_1 \{f_1^{k},H_{|0}\}+\bar{f}^{k}_2  \{f_2^{k},H_{|0}\}\right), \\
&h_G^{k}=\frac{1}{2}\{G^{k},H_{|0}\}, \\ \label{interaction}
&h_I^{k}=e^{-i G^{k}}\bigg[if_1^{k} \{f_2^{k},H_{|0}\}-if_2^{k}\{ f_1^{k},H_{|0}\}+\frac{2\omega_k}{a}f_1^{k}f_2^{k}+m(f_1^{k})^2- m(f_2^{k})^2\bigg],
\end{align}
modulo changes that can be absorbed by the aforementioned redefinition of the homogeneous lapse. At the considered perturbative level, this redefinition amounts to eliminate $\pi_{\phi}^2$ in the above equations by identifying $H_{|0}$ with the zero function \cite{fermihlqc}. It is worth noticing that, in the context of QFT in curved spacetimes, the product of this lapse function and $\tilde{H}_D$ is the Hamiltonian that generates the evolution of the chosen family of fermionic annihilation and creation variables, evolution that generally differs from the original Dirac dynamics generated by $N_{0}H_D$.

There are infinitely many possible families of fermionic annihilation and creation variables defined by means of Equations \eqref{anni3dt} and \eqref{g1,g2,f2}, even when restricting to coefficients that depend on $\vec{k}$ only through $\omega_k$. This freedom not only reflects the infinitely many inequivalent Fock representations of the fermionic degrees of freedom, but also the different possible splittings of the phase space between the fermionic sector and the homogeneous FLRW geometry. Before proceeding to the hybrid quantization of the entire system, it is therefore of the utmost importance to adhere to physical criteria and restrict the allowed families of fermionic annihilation and creation variables. In view of the results presented in the previous sections, a first reasonable condition to impose is that, when a QFT regime in a classical background spacetime is recovered, the quantum Heisenberg evolution of the fermionic annihilation and creation operators can be implemented unitarily. As we have seen, once we set a convention for particles and antiparticles, this criterion on the considered families of variables completely fixes the (symmetry invariant) quantum representation of the field, up to unitary equivalence. In other words, it allows us to determine the Fock space for the representation of the fermionic field. For concreteness, let us set the particle/antiparticle convention to be such that it corresponds to the standard one in Minkowski spacetime as the mass of the fermions goes to zero (situation in which the rescaled Dirac field propagates as if it were in flat spacetime, in conformal time) \cite{uf3}. Then, the families of annihilation and creation variables restricted by the criterion of a unitarily implementable dynamics are those such that, in the asymptotic limit of large $\omega_k$:
\begin{align}\label{unitt3}
f^{k}_{1}=\frac{ma}{2\omega_{k}}e^{iF^{k}_{2}}+\vartheta^{k}\quad {\mathrm{with}}\quad
\sum_{\vec{k}} |\vartheta^{k}|^{2}<\infty .
\end{align}

The above condition greatly restricts the asymptotic behavior of the allowed families of annihilation and creation variables. Furthermore, all such choices lead to unitarily equivalent representations. However, there is still much freedom left, even in the asymptotic regime of large $\omega_k$, as the function $\vartheta^k$ is only constrained to be square summable (including degeneracy). Actually, this freedom can be used to refine the phase space splitting in such a way that the part of the hybrid quantization that concerns the fermionic degrees of freedom displays several desirable properties. In particular, it appears physically sound to demand that the Fock quantization of the contribution $\tilde{H}_D$ to the Hamiltonian constraint result into a well-defined operator on the fermionic vacuum state (and thus on the dense Fock subspace of its associated $n$-particle states). Such an operator can be simply obtained by promoting the variables $a^{(x,y)}_{\vec{k}}$ and $b^{(x,y)}_{\vec{k}}$, on the one hand, and $\bar{a}^{(x,y)}_{\vec{k}}$ and $\bar{b}^{(x,y)}_{\vec{k}}$, on the other hand, respectively to annihilation and creation operators in the fermionic Hamiltonian given in Equation \eqref{newham}. As these variables commute with the (perturbatively corrected) ones that describe the homogeneous FLRW background, the well-definiteness of $\tilde{H}_D$ on the fermionic vacuum is insensitive to whether these FLRW variables are fixed as classical or promoted to quantum operators as well (within the hybrid loop scheme). In turn, if one imposes normal ordering on the products of annihilation and creation operators, it is fairly easy to see that this property on the Fock quantization of $\tilde{H}_D$ depends exclusively on the asymptotic dependence on $\omega_k$ of the terms $h_{I}^k$ defined in Equation \eqref{interaction}. These provide the interactive part of the fermionic Hamiltonian, that is responsible for the annihilation and creation of pairs of particles and antiparticles, and the resulting operator is well defined (with normal ordering) on the vacuum state if and only if they form a square summable sequence, including the degeneracy of the Dirac eigenvalues. Specifically, if one introduces condition \eqref{unitt3}, together with relation \eqref{g1,g2,f2}, in $h^k_{I}$, the dominant contribution to this function in the asymptotic regime of large $\omega_k$ is cancelled out. From a Hamiltonian perspective, this cancelation is the ultimate responsible for the unitarity of the Heisenberg evolution (in the context of QFT in curved spacetimes). However, it does not guarantee that $h_I^k$ forms a square summable sequence over all $\vec{k}$. The necessary and sufficient condition for this to happen, and thus for the Fock quantization of $\tilde{H}_D$ to be well defined on the vacuum, turns out to be that asymptotically \cite{backreaction}
\begin{align}\label{niceham}
\vartheta^{k}=-i\frac{\pi m}{3l_0^3\omega_k^{2}}\pi_{a}e^{iF^{k}_2}+\theta^{k},\qquad {\mathrm{with}} \qquad \sum_{\vec{k}} \omega^2_k|\theta^{k}|^{2}<\infty .
\end{align}

It is worth noticing that this last condition on the allowed families of fermionic annihilation and creation variables restricts even further the admissible phase space splittings between the fermionic sector and the background. Namely, it specifies even further how the assignment of the dynamical content of each sector should be made. 

Finally, hereafter we restrict the discussion exclusively to phases $G^{k}$ such that $\{G^{k},H_{|0}\}=0$, in order not to introduce any artificial asymmetry between the dynamical behavior of the fermionic variables that describe particles and antiparticles [see Equation \eqref{newham}].

\subsection{Hybrid quantization: Fermionic backreaction}

Let us next summarize the main steps that must be followed for the hybrid quantization of our cosmological system, formed by an inflationary FLRW background coupled to a perturbative Dirac field. As mentioned above, one can freely include metric and inflaton perturbations as well, and discuss a similar phase space splitting and choice of Fock representation for them. We nonetheless ignore them in this review, as they do not necessarily affect the fermionic sector at the considered order of truncation in the action. For details on the treatment of metric and inflaton perturbations in the context of hLQC, we refer the reader to Refs. \cite{hybridrev,hybridrev2,h-gaugeinv}. 

The first step towards the hybrid quantization of the full system is to specify a concrete splitting of phase space with physically good properties, along the aforementioned lines. In other words, one starts by identifying a specific set of fermionic annihilation and creation variables [by means of Equations \eqref{anni3dt} and \eqref{g1,g2,f2}] within the family restricted by the asymptotic conditions \eqref{unitt3} and \eqref{niceham}. In addition to the phase space splitting, such choice fixes also the Fock representation for the fermionic degrees of freedom. Indeed, for their quantization, one just needs to promote these fermionic variables to the corresponding annihilation and creation operators, that in turn completely specify the Fock space ${\mathcal{F}}_D$ from their associated cyclic vacuum state. On the other hand, once a preferred phase space splitting has been identified, the FLRW background spacetime is described by the set of variables $\tilde{a}$, $\phi$, and their canonical momenta [see Equations \eqref{modifiedscale} and \eqref{modifiedpi}], that, as mentioned above, Poisson commute with the chosen set of annihilation and creation variables, at the classical level. We then adopt a (discrete) loop quantum cosmology representation for the canonical variables that describe the homogeneous background geometry, with operators defined on a Hilbert space that we call $\mathcal{H}_{\text{kin}}^{\text{grav}}$. For specific details on this representation, see Refs. \cite{fermihlqc,h-gaugeinv}. In this review, we only recall its main features when they are relevant for the quantum dynamics of the fermionic sector. In addition, a standard Schr\"odinger representation is adopted for the homogeneous inflaton and its momentum, with Hilbert space given by $L^2(\mathbb{R},\text{d}\phi)$, such that the inflaton acts by multiplication and the momentum is represented as $-i\partial_{\phi}$. The total representation space for the hybrid quantization of the full canonical set of basic variables of our cosmological system is then the tensor product of all the introduced individual spaces, namely $\mathcal{H}_{\text{kin}}^{\text{grav}}\otimes L^2(\mathbb{R},\text{d}\phi)\otimes{\mathcal{F}}_D$.

This tensor product space is often called the kinematical space, and it is not the fully physical one. Indeed, the whole system is subject to the zero mode of the Hamiltonian constraint, that can be found in Equation \eqref{hamconstr} and classically generates global time reparameterizations. We implement this symmetry at the quantum level by demanding that physical states be annihilated by the representation of the constraint as an operator on the kinematical Hilbert space\footnote{Alternatively, a sufficiently large number of physical states annihilated by the (dual) action of the constraint may live in the dual space of a certain dense subset of the kinematical space.}. Actually, for mathematical convenience, one often rather imposes a rescaled version of this constraint, obtained through multiplication by the volume $\tilde{V}=\tilde{a}^3 l_0^3$ of the homogeneous FLRW sector. In order to find physical states in the system, it is therefore necessary to specify the corresponding constraint operator and deal with the ambiguities that its representation involves, as it is not a linear function of the canonical variables. In this construction, in particular, we impose normal ordering in the fermionic contribution $\tilde{H}_D$ to the constraint. For details about the remaining ambiguities and how to reasonably fix them in the context of loop quantum cosmology, we refer the reader to Ref. \cite{fermihlqc}. 

Let a specific quantum representation of the zero mode of the Hamiltonian constraint (with normal ordering for the fermionic operators) be provided in this way. In order to search for states of physical interest, such that the influence of the perturbations on the FLRW background can be made controllably small, we look for solutions to the quantum constraint starting from the following ansatz for the allowed wavefunctions $\Xi$:
\begin{align}\label{ansatz}
\Xi=\Gamma(\tilde{V},\phi)\psi_D({\mathcal N_D},\phi).
\end{align}
Here, $\tilde{V}$ denotes dependence of the wavefunction on the geometric sector of the homogeneous FLRW background, and $\mathcal N_D$ is a generic label indicating the occupancy numbers in the fermionic Fock space. Hence, in our states we can regard $\Gamma$ as the partial wavefunction that describes the behavior of the FLRW cosmology, whereas $\psi_D$ is the partial state for the fermionic perturbations. We notice that both of these contributions are allowed to depend on the inflaton. It is also worth noting that, in case inhomogeneous perturbations of the metric and inflaton were included, this ansatz can be correspondingly generalized, separating the dependence of the total wavefunction in each different type of perturbation. In what concerns the partial FLRW state $\Gamma$, we further impose as part of our ansatz that it evolves unitarily with respect to its variation in $\phi$ (so that we can normalize it in the Hilbert space $\mathcal{H}_{\text{kin}}^{\text{grav}}$), with a generator that we call $\hat{\tilde{\mathcal{H}}}_{0}$ such that 
\begin{align}
-i\partial_{\phi}\Gamma=\hat{\tilde{\mathcal{H}}}_{0}\Gamma.
\end{align}
We take this generator to be a positive self-adjoint operator and furthermore impose that it gives rise to partial states that are close to exact quantum solutions (i.e., annihilated by the action) of the (rescaled) constraint operator of the homogeneous FLRW background. Explicitly, let us write the constraint operator that we would have for our FLRW model in the absence of perturbations in the form
\begin{align}
-\frac{1}{2}[\partial_{\phi}^2+\hat{\mathcal H}_0^{(2)}]
\end{align}
where $-\hat{\mathcal H}_0^{(2)}$ is the operator that represents the (rescaled) contribution of the inflaton potential and of the homogeneous FLRW geometry [see Equation \eqref{H02}]. Our restriction on $\hat{\tilde{\mathcal{H}}}_{0}$ translates into demanding that the action of $(\hat{\tilde{\mathcal H}}_0)^2  -\hat{\mathcal H}_0^{(2)}-i[\partial_{\phi}, \hat{\tilde{\mathcal H}}_0]$ on $\Gamma$ be small, let us say at most comparable to terms of quadratic order in the perturbative parameter of our system.

For the wave profiles selected by the above ansatz to potentially become physical states, we must impose that they be annihilated by the (rescaled) constraint operator. Namely, the action of the quantum representation of the function \eqref{hamconstr}, rescaled with the homogeneous volume $\tilde{V}$, on states of the form \eqref{ansatz} (and satisfying the aforementioned conditions) must be zero. The resulting constraint equation can be greatly simplified, and viewed as an evolution equation on the partial fermionic wavefunction, if the following approximations are applied \cite{fermihlqc,h-gaugeinv}:
\begin{itemize}
	\item [i)] The partial state $\Gamma$ is such that one can ignore transitions in the FLRW geometry mediated by the zero mode of the Hamiltonian constraint. Then, one can apply a kind of mean-field approximation and take the inner product of the constraint equation with $\Gamma$, with respect to the Hilbert space $\mathcal{H}_{\text{kin}}^{\text{grav}}$.
	\item[ii)] The contribution $\partial^2_{\phi}\psi_D$ can be neglected when compared with $\langle \hat{\tilde{\mathcal H}}_0 \rangle_{\Gamma}\partial_{\phi}\psi_D$. In other words, the contribution to the inflaton momentum of the fermionic partial state is negligible compared with the contribution of the homogeneous FLRW state. The self-consistency of this approximation can be explicitly checked using the constraint equation \cite{h-gaugeinv}.
\end{itemize}
If these approximations hold within our perturbative treatment, then the constraint equation can be recast as the following Schr\"odinger-like one:
\begin{align}\label{schro}
i\partial_{\phi}\psi_D=\left[\frac{\langle \widehat{\tilde{V}\tilde{H}_{D}}  \rangle_{\Gamma}}{ \langle \hat{\tilde{\mathcal H}}_0 \rangle_{\Gamma}}+C_D^{(\Gamma)}(\phi)\right]\psi_D,
\end{align}
where the hat denotes the hybrid loop representation of the underlying function, and \cite{fermihlqc,backreaction}
\begin{align}
C_D^{(\Gamma)}=\frac{\langle  \hat{\mathcal H}_0^{(2)}+i[\partial_{\phi}, \hat{\tilde{\mathcal H}}_0]-(\hat{\tilde{\mathcal H}}_0)^2\rangle_{\Gamma}}{2 \langle \hat{\tilde{\mathcal H}}_0 \rangle_{\Gamma}}.
\end{align}
This function $C_D^{(\Gamma)}$ encodes, in mean value, how much the partial FLRW state $\Gamma$ differs from being an exact solution of the unperturbed system. In this sense, it can be understood as a quantum backreaction term between the fermionic sector and the homogeneous background. Besides, it is worth mentioning that the generator of the fermionic evolution dictated by this Schr\"odinger equation is a $\phi$-dependent operator that only acts on $\mathcal{F}_D$, and depends on the homogeneous background geometry by means of expectation values of FLRW operators on the partial state $\Gamma$. Furthermore, this operator is automatically well defined on the vacuum, given the preferred Fock quantization adopted for the fermionic degrees of freedom. Therefore, the introduced approximations in the hybrid quantum constraint equation allow one to recover a QFT regime in a quantum spacetime, with good physical properties. 

Solutions to the Schr\"odinger equation \eqref{schro} can be obtained by considering its associated Heisenberg dynamics. For that purpose, it is convenient to introduce the following change of evolution parameter:
\begin{align}
\label{etaGamma}
\text{d}{\eta_\Gamma}=  \frac{l_0\langle \hat{\tilde{V}}^{2/3} \rangle_{\Gamma} }{ \langle \hat{\tilde{\mathcal H}}_0 \rangle_{\Gamma} } \text{d}\phi,
\end{align}
which is well defined owing to the positivity of $\hat{\tilde{\mathcal H}}_0$ and the fact that the volume operator is bounded from below by a strictly positive number in the loop quantization of the FLRW background \cite{fermihlqc}. With respect to this parameter, the Heisenberg equations associated with Equation \eqref{schro}, evaluated at time $\eta_{\Gamma}=\eta$, are given by
\begin{align}
\label{equationmotiona}
&d_{\eta_\Gamma}{\hat a}_{\vec{k}}^{(x,y)}(\eta)=- i F_k^{(\Gamma)}{\hat a}_{\vec{k}}^{(x,y)}(\eta)+ G_k^{(\Gamma)} {\hat b}_{\vec{k}}^{(x,y)\dagger}(\eta),\nonumber
\\
&d_{\eta_\Gamma}{\hat b}_{\vec{k}}^{(x,y)\dagger}(\eta)= i F_k^{(\Gamma)}{\hat b}_{\vec{k}}^{(x,y)\dagger}(\eta)- \bar{G}_k^{(\Gamma)} {\hat a}_{\vec{k}}^{(x,y)}(\eta),
\end{align}
where we have introduced the one-parameter family of $\eta_{\Gamma}$-dependent annihilation and creation operators in the Heisenberg picture, with initial data at some $\eta_{\Gamma}=\eta_0$ given by the fermionic annihilation and creation operators ${\hat a}_{\vec{k}}^{(x,y)}$ and ${\hat b}_{\vec{k}}^{(x,y)\dagger}$ of our hybrid quantization. In addition, $d_{\eta_\Gamma}$ denotes the derivative with respect to the parameter $\eta_{\Gamma}$. Besides,
\begin{align}
F_k^{(\Gamma)}=\frac{2\langle \widehat{\tilde{a}^3 h^{k}_D} \rangle_{\Gamma}}{\langle \hat{\tilde a}^{2}\rangle_{\Gamma}}, \qquad G_k^{(\Gamma)}=i\frac{\langle \widehat{\tilde{a}^3 h^{k}_I}\rangle_{\Gamma}}{\langle \hat{\tilde a}^{2}\rangle_{\Gamma}},
\end{align}
where the diagonal mode coefficients $h_D^k$ and the interaction mode coefficients $h_I^k$ of the fermionic Hamiltonian are respectively given in Equations \eqref{hd} and \eqref{interaction} (up to the elimination of any dependence on the inflaton momentum by identifying $H_{|0}$ with zero, as we have explained). It is worth noting that, strictly speaking, these Heisenberg equations for the fermionic modes can be derived without the second approximation (ii) introduced above. In fact, given only the first approximation (i), it suffices that there exists a regime for all the modes in which the annihilation and creation operators find a straightforward counterpart in the Grassmann variables that they represent \cite{h-gaugeinv,hybridrev2}.

The resulting evolution from time $\eta_0$ to time $\eta$ of the annihilation and creation operators is a Bogoliubov transformation that can be easily seen to take the form \cite{fermihlqc}
\begin{align}
\label{qBogoliubov}
&{\hat a}_{\vec{k}}^{(x,y)}(\eta)= \alpha_{k}(\eta,\eta_0){\hat a}_{\vec{k}}^{(x,y)}+\beta_{k}(\eta,\eta_0){\hat b}_{\vec{k}}^{(x,y)\dagger},\nonumber\\
&{\hat b}_{\vec{k}}^{(x,y)\dagger}(\eta)=-\bar{\beta}_{k}(\eta,\eta_0){\hat a}_{\vec{k}}^{(x,y)}+\bar{\alpha}_{k}(\eta,\eta_0){\hat b}_{\vec{k}}^{(x,y)\dagger},
\end{align}
where $\alpha_k(\eta_0,\eta_0)=1$, $\beta_k(\eta_0,\eta_0)=0$, and, for all $\eta$,
\begin{align}
|\alpha_{k}(\eta,\eta_0)|^2+|\beta_{k}(\eta,\eta_0)|^2=1.
\end{align}
In particular, recall that we are focusing our attention on phase space splittings and Fock representations of the fermionic degrees of freedom characterized by Equations \eqref{unitt3} and \eqref{niceham} in the asymptotic regime of large $\omega_k$. These lead to interacting contributions to the fermionic Hamiltonian such that the beta coefficients of this transformation have the asymptotic behavior $\beta_k=\mathcal{O}[\text{Max}(\theta^k,\omega_k^{-3})]$, where the function $\text{Max}[.,.]$ picks out the argument of dominant asymptotic order \cite{backreaction}. Taking into account the second condition in Equation \eqref{niceham}, it is clear that these beta coefficients form a square summable sequence, over all $\vec{k}$. It then follows that our quantum Heisenberg dynamics is unitarily implementable on the fermionic Fock space.

The unitary operator that implements the considered Heisenberg evolution can be explicitly constructed \cite{fermihlqc,backreaction}. Furthermore, it can be checked that its action on the fermionic vacuum state picked out by the hybrid quantization provides a solution to the Schr\"odinger equation \eqref{schro} if the backreaction function is of the form
\begin{align}
C_{D}^{(\Gamma)}(\phi)=\frac{l_0\langle \hat{\tilde V}^{2/3}\rangle_{\Gamma}}{\langle \hat{\tilde{\mathcal H}}_0 \rangle_{\Gamma}} \sum_{\vec{k},(x,y)} \Big[\Im{}(\eta_{\Gamma}) + d_{\eta_{\Gamma}}c_k^{(x,y)}\Big].
\end{align}
Here, $c_k^{(x,y)}$ are arbitrary real phases of the evolution operator and $\Im{}$ is certain function with an asymptotic behavior which depends on that of the interaction contribution $G_k^{(\Gamma)}$ (or, equivalently, on that of $\beta_k$). For our allowed families of Fock representations of the fermionic degrees of freedom, one can see that this function is of dominant asymptotic order $\mathcal{O}[\text{Max}(\omega_k|\theta^k|^2,\omega_k^{-5})]$. Therefore, our backreaction term relating the FLRW background and the fermionic perturbations in the regime of QFT in a quantum spacetime, attained by the hybrid quantization of the system, turns out to be an absolutely convergent quantity. Thus, one needs not perform any regularization or {\em substraction of infinities} to render this backreaction term finite, something that could be done by using the arbitrary phases $c_k^{(x,y)}$ that one can freely add to the evolution operator, but that would imply an unjustified adjustment of an infinite number of quantities.

We end this section with the following remark. The first work \cite{fermihlqc} that was carried out about the introduction of fermionic perturbations in hLQC employed a choice of annihilation and creation variables analogous to that introduced by D'Eath and Halliwell in Ref. \cite{H-D}. As mentioned in the previous section, this choice is within the family of unitarily equivalent quantizations that allow for a unitarily implementable dynamics, in the context of QFT in curved spacetimes. The analysis in Ref. \cite{fermihlqc} shows that, with this choice of fermionic variables, the backreaction term $C_D^{(\Gamma)}$ in the hybrid quantum theory fails to be an absolutely convergent quantity and needs to be regularized. This is an example of how the requirement of a unitarily implementable evolution alone is not enough to guarantee nice properties of the hybrid quantization of the full system. In fact, the conditions that are needed for the absolute convergence of $C_D^{(\Gamma)}$ are very similar, but slightly weaker, than those for a proper definition of the fermionic Hamiltonian on the vacuum. Indeed, it is enough that Equation \eqref{niceham} is fulfilled and $\omega_k|\theta^k|^2$ is a summable sequence over all $\vec{k}$ \cite{backreaction}. In this respect, any choice of fermionic variables that leads to a good definition of the Hamiltonian operator on the vacuum automatically guarantees that the quantum fermionic backreaction is finite, without the need of regularization.

\section{Fermionic Hamiltonian diagonalization: Choice of vacuum state}\label{diagonalization}

The restrictions on the definition \eqref{anni3dt} of fermionic annihilation and creation variables in hybrid quantum cosmology to guarantee: (i) unitarity of the QFT evolution [Equation \eqref{unitt3}], and (ii) a well-defined Hamiltonian on the vacuum [Equation \eqref{niceham}], allow us to considerably reduce the possible choices of such variables, at least in the asymptotic regime of large $\omega_k$. However, and even though all of the resulting Fock representations are unitarily equivalent, there is still much freedom in identifying a particular set of annihilation and creation variables (even in the asymptotic regime). In other words, there remains ambiguity in the complete characterization of the phase space splitting between the FLRW background and the fermionic sector, as well as of the fermionic vacuum state of the theory. In this section we motivate and adhere to the physical criterion of Hamiltonian diagonalization to try and fix this remaining choice, following a procedure that is specifically adapted to the spatially local structure of the fermionic dynamics.

\subsection{Hamiltonian diagonalization in hLQC}

The previously imposed criteria on the allowed choices of fermionic annihilation and creation variables [Equations \eqref{unitt3} and \eqref{niceham}] are tailored to have the net effect of diminishing the dominant asymptotic order, in the regime of large $\omega_k$, of the interaction parts $h_I^{k}$ of the fermionic Hamiltonian $\tilde{H}_D$ [given in Equation \eqref{newham}]. Ultimately, this fact is responsible for the unitarity of the fermionic evolution, as well as for a proper definition of the fermionic Hamiltonian on the vacuum. As we have already commented, these interaction parts annihilate and create pairs of particles and antiparticles in the quantum fermionic dynamics. From a physical perspective, one would think that a choice of phase space splitting where the assignment of the dynamical contribution of the FLRW background to the system is such that the fermionic states undergo no annihilation and creation of pairs, would be one that is naturally adapted to the dynamics of the entire cosmological system. This choice should then be such that $h_I^{k}=0$, so that the Fock quantization of the resulting fermionic Hamiltonian $\tilde{H}_D$ would have a diagonal action on the $n$-particle states associated with the selected set of annihilation and creation operators.

According to this line of reasoning, we refer to any choice of fermionic annihilation and creation variables that lead to vanishing interaction terms $h_I^{k}$, for all $\vec{k}$, as variables for the Hamiltonian diagonalization. In general, restricting to cases with $f_2^k \neq 0$ (which include all those choices that respect the standard convention for particles and antiparticles in the massless limit \cite{uf3}), one can readily check that the diagonalization condition $h_I^{k}=0$ is fulfilled for all $\vec{k}$ if and only if
\begin{align}\label{diagallk}
a\left\{\varphi_{k},H_{|0}\right\}+2i\omega_k \varphi_{k}+iam\varphi_{k}^{2}-iam=0,
\end{align}
where $f_{1}^{k}=f_{2}^{k}\varphi_{k}$. This is a semilinear partial differential equation which has locally unique solutions, as long as the section of initial conditions is transversal to the flow of the Hamiltonian vector field $\{.,H_{|0}\}$ \cite{deqs}. Naturally, there are several possible families of such solutions for all $\vec{k}$, and each of them completely characterizes (up to the two phases $G^{k}$ and $F_2^{k}$) a different set of fermionic variables for the Hamiltonian diagonalization, in virtue of Equation \eqref{g1,g2,f2}. Explicitly, it holds that
\begin{align}\label{f2fi}
| f_{2}^{k}| ^2=\frac{1}{1+|\varphi_{k}|^2}.
\end{align}

Let us point out that, in fact, the structure of the differential Equation \eqref{diagallk} allows for solutions that, besides on the FLRW geometry, can depend also on the inflaton and its canonical momentum. Remarkably, nonetheless, the definition of fermionic annihilation and creation variables resulting from any such choice of coefficients for the Hamiltonian diagonalization can still be completed to become canonical in the entire cosmological system, following an analogous procedure to that discussed in the previous section. We adopt this extended framework for the definition of the fermionic variables in this and the next subsection.

Finally, it is worth noticing that relation \eqref{f2fi} can be used to show that the mode coefficients $h_D^{k}$ of the resulting diagonal Hamiltonian, for each choice of fermionic variables characterized by a set of solutions to Equation \eqref{diagallk} (for all $\vec{k}$), acquire the form \cite{fermidesitter}
\begin{align}\label{diagham}
2h_{D}^{k}=a^{-1}\omega_k +m\text{Re}\left(\varphi_{k}\right)-\left\{F_{2}^{k},H_{|0}\right\}.
\end{align}

\subsection{Asymptotic diagonalization}

In the following, we try and fix a preferred solution to Equation \eqref{diagallk}, using our previous knowledge on the restrictions that unitary evolution and a proper definition of the fermionic Hamiltonian impose, and by looking into the details of this Hamiltonian in the asymptotic regime of large $\omega_k$. For fermionic variables that admit a unitarily implementable evolution, namely when Equation \eqref{unitt3} holds, the interaction coefficients $h_I^k$ of the fermionic Hamiltonian $\tilde{H}_D$ behave asymptotically as
\begin{align}\label{inter2}
h_I^k=\frac{2\omega_k}{a}\vartheta^k+i\frac{2\pi m}{3l_0^3\omega_k}\frac{\pi_a}{a} e^{iF_2^k}+\mathcal{O}[\text{Max}(\vartheta^k,\omega_k^{-2})],
\end{align}
where the second summand results from the second Poisson bracket in Equation \eqref{interaction}. As we pointed out above, we explicitly see here that demanding condition \eqref{niceham} for a well-defined action of the Fock quantization of the Hamiltonian on the fermionic vacuum is equivalent to diminishing the dominant asymptotic order, in inverse powers of $\omega_k$, of the interaction coefficients. Once this condition is imposed, one arrives at an asymptotic behavior for $h_I^k$ with an analogous structure as in Equation \eqref{inter2}. More concretely, its dominant contribution is given by $2\omega_k \theta^k/a$ plus certain specific terms that are proportional to $\omega_k^{-2}$. One can cancel this contribution again by conveniently fixing the dominant asymptotic behavior of $\theta^{k}$. The resulting interaction coefficient $h_I^k$ displays, once more, a similar asymptotic structure, but with the role of $\theta^{k}$ played by its subdominant contributions and the remaining summands having a lower asymptotic order in powers of $\omega_{k}$. Owing to the asymptotic structure of the Hamiltonian, this pattern repeats itself at each asymptotic order in inverse powers of $\omega_k$, if one admits an asymptotic expansion of this form and imposes the cancellation of the previous dominant terms in the interaction coefficient $h_I^k$.

Motivated by these properties of the fermionic Hamiltonian, and taking into account that asymptotically
\begin{align}
f_{2}^{k}=e^{iF_2^k}+\mathcal{O}(\omega_k^{-2})
\end{align}
if the unitary evolution condition \eqref{unitt3} is fulfilled, we propose the following asymptotic series as an ansatz for a Hamiltonian diagonalization in the regime of large $\omega_k$:
\begin{align}\label{asympdiag}
\varphi_k\sim\frac{1}{2\omega_k}\sum_{n=0}^{\infty}\left(-\frac{i}{2\omega_k}\right)^{n}\gamma_{n}, \qquad \gamma_{0}=ma.
\end{align}
Here, the symbol $\sim$ indicates the equality of the asymptotic expansions, and $\gamma_{n}$ are functions of the homogeneous FLRW background canonical variables. These are completely fixed in an iterative way if one introduces our ansatz in the interaction coefficients $h_I^k$ of the fermionic Hamiltonian, and imposes that each contribution in inverse powers of $\omega_k$ be equal to zero. Specifically, one then obtains \cite{fermidiagonalization,fermidesitter}
\begin{align}\label{recurrence}
\gamma_{n+1}=a\left\{H_{|0},\gamma_n\right\}+ma\sum_{m=0}^{n-1}\gamma_{m}\gamma_{n-(m+1)},\qquad \forall n\geq 0,
\end{align}
where $\gamma_{-n} \equiv 0$ for all $n>0$. This is a deterministic recurrence relation that can be used to fix all of the functions $\gamma_{n}$, starting from the initial datum $\gamma_0$.

It is worth mentioning that the proposed function $\varphi_k$ for the asymptotic diagonalization of the fermionic Hamiltonian, given by Equations \eqref{asympdiag} and \eqref{recurrence}, provides a very specific solution to the general equation \eqref{diagallk}, in the sector of unboundedly large $\omega_k$. Such an asymptotic solution can be thought of as a physically preferred solution, inasmuch as it has been obtained by exclusively adhering to local features of the fermionic Hamiltonian. However, despite the strong asymptotic restriction that our requirement sets on the admissible solutions to Equation \eqref{diagallk}, there may exist many such solutions for all $\vec{k}$ that, viewed as functions of $\omega_k$, display the same, preferred asymptotic behavior. It seems therefore most convenient to investigate whether the imposition of certain smoothness conditions on the dependence of any such $\varphi_k$ on $\omega_k$ (e.g., continuity or analiticity) can allow us to fix this solution completely. Indeed, if we were able to ensure this uniqueness, by taking into account relations \eqref{g1,g2,f2}, we would solve the last remaining ambiguity in the choice of fermionic variables for the hybrid description of the system, up to the phases $G^{k}$ and $F_2^{k}$. In other words, once these phases were chosen, we would succeed in specifying a Fock representation (with its associated vacuum state) of the fermionic degrees of freedom, together with a particular splitting of the fermionic and FLRW sectors of the phase space and of their contribution to the dynamics of the entire system. Actually, we have already restricted the phase $G^{k}$ in a substantial way by demanding it to be a dynamically irrelevant constant in the classical linearized system, after imposing $\{G^k,H_{|0} \}=0$ in order to eliminate asymmetries in the evolution of particles and antiparticles. As for the other phase, $F_2^k$, it can be naturally selected by demanding that the original dynamics that is extracted from the Dirac field by means of the background-dependent transformation \eqref{anni3dt} be minimal, along the lines detailed in Ref. \cite{diagonalization}.

We end this subsection by noting that, if one specifies a preferred choice of fermionic annihilation and creation variables for the Hamiltonian diagonalization, then one is not only fixing the relevant structures for the hybrid quantization of the cosmological system, but also a concrete Fock representation of the Dirac field, within the framework of QFT in curved spacetimes. This regime is attained when the homogeneous background obeys the Friedmann equations, whereas the fermionic annihilation and creation variables evolve with the dynamics dictated by the fermionic contribution to the Hamiltonian \eqref{hamconstr}. If the interaction terms in this Hamiltonian are zero, then this fermionic dynamics can be straightforwardly solved, namely, the annihilation and creation variables just evolve via multiplication of their initial data (at an arbitrary initial time) by a complex phase. With these solutions at hand, if one takes the inverse of the defining transformation of these variables given by Equation \eqref{anni3dt}, and introduces it in the mode expansion of the Dirac field, one immediately obtains a complete basis of solutions for the Dirac equation, in the sense of Equation \eqref{ovdf}. The constant coefficients of the elements of this basis in the expansion of the field are the annihilation and creation initial data, that select a unique Fock representation (and its associated vacuum state) once they are promoted to operators.

\subsection{Uniqueness of the vacuum: Minkowski and de Sitter spacetimes}

In this subsection we focus our discussion on the aforementioned regime of QFT in curved spacetime, and explain how our ansatz for asymptotic diagonalization succeeds in the selection of natural vacuum states in Minkowski and de Sitter spacetimes. In fact, the asymptotic expansion given in Equations \eqref{asympdiag} and \eqref{recurrence} allows us to determine a complete basis of solutions for the Dirac equation [as in relation \eqref{ovdf}] that turns out to correspond to the choice of the Poincar\'e or the Bunch-Davies vacuum, respectively, when the background cosmology is fixed as the Minkowski or the de Sitter spacetime. Taking into account that, when the homogeneous background obeys the Friedmann equations, we have that $a\{.,H_{|0}\}$ is simply the derivative with respect to the conformal time, in the considered situations in QFT the recurrence relation \eqref{recurrence} becomes
\begin{align}\label{recurrqft}
\gamma_{n+1}=-\gamma_{n}'+ma\sum_{m=0}^{n-1}\gamma_{m}\gamma_{n-(m+1)},\qquad \forall n\geq 0.
\end{align}

Let us start by considering the case of a background given by the classical Minkowski spacetime. This particularization is easily implemented by setting the scale factor as the unit constant, the inflaton as an arbitrary constant, and its potential equal to zero. Then, we immediately have that $\gamma_{0}=m$, while any other $\gamma_{n}$, determined by the recursion relation \eqref{recurrqft}, has a vanishing time derivative. That iterative equation can be solved by introducing the generating function $G(x)=\sum_{n=0}^{\infty}\gamma_n x^n$, that leads to a quadratic equation with only one solution consistent with the initial datum $\gamma_{0}=m$:
\begin{align}
G(x)=\frac{1}{2mx^2}\left[1-\sqrt{1-4m^2x^2}\right].
\end{align}
Around $x=0$, this is an analytic function with power series in $x$ characterized by the coefficients $\gamma_{n}$, by construction. Then, comparing this series to our ansatz \eqref{asympdiag} and employing the uniqueness of the asymptotic expansion, we can directly identify the function $\varphi_k$  that leads to an asymptotic diagonalization with the following analytic function:
\begin{align}
\varphi_k=\frac{1}{2\omega_k}G\left(\frac{-i}{2\omega_k}\right)=\frac{\omega_k}{m}\left[\sqrt{1+\frac{m^2}{\omega_k^{2}}}-1\right].
\end{align}
Using the relations \eqref{g1,g2,f2} and \eqref{f2fi}, that arise from the requirement that the fermionic annihilation and creation variables be defined by means of a canonical transformation, one eventually obtains
\begin{align}\label{fsmink}
|f_{1}^{k}|=\sqrt{\frac{\xi_k-\omega_k}{2\xi_k}},\qquad |f_{2}^{k}|=\sqrt{\frac{\xi_k+\omega_k}{2\xi_k}},
\end{align}
where $\xi_k=\sqrt{\omega_k^2+m^2}$.  Since the selected function $\varphi_k$ that permits the Hamiltonian diagonalization is completely independent of time in this case, according to our comments above, it is natural to demand that the phase $F_{2}^{k}$ be simply an arbitrary constant, as well as $G^{k}$. From Equation \eqref{diagham}, one can straightforwardly check that the diagonal coefficients of the resulting fermionic Hamiltonian are then $h_D^k=\xi_k$. A simple inspection of this result, together with Equation \eqref{fsmink}, immediately reveals that our criterion of asymptotic diagonalization selects indeed the basis of solutions to the Dirac equation in Minkowski spacetime that corresponds to the Poincar\'e Fock representation of the field. Namely, it is the quantization that, with a standard convention for particles and antiparticles, separates between positive and negative mass-shell frequencies $\xi_k$.

Let us show, in addition, how our criterion of asymptotic diagonalization recovers the common notion of Bunch-Davies vacuum for the Dirac field in de Sitter spacetime. In a flat slicing, this spacetime can be understood as a cosmological solution of Friedmann equations obtained by setting the inflaton potential equal to the constant $3 H^2_{\Lambda}/(8\pi)$ and the inflaton momentum equal to zero. Here, $H_{\Lambda}$ is the constant Hubble parameter. In conformal time, the expanding scale factor then behaves as
\begin{align}
a=-\frac{1}{\eta H_{\Lambda}}, \qquad -\infty<\eta<0.
\end{align}
The analysis of the restrictions that the asymptotic diagonalization imposes on the choice of a fermionic vacuum is easier if one first considers the general differential equation for $\varphi_k$ in our de Sitter background. It reads
\begin{align}
\label{eqdesitter}
\varphi_{k}^{\prime}+2i\omega_k \varphi_{k} -i\frac{m}{\eta H_{\Lambda}}\varphi_{k}^{2}+i\frac{m}{\eta H_{\Lambda}}=0.
\end{align}
The general solution of this equation can be found by introducing a mode-dependent complex time $T_k=-2i\omega_k \eta$ and the following change of variables \cite{fermidesitter}:
\begin{align}\label{varphiv}
\varphi_{k}=1+i\frac{H_{\Lambda}}{m}T_k\frac{d}{dT_k}\left(\log{v_{k}}\right).
\end{align}
The function $v_{k}$ turns out to satisfy a confluent hypergeometric equation in the complex variable $T_k$, that has the general solution
\begin{eqnarray}\label{generalv}
v_{k}&=&A\,{}_{1}F_{1}\left(-imH_{\Lambda}^{-1};1-2imH_{\Lambda}^{-1};T_k\right)\nonumber\\
&+&B\,T_k^{2imH_{\Lambda}^{-1}}{}_{1}F_{1}\left(imH_{\Lambda}^{-1};1+2imH_{\Lambda}^{-1};T_k\right),
\end{eqnarray}
where $A$ and $B$ are integration constants, and ${}_{1}F_{1}(.;.;z)$ is the hypergeometric function of type $(1,1)$, that is absolutely convergent for all values of its complex argument $z$ \cite{spfunc}.

The asymptotic expansion determined by our criterion of Hamiltonian diagonalization, given in Equations \eqref{asympdiag} and \eqref{recurrqft}, actually picks out a specific solution of the mentioned hypergeometric equation, up to an irrelevant multiplicative constant, namely a particular ratio between the integration constants $A$ and $B$. Indeed, the iterative relation \eqref{recurrqft}, particularized to our de Sitter background, can be seen to lead to coefficients $\gamma_{n}$ such that
\begin{align}
\varphi_{k,\lambda}\sim i \frac{1}{T_k}\sum_{n=0}^{\infty}\left(-\frac{1}{T_k}\right)^{n}C_{n},
\end{align}
where $C_{n}$ are constants that are completely specified by a complicated nonlinear recurrence relation, with the initial value $C_0=mH_{\Lambda}^{-1}$ \cite{fermidesitter}. We need not solve this relation, since the $T_k$-dependence of the above asymptotic series, together with the known value of $C_0$, is enough to restrict the associated expansion of $v_k$ in Equation \eqref{varphiv} so that
\begin{align}
v_{k}\sim T_{k}^{imH_{\Lambda}^{-1}}\sum_{n=0}^{\infty}\left(-\frac{1}{T_k}\right)^{n}v_{n},\qquad \text{with}\qquad v_1=\left(\frac{m}{H_{\Lambda}}\right)^2 v_0.
\end{align}
By introducing this ansatz for the asymptotic behavior of $v_k$ in the confluent hypergeometric equation that it must satisfy, one can determine the coefficients $v_{n}$ of the asymptotic expansion exclusively in terms of $v_0$, yielding
\begin{align}\label{asympv}
v_{k}\sim v_{0}T_{k}^{imH_{\Lambda}^{-1}}{}_{2}F_{0}\left(imH_{\Lambda}^{-1},-imH_{\Lambda}^{-1};-;-T_k^{-1}\right),
\end{align}
where ${}_{2}F_{0}(.,.;-;z)$ is the hypergeometric function of type $(2,0)$, that has a zero radius of convergence. Even though it formally diverges, its series is known to provide the asymptotic expansion of a very particular type of solution to the confluent hypergeometric equation, namely the Tricomi solution \cite{spfunc}. In fact, using the asymptotic properties of the hypergeometric functions in Equation \eqref{generalv}, it is possible to prove that the Tricomi solution is the unique one that has an asymptotic expansion of the form \eqref{asympv}. Therefore, after substituting this solution in relation \eqref{varphiv}, we immediately see that our procedure of asymptotic Hamiltonian diagonalization allows us to obtain again a unique solution to the general equation \eqref{diagallk}, for all wave vectors $\vec{k}$.

In more detail, the Tricomi solution for $v_k$ selected by our criterion of asymptotic diagonalization leads, after several manipulations, to the following function $\varphi_k$ in Equation  \eqref{varphiv} \cite{fermidesitter}:
\begin{align}\label{varphidesitter}
\varphi_{k}(\eta)=\frac{H^{(1)}_{-\mu}(\omega_k\eta)-iH^{(1)}_{1-\mu}(\omega_k\eta)}{H^{(1)}_{-\mu}(\omega_k\eta)+iH^{(1)}_{1-\mu}(\omega_k\eta)}, \qquad \mu=i\frac{m}{H_{\Lambda}}+\frac{1}{2},
\end{align}
where $H_{\nu}^{(1)}$ denotes the Hankel function of the first kind \cite{abram}. In turn, using relations \eqref{g1,g2,f2} and \eqref{f2fi}, this result determines the coefficients $f_1^{k}$, $f_2^k$, $g_1^k$, and $g_2^k$ that define the fermionic annihilation and creation variables, up to the phases $F_2^k$ and $G^{k}$. We recall that the latter is fixed as an arbitrary constant. As for the former, namely $F_2^k$, one can select it following the ideas put forward in the previous subsection, so that it minimizes the amount of dynamics extracted by the time-dependent canonical transformation \eqref{anni3dt}. In any case, the details about the time dependence of this phase are irrelevant for the basis of solutions to the Dirac equation that $\varphi_k$ selects (in the context of QFT in a fixed curved spacetime), as one can straightforwardly check using Equations \eqref{anni3dt} and \eqref{diagham}. Then, after taking into account the diagonal evolution of the annihilation and creation variables dictated by $h_D^{k}$ in Equation \eqref{diagham}, the resulting complete set of solutions for the Dirac equation in which the field decomposes is given by very specific linear combinations of Hankel functions of the first (in the case of antiparticles) and second (in the case of particles) kinds, different for each chirality and helicity \cite{fermidesitter}. We recall that any such basis decomposition fixes a particular Fock representation of the field. In this case, the basis of solutions turns out to be precisely the one that has been naturally associated in the literature with the fermionic analog of the Bunch-Davies vacuum in de Sitter spacetime \cite{fermidesitter,bdf1,bdf2}. Therefore, our criterion of asymptotic Hamiltonian diagonalization provides again the vacuum state that is physically accepted as preferred, in this case in a de Sitter background.

We end this section with a final remark. Beyond the well-known background spacetimes analyzed here, namely Minkowski and de Sitter, quantum fields in FLRW cosmologies suffer from the lack of a natural choice of vacuum state, at least if one appeals only to the symmetries of the system to select it. In the case of scalar fields, a common approach to mitigate this issue is the introduction of adiabatic states (see e.g. Refs. \cite{adiabatic1,adiabatic2,adiabatic3,luders}). Their construction is based on an iterative procedure to solve the field equations such that the resulting quantization displays certain local Poincar\'e-like features. In the case of Dirac fields, there have been at least two notable attempts to generalize the notion of adiabatic states \cite{hollands,barbero}. The proposal in Ref. \cite{barbero} follows closely the construction procedures previously established for scalar fields. In particular, this work introduces an algorithm to iteratively solve the Dirac equation that, at each consecutive step, is able to approach the actual mode solutions up to contributions that are more and more subdominant in the asymptotic regime of large $\omega_k$. An adiabatic state of $n$th-order is defined by truncating this procedure at the $n$th-step of the iteration and setting the value of the resulting approximate mode solutions, at an arbitrary time, as the initial data that specify the basis of solutions for the Fock representation of the field. Ref. \cite{fermiadiab} analyzed the relation between these adiabatic states and our family of unitarily equivalent quantizations of the Dirac field, including those that satisfy the condition of asymptotic diagonalization. It was shown that the representations associated with adiabatic states of all orders belong to the same equivalence class. In particular, the zeroth-order state already corresponds to a representation that admits unitarily implentable evolution, once the time dependence attributed to the FLRW background has been conveniently extracted. In fact, this unitarity guarantees that any two states defined with adiabatic initial data at different times are unitarily equivalent, so that the choice of initial time for the definition of the adiabatic states is not a relevant ambiguity. Furthermore, the first-order adiabatic state directly leads to a Fock quantization of the Dirac field in the family selected by imposing that the fermionic Hamiltonian for the annihilation and creation variables be well defined on the vacuum. The question of whether higher-order adiabatic states give rise to representations that behave, in the asymptotic regime of large $\omega_k$, increasingly closer to the one(s) selected by our criterion of asymptotic Hamiltonian diagonalization is yet an open issue.

\section{Conclusions}\label{conclu}

In this work, we have reviewed some recent investigations, carried out by us and our collaborators, about the physical motivation and use of certain criteria capable to ensure the uniqueness of the Fock quantization of fields in cosmological systems, specialized to the case of fermions described by Dirac fields. The presented results have been applied to the study of the hybrid quantization of the primordial Universe with perturbations, that contain all the fermionic degrees of freedom described by a Dirac field (and may also include other matter field perturbations and metric perturbations).

We have first considered the Fock quantization of the CARs for Dirac fields in conformally ultrastatic three-dimensional spacetimes, as well as in cosmological FLRW spacetimes in four dimensions, with spherical or toroidal spatial hypersurfaces. We have characterized the set of vacua that are invariant under the physical symmetries of the Dirac equation in these spacetimes. These symmetries include the continuous isometries of the spatial hypersurfaces, enlarged with the spin rotations generated by the helicity in the case of FLRW cosmologies with sections of toroidal topology.

For all the Fock representations associated with the above set of invariant vacua, we have proven that there exists a subset that admits unitary implementability of
the dynamics on the Fock space. This evolution comes from the Dirac equation, after extracting from the fermionic field some of its time variation that can be attributed to the dependence on the variables that describe the spacetime background. In the Heisenberg picture, the extracted part is regarded as explicitly time dependent, and therefore is not included in the proper quantum dynamics of the annihilation and creation operators. In fact, this extraction is necessary in order to achive the unitary implementability of the quantum evolution. It must be restricted, nonetheless, by the condition that the evolution remaining from the original Dirac equation be not trivialized.

After determining all the Fock representations that are allowed by the criteria of invariance under the symmetries of the equations of motion and of a nontrivial unitary implementability of the dynamics, we have shown that all these representations are unitarily equivalent for each of the spacetime scenarios that we have considered, provided that one fixes a convention to distinguish between particles and antiparticles of the Dirac field. In other words, our well-motivated conditions of unitarity and invariance guarantee the uniqueness of the Fock representation, up to unitary equivalence.

This uniqueness result has the immediate consequence of specifying also a unique concept of quantum dynamics for the fermionic annihilation and creation operators, modulo unitary redefinitions. Indeed, our analysis allows us to fully characterize the functions of the spacetime background that need to be removed from the time dependence of the fields, at least in the ultraviolet limit of large eigenvalues of the Dirac operator on the spatial hypersurfaces. This characterization can alternatively be understood as the determination of which field excitations are the particles and antiparticles that preserve their coherence over time.

We have provided an optimal description, with an eye to its quantization, of the phase space of a homogeneous and isotropic cosmology coupled to a homogeneous scalar field (that acts as an inflaton in General Relativity) and with fermionic perturbations, when the Einstein-Dirac action is truncated at quadratic perturbative order. For the fermionic sector of the phase space, we have used our previous result about the Fock representation of a Dirac field in order to select a quantization of the fermionic degrees of freedom, up to unitary modifications. Hence, for the fermionic field, we have chosen certain annihilation and creation variables that are related with the Dirac modes through a canonical transformation that depends on the homogeneous and isotropic background, and that supports a unitarily implementable Heisenberg dynamics when the background is viewed as a fixed entity. This leads to a specific splitting of the phase space between the background degrees of freedom and the fermionic content. A particular consequence is the modification of the contribution to the global Hamiltonian constraint associated with the fermionic perturbations. We have taken advantage of this modification and, going beyond the criterion of unitary dynamics for the selection of the fermionic variables, we have employed the remaining freedom in the background dependence of this choice to obtain other desirable properties in the quantization of our system. One property that we have investigated is a proper definition of the fermionic Hamiltonian operator on the set of finite particle/antiparticle states constructed from the vacuum. On the other hand, the discussed splitting of the phase space implies also a change in the canonical variables that describe the homogeneous background, so as to preserve the symplectic canonical structure of the system at the perturbative order of our truncation. The corresponding change in the background variables amounts to the correction of the original ones with terms that are quadratic in the perturbations.

Using the above description of the phase space, the only nontrivial constraint that needs to be imposed quantum mechanically is the zero mode of the Hamiltonian constraint.
This global constraint interrelates the different physically relevant sectors of the phase space, namely, the geometric FLRW sector, for which a loop representation is adopted, the inflaton, with a Schr\"odinger-like representation, and the fermionic perturbations, for which one takes a Fock representation in the selected family (in addition, it is possible to include scalar and tensor perturbations, described by perturbative gauge invariants, with Fock representations that can be picked out as well with our proposed criteria). In order to single out this preferred family of Fock representations, in addition to the invariance under the symmetries of the field equations and the unitary implementability of the Heisenberg dynamics, we have chosen a convention for the distinction between particles and antiparticles which smoothly connects with the standard convention of QFT when the mass of the field vanishes.

We have shown how to impose the operator that represents the zero mode of the Hamiltonian constraint of our perturbed cosmological system on states for which the dependence on the different sectors of the phase space, except the inflaton, becomes separable. In this way, we have been able to find mild conditions under which the  imposition of the constraint turns out to be essentially equivalent to a certain master constraint equation on the fermionic perturbations. Given our perturbative hierarchy, these conditions amount to have negligible FLRW geometry transitions mediated by the zero mode of the Hamiltonian for the partial wave function that describes such an homogeneous geometry in our state. The resulting equation is special inasmuch as the dependence on the FLRW geometry only persists by the inclusion of expectation values over that geometry. With an additional approximation on the variation of the partial wave function of the fermionic perturbations with respect to the inflaton, that can be checked at least for self-consistency, one can deduce from this master constraint on the perturbations a Schr\"odinger equation for the partial state that describes the fermionic content. This Schr\"odinger equation involves the quantum backreaction that the fermionic perturbations produce on the FLRW geometry. 

Besides, within our approximations, it is possible to solve the quantum dynamics dictated by the commented master contraint equation on the perturbative fermionic modes. We have reviewed how such dynamics can be implemented on our Fock space. The resulting evolution depends, in particular, on the FLRW geometry, but this is so exclusively through expectation values, that turn out to be different for each fermionic mode and that are well defined thanks to the loop representation adopted in the scheme of hLQC. These facts, together with ultraviolet properties, guarantee the unitary implementability of the fermionic Heisenberg dynamics. One can construct the associated unitary evolution operator, that is generated by the fermionic Hamiltonian that appears in the Schr\"odinger equation that has been derived. Actually, we have seen that the requirement that this Hamiltonian be well defined on the vacuum is enough to guarantee a finite fermionic backreaction on the FLRW background, without the need of any regularization. On the other hand, the unitarity of the fermionic dynamics translates into a finite production of pairs of particles and antiparticles in the evolved vacuum. 

We have gone one step beyond and employed the still remaining freedom in the determination of the Fock representation of the fermionic degrees of freedom, and their splitting from the FLRW sector of the phase space, to demand an additional feature in the fermionic Hamiltonian, namely, that it become diagonal in terms of the fermionic annihilation and creation variables in the asymptotic region of large wave numbers of the modes, in the sense that it do not contain interactions in that region that produce pairs of particles and antiparticles. We have seen that this condition indeed fixes asymptotically the choice of vacuum state. Furthermore, we have argued in favor of the uniqueness of the vacuum selected by means of this asymptotic Hamiltonian diagonalization when extended to all wave numbers by suitable smoothness conditions. In this respect, we have demonstrated the uniqueness in the case of standard QFT in Minkowski and de Sitter spacetimes, treated as fixed backgrounds, showing in addition that the vacua that are picked out by the diagonalization procedure are the Poincar\'e and the Bunch-Davies vacua, respectively. For more general backgrounds, either of classical or quantum nature, our proposal can potentially serve to attain a well-defined and unique choice of vacuum state with especially good physical and mathematical properties.

Finally, we have commented on the relation between adiabatic states and the vacua selected by our criteria. For iterative constructions of fermionic adiabatic states, all of them turn out to belong to the unitary equivalence class of Fock states that incorporate symmetry invariance and allow for a unitarily implementable dynamics. Besides, states of first or higher adiabatic order belong to the family of Fock states picked out by the additional requirement of a fermionic Hamiltonian with a well-defined action on the dense set of finite particle/antiparticle states, and therefore it is ensured that they lead also to a finite fermionic backreation. As for the issue of Hamiltonian diagonalization, it is an open question whether higher-order adiabatic states give rise to representations in which the vacuum state increasingly approaches our choice in the asymptotic regime of large wave numbers.

\vspace{6pt} 
	
\authorcontributions{Conceptualization: J.C.; B.E.N.; G.A.M.M.; S.P., and J.M.V. Original Draft Preparation: J.C.; B.E.N.; G.A.M.M., and J.M.V.}

\funding{This work was supported by the Spanish MINECO grant number FIS2017-86497-C2-2-P, by the European COST (European Cooperation in Science and Technology) Action number CA16104 GWverse, and by the Mexican grant number DGAPA-UNAM IN113618.}
	
%%%%%%%%%%%%%%%%%%%%%%%%%%%%%%%%%%%%%%%%%%
\acknowledgments{The authors are grateful to M. Mart\'{\i}n-Benito, H. Sahlmann, and T. Thiemann for discussions.}
	
%%%%%%%%%%%%%%%%%%%%%%%%%%%%%%%%%%%%%%%%%%
\conflictsofinterest{The authors declare no conflict of interest.} 
	
%%%%%%%%%%%%%%%%%%%%%%%%%%%%%%%%%%%%%%%%%%
	
%=====================================
\reftitle{References}

\end{document}